\documentclass[11pt,english,onecolumn]{IEEEtran}
\usepackage[T1]{fontenc}
\usepackage[latin9]{inputenc}
\usepackage{float}
\usepackage{units}
\usepackage{amsthm}
\usepackage{amsmath}
\usepackage{amssymb}
\usepackage{graphicx}
\usepackage{setspace}
\PassOptionsToPackage{normalem}{ulem}
\usepackage{ulem}
\makeatletter

\floatstyle{ruled}
\newfloat{algorithm}{tbp}{loa}
\providecommand{\algorithmname}{Algorithm}
\floatname{algorithm}{\protect\algorithmname}

\theoremstyle{plain}
\newtheorem{thm}{\protect\theoremname}
\theoremstyle{definition}
\newtheorem{defn}[thm]{\protect\definitionname}
\theoremstyle{remark}
\newtheorem{rem}[thm]{\protect\remarkname}
\theoremstyle{plain}
\newtheorem{cor}[thm]{\protect\corollaryname}
\theoremstyle{plain}
\newtheorem{lem}[thm]{\protect\lemmaname}
\theoremstyle{plain}
\newtheorem{prop}[thm]{\protect\propositionname}

\usepackage{hyperref}
\usepackage{breqn}
  
\DeclareMathOperator*{\argmax}{arg\,max} \DeclareMathOperator*{\argmin}{arg\,min}
\DeclareMathOperator{\supp}{supp} 
\DeclareMathOperator{\interior}{int} 
\allowdisplaybreaks

\global\long\def\P{\mathbb{P}}
\global\long\def\E{\mathbb{E}}

\global\long\def\RE[#1]{\mathsf{E}_r\left(#1\right)}
\global\long\def\RES[#1]{\mathsf{E}^*_r\left(#1\right)} % starred exponent
\global\long\def\RR{\mathsf{R}}

\newcommand{\EE}[1][]{\mathsf{E}_{e#1}}
\global\long\def\teq{\triangleq}

\global\long\def\s[#1]{\textnormal{\scriptsize #1}}
\global\long\def\st[#1]{\textnormal{\tiny #1}}

\global\long\def\trre[#1,#2]{\overset{{\scriptstyle (#2)}}{#1}} % transition explained with reason
\author{
\authorblockN{Nir Weinberger and Neri Merhav}

\authorblockA{Dept. of Electrical Engineering\\
     	    Technion - Israel Institute of Technology\\
Technion City, Haifa 3200004, Israel
} \\
\authorblockA{\{nirwein@tx, merhav@ee\}.technion.ac.il}
}
\makeatother

\usepackage{babel}
\providecommand{\corollaryname}{Corollary}
\providecommand{\definitionname}{Definition}
\providecommand{\lemmaname}{Lemma}
\providecommand{\propositionname}{Proposition}
\providecommand{\remarkname}{Remark}
\providecommand{\theoremname}{Theorem}

\begin{document}

\title{Optimum Trade-offs Between the Error Exponent and the Excess-Rate
Exponent of Variable-Rate Slepian-Wolf Coding}
\maketitle
\begin{abstract}
We analyze the optimal trade-off between the error exponent and the
excess-rate exponent for variable-rate Slepian-Wolf codes. In particular,
we first derive upper (converse) bounds on the optimal error and excess-rate
exponents, and then lower (achievable) bounds, via a simple class
of variable-rate codes which assign the same rate to all source blocks
of the same type class. Then, using the exponent bounds, we derive
bounds on the optimal rate functions, namely, the minimal rate assigned
to each type class, needed in order to achieve a given target error
exponent. The resulting excess-rate exponent is then evaluated. Iterative
algorithms are provided for the computation of both bounds on the
optimal rate functions and their excess-rate exponents. The resulting
Slepian-Wolf codes bridge between the two extremes of fixed-rate coding,
which has minimal error exponent and maximal excess-rate exponent,
and average-rate coding, which has maximal error exponent and minimal
excess-rate exponent.\end{abstract}

\section{Introduction}

\renewcommand\[{\begin{equation}}
\renewcommand\]{\end{equation}}
\thispagestyle{empty}The problem of distributed encoding of correlated sources has been
studied extensively since the seminal paper of Slepian and Wolf \cite{slepian1973noiseless}.
That paper addresses the case, where a memoryless source $\{(X_{i},Y_{i})\}$
needs to be compressed by two separate encoders, one for $\{X_{i}\}$
and one for $\{Y_{i}\}$. In a nutshell, the most significant result
of \cite{slepian1973noiseless} states that if $\{Y_{i}\}$ is known
at the decoder side, then $\{X_{i}\}$ can be compressed at the rate
of the \emph{conditional} entropy of $\{X_{i}\}$ given $\{Y_{i}\}$.
Since this is the minimal rate even for the case where $\{Y_{i}\}$
is known also to the encoder, then no rate loss is incurred by the
lack of knowledge of $\{Y_{i}\}$ at the encoder. Early research
has focused on asymptotic analysis of the decoding error probability
for the ensemble of random binning codes. Gallager \cite{gallager1976source}
has adapted his well known analysis techniques from random channel
coding \cite[Sections 5.5-5.6]{gallager1968information} to the random
binning ensemble of distributed source coding. Later, it was shown
in \cite{csiszar1980towards} and \cite{csiszar1977new} that the
universal \emph{minimum entropy }decoder also\emph{ }achieves the
same exponent. Expurgated error exponents were given in \cite{csiszar1981graph},
assuming optimal decoding (non-universal). In \cite[Appendix I]{Ahlswede_permutations},
Ahlswede has shown the achievability of random binning and expurgated
bounds via codebooks generated by permutations of good channel codes.
The expurgated exponent analysis was then generalized to coded side
information in \cite{csiszar1982linear} (with linear codes) and \cite{oohama1994universal}. 

In all the above papers, fixed-rate coding was assumed, perhaps because,
as is well known, Slepian-Wolf (SW) coding is, in some sense, analogous
to channel coding (without feedback) \cite{csiszar1981graph,Ahlswede_permutations,chen_linear_codebook,chen_mismatch},
for which variable-rate is usually of no use. More recently, it was
recognized that variable-rate SW coding may have improved performance.
For example, it was shown in \cite{he_redundacny2009,Kuzuoka_redundancy,redundancy_Baron_feedback}
that variable-rate SW codes might have lower redundancy (additional
rate beyond the conditional entropy, for a given error probability).
Other results on variable-rate coding can be found in \cite{kelly2011improved,Kuzuoka_spectrum,merhav2013erasure}.
In another line of work, which is more relevant to this paper, it
was observed that variable-rate coding under an average rate constraint
\cite{ChenUnpublished,Chen2007reliability,chen2008universal} outperforms
fixed-rate coding in terms of error exponents. The intuitive reason
is that the empirical probability mass function (PMF) of the source
tends to concentrate exponentially fast around the true PMF, and so
in order to asymptotically satisfy an average rate constraint $\RR$,
it is only required that the rates allocated to typical source blocks
would have rate less than $\RR$ (see \cite[Thm. 1]{ChenUnpublished}).
Other types, distant from the type of the source, can be assigned
with arbitrary large rates, and thus effectively may be sent uncoded. 

The expected value of the rate is, however, a rather soft requirement,
and it provides a meaningful performance measure only in the case
of many system uses, where the random rate concentrates around its
expected value. Consider, for example, an on-line compression scheme,
in which the codeword is buffered at the encoder before transmitted
\cite{Humblet81generalizationof,Jelink_buffer}. If the instantaneous
codeword length is larger than the buffer size, then the buffer overflows.
If the decoder is aware of this event (using a dedicated feed-forward
channel, e.g.) then this is an \emph{erasure} event, and so, it is
desirable to minimize this probability, while maintaining some given
error probability.\emph{ }In a different case, the buffer length might
be larger than the maximal codeword length, but the buffer is also
used for other purposes (e.g., sending status data). If the data codewords
have priority over all other uses, then it is desirable to minimize
the occasions of blocking other usage of the buffer. This motivates
us to take a somewhat different approach and address a more refined
figure of merit for the rate. Specifically, we will be interested
in the probability that the rate exceeds a certain threshold. While
the aforementioned average-rate coding increases error exponents,
its excess-rate probability is clearly inferior to that of fixed-rate
coding. 

It should be mentioned that in many other problems in information
theory, instead of considering the average value of some cost (which
for SW coding is the rate) more refined figures of merit are imposed,
such as the excess probability or higher moments. Beyond lossless
compression, which was mentioned above, other examples include excess
distortion \cite{Csiszar82_distortion,marton1974}, variable-rate
channel coding with feedback \cite{channel_feedback_variable_rate},
list size of a list decoder \cite{telatar1998exponential}, and estimation
\cite{Neri_modulation_estimation} (see also \cite{Neri_exp_moments}
for intimately related problem of minimizing exponential moments of
a cost function, and many references therein). 

In this paper, we systematically analyze the trade-off between excess-rate
exponent and the error exponent. Based on the analogy of SW and channel
coding, we provide upper (converse) bounds on the error and excess-rate
exponents of a general SW code. Then, we derive lower (achievability)
bounds via a special class of SW codes, which assign the same coding
rate to all source blocks of the same type class. The bounds on error
exponents may be considered as a generalization of the error exponents
of \cite{ChenUnpublished,Chen2007reliability,chen2008universal} to
the case where excess-rate performance is of importance. As will be
seen, this requires a joint treatment of \emph{all} possible types
of the source at the same time, and not just the type of the source,
as in average-rate coding. Both bounds will initially be expressed
via fixed-composition reliability functions of channel codes (to be
defined in the sequel), and only afterwards, specific known bounds
(random coding, expurgated and sphere packing) on the reliability
functions will be applied. This links the question whether assigning
equal rates to source blocks of the same type class is asymptotically
optimal, to the unsettled gap between the infimum and supremum reliability
functions (to be also defined in the sequel) \cite[Problem 10.7]{csiszar2011information}.
Whenever it can be verified that no gap exists, then assigning equal
rates to types is optimal. However, similarly as in channel coding,
above the critical rate, where the reliability function is known exactly,
the upper and lower bounds of SW exponents coincide for small error
exponents, and then assigning equal rates to types class is optimal.
Next, for every type class, bounds on the minimal encoding rate, required
to meet a prescribed value of error exponent, will be found, and corresponding
bounds on the resulting excess-rate performance of the system will
be derived. Since the computation of both the rate for a given type,
and the excess-rate exponent, lead to optimization problems that lack
closed-form solutions, we will provide explicit iterative algorithms
that converge to the optimal solutions. 

The outline of the remaining part of the paper is as follows. In Section
\ref{sec:Problem-Formulation}, we establish notation conventions
and formulate SW codes. We also formulate channel codes and provide
background of known results, which are useful for the analysis of
SW codes. In Section \ref{sec:Exponents}, we derive upper and lower
bounds on the error exponent and excess-rate exponent of general SW
codes, and discuss the trade-off between the two exponents. Then,
in Section \ref{sec:Optimal rate functions}, we characterize the
optimal rate allocation (in a sense that will be made precise), under
an error exponent constraint, and in Section \ref{sec:Excess-Rate-Performance},
we analyze the resulting excess-rate exponent. In Section \ref{sec:Computational algorithms},
we discuss computational aspects of the bounds on the optimal rate
allocation, as well as the bounds on the optimal excess-rate exponent.
Section \ref{sec:A-Numerical-Example} demonstrates the results via
a numerical example, and Section \ref{sec:Summary} summarizes the
paper, along with directions for further research. Almost all proofs
are deferred to Appendix \ref{sec:Proofs}. In Appendix \ref{sec:The-Fixed-Composition},
we provide several general results on the reliability function of
channel coding, which are required in order to fully understand the
proofs in Appendix \ref{sec:Proofs}. In Appendix \ref{sec:Tightness-of-the-Random-Binning}
and Appendix \ref{sec: wakly correlated sources}, we provide some
side results, and Appendix \ref{sec: Useful Lemmas} we provide some
useful Lemmas.

\section{\label{sec:Problem-Formulation}Problem Formulation }

\subsection{Notation Conventions}

Throughout the paper, random variables will be denoted by capital
letters, specific values they may take will be denoted by the corresponding
lower case letters, and their alphabets will be denoted by calligraphic
letters. Random vectors and their realizations will be denoted, respectively,
by capital letters and the corresponding lower case letters, both
in the bold face font. Their alphabets will be superscripted by their
dimensions. For example, the random vector $\mathbf{X}=(X_{1},\ldots,X_{n})$,
($n$ positive integer) may take a specific vector value $\mathbf{x}=(x_{1},\ldots,x_{n})$
in ${\cal X}^{n}$, the $n$th order Cartesian power of ${\cal X}$,
which is the alphabet of each component of this vector. For any given
vector $\mathbf{x}$ and set of indices ${\cal I}\teq\{i_{1},\ldots,i_{I}\}$
we will denote $\mathbf{x}({\cal I})=(x_{i_{1}},\ldots,x_{i_{I}})$,
and for $1\leq i\leq j\leq n$ we will denote $\{i:j\}\teq\{i,i+1,\ldots,j\}$. 

The source to be compressed will be denoted by the letter $P$, subscripted
by the names of the relevant random variables/vectors and their conditionings,
if applicable. We will follow the standard notation conventions, e.g.,
$P_{X}(x)$ will denote the $X$-marginal of $P$, $P_{Y|X}(y|x)$
will denote the conditional distribution of $Y$ given $X$, $P_{XY}(x,y)$
will denote the joint distribution, and so on. The arguments will
be omitted when we address the entire PMF, e.g., $P_{X}$, $P_{Y|X}$
and $P_{XY}$. Similarly, generic sources will be denoted by $Q$,
$\tilde{Q}$, $Q^{*}$, and in other forms, again subscripted by the
relevant random variables/vectors/conditionings. The joint distribution
induced by a PMF $Q_{X}$ and conditional PMF $Q_{Y|X}$ will be denoted
by $Q_{X}\times Q_{Y|X}$, and its $Y$-marginal will be denoted by
$(Q_{X}\times Q_{Y|X})_{Y}$, or simply by $Q_{Y}$ when understood
from context. An exceptional case will be the `hat' notation. For
this notation, $\hat{Q}_{\mathbf{x}}$ will denote the empirical distribution
of a vector $\mathbf{x}\in{\cal {\cal X}}^{n}$, i.e., the vector
of relative frequencies $\hat{Q}_{\mathbf{x}}(x)$ of each symbol
$x\in{\cal X}$ in $\mathbf{x}$. The type class of $\mathbf{x}\in{\cal X}^{n}$,
which will be denoted by ${\cal T}_{n}(\hat{Q}_{\mathbf{x}})$, is
the set of all vectors $\mathbf{x}'$ with $\hat{Q}_{\mathbf{x}'}=\hat{Q}_{\mathbf{x}}$.
The set of all type classes of vectors of length $n$ over ${\cal X}$
will be denoted by ${\cal P}_{n}({\cal X})$, and the set of all possible
types over ${\cal X}$ will be denoted by ${\cal P}({\cal X})\teq\bigcup_{n=1}^{\infty}{\cal P}_{n}({\cal X})$.
Similar notation for type classes will also be used for generic types
$Q_{X}\in{\cal P}({\cal X})$, i.e. ${\cal T}_{n}(Q_{X})$ will denote
the set of all vectors $\mathbf{x}$ with $\hat{Q}_{\mathbf{x}}=Q_{X}$.
In the same manner, the empirical distribution of a pair of vectors
$(\mathbf{x},\mathbf{y})$ will be denoted by $\hat{Q}_{\mathbf{x}\mathbf{y}}$
and the joint type class will be denoted by ${\cal T}_{n}(\hat{Q}_{\mathbf{x}\mathbf{y}})$.
The joint type classes over the Cartesian product alphabet ${\cal X}\times{\cal Y}$
will be denoted by ${\cal P}_{n}({\cal X}\times{\cal Y})$, and ${\cal P}({\cal X}\times{\cal Y})\teq\bigcup_{n=1}^{\infty}{\cal P}_{n}({\cal X}\times{\cal Y})$.
For a joint type $Q_{XY}\in{\cal P}({\cal X}\times{\cal Y})$, ${\cal T}_{n}(Q_{XY})$
will denote the set of all pair of vectors $(\mathbf{x},\mathbf{y})$
with $\hat{Q}_{\mathbf{x}\mathbf{y}}=Q_{XY}$. The empirical conditional
distribution induced by $(\mathbf{x},\mathbf{y})$ will be denoted
by $\hat{Q}_{\mathbf{x}|\mathbf{y}}$, and the conditional type class,
namely, the set $\{\mathbf{x}':\hat{Q}_{\mathbf{x}'\mathbf{y}}=\hat{Q}_{\mathbf{x}\mathbf{y}}\}$,
will be denoted by ${\cal T}_{n}(\hat{Q}_{\mathbf{x}|\mathbf{y}})$,
or more generally ${\cal T}_{n}(Q_{X|Y})$ for a generic empirical
conditional probability. The probability simplex for ${\cal X}$ will
be denoted by ${\cal Q}({\cal X})$, and the simplex for the alphabet
${\cal X}\times{\cal Y}$ will be denoted by ${\cal Q}({\cal X}\times{\cal Y})$. 

The support of a PMF $Q_{X}$ will be denoted by $\supp(Q_{X})\triangleq\left\{ x:Q_{X}(x)\neq0\right\} \subseteq{\cal X}$.
For two PMFs $P_{X},Q_{X}$ over the same finite alphabet ${\cal X}$,
we will denote the variation distance (${\cal L}_{1}$ norm) by
\[
||P_{X}-Q_{X}||\teq\sum_{x\in{\cal X}}|P_{X}(x)-Q_{X}(x)|.
\]
When optimizing a function of a distribution $Q_{X}$ over the entire
probability simplex ${\cal Q}({\cal X})$, the explicit display of
the constraint will be omitted. For example, for a function $f(Q)$,
we will write $\min_{Q}f(Q)$ instead of $\min_{Q\in{\cal Q}({\cal X})}f(Q)$.
The same will hold for optimization of a function of a distribution
$Q_{XY}$ over the probability simplex ${\cal Q}({\cal X}\times{\cal Y})$. 

The expectation operator with respect to (w.r.t.) a given distribution,
e.g. $Q_{XY}$, will be denoted by $\E_{Q_{XY}}[\cdot]$ where, again,
the subscript will be omitted if the underlying probability distribution
is clear from the context. The entropy of a given distribution, e.g.
$Q_{X}$, will be denoted by $H(Q_{X})$, and the binary entropy function
will be denoted by $h_{B}(q)$ for $0\leq q\leq1$. The average conditional
entropy of $Q_{Y|X}$ w.r.t. $Q_{X}$ will be denoted by $H(Q_{Y|X}|Q_{X})\teq\sum_{x\in{\cal X}}Q_{X}(x)H(Q_{Y|X}(\cdot|x))$,
and the mutual information of a joint distribution $Q_{XY}$ will
be denoted by $I(Q_{XY})$. The information divergence between two
distributions, e.g. $P_{XY}$ and $Q_{XY}$, will be denoted by $D(P_{XY}||Q_{XY})$
and the average divergence between $Q_{Y|X}$ and $P_{Y|X}$ w.r.t.
$Q_{X}$ will be denoted by $D(Q_{Y|X}||P_{Y|X}|Q_{X})\teq\sum_{x\in{\cal X}}Q_{X}(x)D(Q_{Y|X}(\cdot|x)||P_{Y|X}(\cdot|x))$.
In all the information measures above, the PMF may also be an empirical
PMF, for example, $H(\hat{Q}_{\mathbf{x}})$, $D(\hat{Q}_{\mathbf{y}|\mathbf{x}}||P_{Y|X})$
and so on. 

We will denote the Hamming distance of two vectors $\mathbf{x}\in{\cal X}^{n}$
and $\mathbf{z}\in{\cal X}^{n}$ by $d_{\s[H]}(\mathbf{x},\mathbf{z})$.
The length of a string $b$ will be denoted by $|b|$, and the concatenation
of the strings $b_{1},b_{2},\ldots$ will be denoted by $(b_{1},b_{2},\ldots)$.
For a set ${\cal A}$, we will denote its complement by ${\cal A}^{c}$,
its closure by ${\cal \overline{A}}$, its interior by $\interior({\cal A})$,
and its boundary by $\partial{\cal A}$. If the set ${\cal A}$ is
finite, we will denote its cardinality by $|{\cal A}|$. The probability
of the event ${\cal A}$ will be denoted by $\P({\cal A})$, and $\mathbb{I}(A)$
will denote the indicator function of this event. 

For two positive sequences, $\{a_{n}\}$ and $\{b_{n}\}$ the notation
$a_{n}\doteq b_{n}$ will mean asymptotic equivalence in the exponential
scale, that is, $\lim_{n\to\infty}\frac{1}{n}\log(\frac{a_{n}}{b_{n}})=0$.
Similarly, $a_{n}\dot{\leq}b_{n}$ will mean $\limsup_{n\to\infty}\frac{1}{n}\log(\frac{a_{n}}{b_{n}})\leq0$,
and so on. The function $[t]_{+}$ will be defined as $\max\{t,0\}$,
and $\left\lceil t\right\rceil $ will denote the ceiling function.
For two integers, $a,b$,  we denote by $a\mod b$ the modulo of $a$
w.r.t. $b$. Unless otherwise stated, logarithms and exponents will
be understood to be taken to the natural base.

\subsection{Slepian-Wolf Coding\label{sub:System Model} }

Let $\{(X_{i},Y_{i})\}_{i=1}^{n}$ be $n$ independent copies of a
pair of random variables $(X,Y)$. We assume that $X\in{\cal X}$
and $Y\in{\cal Y}$, where ${\cal X}$ and ${\cal Y}$ are finite
alphabets, are distributed according to $P_{XY}(x,y)=\P(X=x,Y=y)$.
It is assumed that $\supp(P_{X})={\cal X}$ and that $\supp(P_{Y})={\cal Y}$,
otherwise, remove the irrelevant letters from their alphabet. We say
that the conditional distribution $P_{Y|X}$ is not \emph{noiseless}
if there exists a pair of input letters $x,x'\in{\cal X}$ and an
output letter $y\in{\cal Y}$ such that $P_{Y|X}(y|x)\cdot P_{Y|X}(y|x')>0$,
and assume this property for $P_{XY}$, and so, $H(X|Y)>0$. 

A SW code ${\cal S}_{n}$ for sequences of length $n$ is defined
by a prefix code with encoder 
\[
s_{n}:{\cal X}^{n}\to\left\{ 0,1\right\} ^{*}
\]
and a decoder

\[
\sigma_{n}:\left\{ 0,1\right\} ^{*}\times{\cal Y}^{n}\to{\cal X}^{n},
\]
where $\left\{ 0,1\right\} ^{*}$ is the set of all finite length
binary strings. The encoder maps a source block $\mathbf{x}$ into
a binary string $s_{n}(\mathbf{x})\in\left\{ 0,1\right\} ^{*}$, where
for $b\in\left\{ 0,1\right\} ^{*}$, the inverse image of $s_{n}$
is defined as
\[
s_{n}^{-1}(b)\teq\left\{ \mathbf{x}\in{\cal X}^{n}:s_{n}(\mathbf{x})=b\right\} 
\]
and it is called a \emph{bin}. The decoder $\sigma_{n}$, which observes
$b=s_{n}(\mathbf{x})$ and the side information\textbf{ $\mathbf{y}$},
has to decide on the particular source block $\mathbf{x}\in s_{n}^{-1}(b)$
to obtain a decoded source block $\mathbf{\hat{x}}\teq\sigma_{n}(s_{n}(\mathbf{x}),\mathbf{y})$.
A sequence of SW codes $\{{\cal S}_{n}\}_{n\geq1}$, indexed by the
block length $n$ will be denoted by ${\cal S}$. 

The error probability, for a given code ${\cal S}_{n}=\{s_{n},\sigma_{n}\}$,
is denoted by $p_{e}({\cal S}_{n})\teq\P(\hat{\mathbf{X}}\neq\mathbf{X})$.
The \emph{infimum error exponent} achieved for a sequence of codes
${\cal S}$ is defined as
\begin{equation}
{\cal E}_{e}^{-}({\cal S})\triangleq\liminf_{n\to\infty}-\frac{1}{n}\log p_{e}({\cal S}_{n})\label{eq: error exponent definition given code}
\end{equation}
and the \emph{supremum error exponent} achieved is defined as 
\begin{equation}
{\cal E}_{e}^{+}({\cal S})\triangleq\limsup_{n\to\infty}-\frac{1}{n}\log p_{e}({\cal S}_{n}).\label{eq: error exponent definition given code sup}
\end{equation}
While clearly, ${\cal E}_{e}^{-}({\cal S})\leq{\cal E}_{e}^{+}({\cal S})$,
it is guaranteed that $p_{e}({\cal S}_{n})\dot{\geq}\exp\left[-n{\cal E}_{e}^{-}({\cal S})\right]$
for all sufficiently large block lengths, while $p_{e}({\cal S}_{n})\doteq\exp\left[-n{\cal E}_{e}^{-}({\cal S})\right]$
may hold only for some sub-sequence of block lengths. Thus, from a
practical perspective, ${\cal E}_{e}^{-}({\cal S})$ is more robust
to the choice of block length. 

For a given $Q_{X}\in{\cal P}({\cal X})$, we define, with a slight
abuse of notation, the \emph{conditional infimum error exponent} as
\[
{\cal E}_{e}^{-}({\cal S},Q_{X})\teq\liminf_{n\to\infty}-\frac{1}{n}\log\P(\hat{\mathbf{X}}\neq\mathbf{X}|\mathbf{X}\in{\cal T}_{n}(Q_{X}))
\]
where we use the convention that $\P(\hat{\mathbf{X}}\neq\mathbf{X}|\mathbf{X}\in{\cal T}_{n}(Q_{X}))\teq0$
if ${\cal T}_{n}(Q_{X})$ is empty. ${\cal E}_{e}^{+}({\cal S},Q_{X})$
is defined analogously. 

The coding rate of $\mathbf{x}\in{\cal X}^{n}$ is defined as $r(\mathbf{x})\triangleq\frac{\left|s_{n}(\mathbf{x})\right|}{n\cdot\log_{2}e}$.
A SW code is termed a \emph{fixed-rate} code of rate $\RR_{0}$ if
$r(\mathbf{x})=\RR_{0}$ for all $\mathbf{x}\in{\cal X}^{n}$. Otherwise
it is called a \emph{variable-rate} code\emph{,} and has an \emph{average
rate} $\E[r(\mathbf{X})]$. We define the \emph{conditional rate}
of  $Q_{X}\in{\cal P}({\cal X})$ as 
\begin{equation}
\overline{R}(Q_{X};{\cal S})\teq\limsup_{n\to\infty}\E[r(\mathbf{X})|\mathbf{X}\in{\cal T}_{n}(Q_{X})]\label{eq: conditional rates def}
\end{equation}
where $\E[r(\mathbf{X})|\mathbf{X}\in{\cal T}_{n}(Q_{X})]\teq0$ if
${\cal T}_{n}(Q_{X})$ is empty. Since $r(\mathbf{x})=\log{\cal X}$
allows the encoding of $\mathbf{x}$ with zero error, it will be assumed
that $\overline{R}(Q_{X};{\cal S})$ is finite. For a given target
rate $\mathsf{R}$, the excess-rate probability, of a code ${\cal S}_{n}$,
is denoted by $p_{r}({\cal S}_{n},\RR)\teq\P\left\{ r(\mathbf{X})\geq\RR\right\} $,
and the \emph{excess-rate exponent function, }achieved for a sequence
of codes ${\cal S}$, is defined as%
\footnote{In the definition of achievable excess-rate exponent, we use only
limit inferior. It should be observed that for an operational meaning,
the error exponent and excess-rate exponent should be \emph{jointly}
approached by a sub-sequence of block lengths. When the limit inferior
is used for both the definition of the error exponent and the definition
of the excess-rate exponent, any sufficiently large block length will
approach the asymptotic limit of both exponents. When one of the exponents
is defined as limit inferior, and the other is defined as limit superior,
then there exists a \emph{sub-sequence} of block lengths with the
required limits, but these block lengths may be arbitrarily distant.
Finally, there is no operational meaning to defining both exponents
with limit superior, since the two sub-sequences of block lengths
which achieve each of the exponents might be completely disjoint.%
}
\begin{equation}
{\cal E}_{r}({\cal S},\RR)\triangleq\liminf_{n\to\infty}-\frac{1}{n}\log p_{r}({\cal S}_{n},\RR).\label{eq: excess rate exponent function definition}
\end{equation}
For a given $Q_{X}\in{\cal P}({\cal X})$, we define, with a slight
abuse of notation, the \emph{conditional excess-rate exponent} as
\[
{\cal E}_{r}({\cal S},\RR,Q_{X})\triangleq\liminf_{n\to\infty}-\frac{1}{n}\log\P(r(\mathbf{X})\geq\RR|\mathbf{X}\in{\cal T}_{n}(Q_{X}))
\]
where $\P(r(\mathbf{X})\geq\RR|\mathbf{X}\in{\cal T}_{n}(Q_{X}))\teq0$
if ${\cal T}_{n}(Q_{X})$ is empty.

In the remaining part of the paper, we will mainly be interested in
the following sub-class of variable-rate SW codes. 
\begin{defn}
A SW code ${\cal S}_{n}$ is termed \emph{type-dependent, variable-rate}
code,\emph{ }if $r(\mathbf{x})$ depends on $\mathbf{x}$ only via
its type (empirical PMF). Namely, $\hat{Q}_{\mathbf{x}}=\hat{Q}_{\mathbf{\tilde{\mathbf{x}}}}$
implies $r(\mathbf{x})=r(\tilde{\mathbf{x}})$. Any finite function
$\rho(\cdot):{\cal Q}({\cal X})\to\mathbb{R}^{+}$ is called a \emph{rate
function}. A rate function is termed\emph{ regular} if there exists
a constant $d>0$ and a set \textbf{${\cal V}\teq\{Q_{X}\in{\cal Q}({\cal X}):D(Q_{X}||P_{X})<d\}$,
}such that $\rho(\cdot)$ is continuous in ${\cal V}$, and equals
some constant $\RR_{0}$ for $Q_{X}\in{\cal V}^{c}$. 

The main objective of the paper is to derive the optimal trade-off
between the error exponent and the excess rate exponent, i.e., to
find the maximal achievable excess rate exponent, under a constraint
on the error exponent. The subclass of type-dependent, variable-rate
SW codes will be shown to achieve the optimal trade-off in a certain
range of exponents, and the question of their optimality in other
ranges will be discussed. 
\end{defn}

\subsection{Channel Coding\label{sub:Channel-Coding defs}}

In SW coding, the collection of source words that belong to the same
bin, can be considered a channel code, and given the bin index, the
SW decoder acts just as a channel decoder (with the exception that
the prior probabilities of the source blocks in the bin may not necessarily
be uniform). Thus, error exponents of SW codes are intimately related
to error exponents of channel codes (e.g. \cite{csiszar1981graph,Chen2007reliability}).
Accordingly, we next define a few terms associated with channel codes,
which will be needed in the sequel. 

Consider a discrete memoryless channel $\{W(y|x),~x\in{\cal X},~y\in{\cal Y}\}$
with input alphabet ${\cal X}$ and output alphabet ${\cal Y}$, which
are both finite. A channel code ${\cal C}_{n}$ of block length $n$
is defined by an encoder 
\[
f_{n}:\left\{ 1,\ldots,\bigl\lceil e^{n\RR}\bigr\rceil\right\} \to{\cal X}^{n}
\]
and a decoder

\[
\varphi_{n}:{\cal Y}^{n}\to\{1,\ldots,\bigl\lceil e^{n\RR}\bigr\rceil\},
\]
where $\RR$ is the rate of the code. We say that the channel code
is a \emph{fixed composition} code, if all codewords $\{f_{n}(m)\},\:1\leq m\leq\bigl\lceil e^{n\RR}\bigr\rceil,$
belong to a single type class ${\cal T}_{n}(Q_{X})$. A sequence of
channel codes will be denoted by ${\cal C}=\{{\cal C}_{n}\}_{n\geq1}$.
The error probability for a given channel code ${\cal C}_{n}=\{f_{n},\varphi_{n}\}$
is denoted by $p_{e}({\cal C}_{n})\teq\P(\varphi_{n}(f_{n}(M))\neq M)$,
where $M$ is a uniform random variable over the set $\left\{ 1,\ldots,\bigl\lceil e^{n\RR}\bigr\rceil\right\} $.
The infimum error exponent, achieved for a given sequence of channel
codes ${\cal C}$ is defined as
\begin{equation}
{\cal E}_{c}^{-}({\cal C})\triangleq\liminf_{n\to\infty}-\frac{1}{n}\log p_{e}({\cal C}_{n})\label{eq:channel error exponent inf}
\end{equation}
and the supremum error exponent is defined as

\begin{equation}
{\cal E}_{c}^{+}({\cal C})\triangleq\limsup_{n\to\infty}-\frac{1}{n}\log p_{e}({\cal C}_{n}).\label{eq:channel error exponent sup}
\end{equation}

A number $\mathsf{E}_{e}>0$ is an achievable infimum (supremum) error
exponent for the type $Q_{X}\in{\cal Q}({\cal X})$ and the channel
$W$ at rate $\RR$, if for any $\delta>0$ there exists a sequence
of types $Q_{X}^{(n)}\in{\cal P}_{n}({\cal X})$ such that $Q_{X}^{(n)}\to Q_{X}$
and there exists a sequence of fixed composition channel codes ${\cal C}_{n}\subseteq{\cal T}_{n}(Q_{X}^{(n)})$
with 
\begin{equation}
\liminf_{n\to\infty}\frac{\log|{\cal C}_{n}|}{n}\geq\RR-\delta\label{eq: channel reliability function rate definition}
\end{equation}
and ${\cal E}_{c}^{-}({\cal C})\geq\mathsf{E}_{e}-\delta$ (respectively,
${\cal E}_{c}^{+}({\cal C})\geq\mathsf{E}_{e}-\delta$). For a given
rate $\RR$, a type $Q_{X}$, and the channel $W$, we let $\underline{E}_{e}^{*}(\RR,Q_{X},W)$
($\overline{E}_{e}^{*}(\RR,Q_{X},W)$) be the largest achievable infimum
(respectively, supremum) error exponent over all possible sequences
of codes ${\cal C}$ for the type $Q_{X}$. The functions $\underline{E}_{e}^{*}(\RR,Q_{X},W)$
and $\overline{E}_{e}^{*}(\RR,Q_{X},W)$ may be interpreted as \emph{infimum/supremum}
\emph{fixed-composition reliability functions} of the channel $W$,
when the type of the codewords must tend to $Q_{X}$. 

We define by $C_{0}^{-}(Q_{X},W)$ (respectively, $C_{0}^{+}(Q_{X},W)$)
the maximum of all rates such that $\underline{E}_{e}^{*}(\RR,Q_{X},W)$
(respectively, $\overline{E}_{e}^{*}(\RR,Q_{X},W)$) is infinite,
which can be regarded as the \emph{zero-error capacity} of the channel
$W$ of fixed composition codes with codebook types which tends to
$Q_{X}$. Fekete's Lemma \cite[Lemma 11.2]{csiszar2011information}
implies that $C_{0}^{-}(Q_{X},W)=C_{0}^{+}(Q_{X},W)$ and thus we
will denote henceforth both quantities by $C_{0}(Q_{X},W)$. Notice
that when $W$ is not noiseless and $Q_{X}\in\interior{\cal Q}({\cal X})$,
we have $C_{0}(Q_{X},W)=0$, namely, $C_{0}(Q_{X},W)$ may be strictly
positive only for types which belong to $\partial{\cal Q}({\cal X})$.
For any $Q_{X}\in{\cal Q}({\cal X})$, we define 
\begin{equation}
E_{0}^{-}(Q_{X},W)\teq\lim_{\RR\downarrow C_{0}(Q_{X},W)}\underline{E}_{e}^{*}(\RR,Q_{X},W)\label{eq: maximal inf error exponent channel coding}
\end{equation}
and $E_{0}^{+}(Q_{X},W)$ is defined analogously.

Unfortunately, it is a long-standing open problem to find the exact
values of $\underline{E}_{e}^{*}(\RR,Q_{X},W)$ and $\overline{E}_{e}^{*}(\RR,Q_{X},W)$
for an arbitrary rate $\RR\in[0,I(Q_{X}\times W)]$, and it is not
even known if $\underline{E}_{e}^{*}(\RR,Q_{X},W)=\overline{E}_{e}^{*}(\RR,Q_{X},W)$
\cite[Problem 10.7]{csiszar2011information}. However, the following
bounds on the fixed composition reliability function are well known
when $Q_{X}\in{\cal P}_{n}({\cal X})$. The random coding bound \cite[Theorem 10.2]{csiszar2011information}
is a lower bound on the infimum fixed-composition reliability function,
given by 
\begin{equation}
\underline{E}_{e}^{*}(\RR,Q{}_{X},W)\geq E_{\s[rc]}(\RR,Q{}_{X},W)\teq\min_{Q_{Y|X}}\left\{ D(Q_{Y|X}||W|Q_{X})+\left[I(Q_{X}\times Q_{Y|X})-R\right]_{+}\right\} .\label{eq: channel random coding bound}
\end{equation}
Similarly, the expurgated lower bound \cite[Problem 10.18]{csiszar2011information}
is given by

\begin{equation}
\underline{E}_{e}^{*}(\RR,Q{}_{X},W)\geq E_{\s[ex]}(\RR,Q{}_{X},W)\teq\min_{Q_{X\tilde{X}}:\: Q_{\tilde{X}}=Q_{X},\: I(Q_{X\tilde{X}})\leq\RR}\left\{ B(Q_{X\tilde{X}},W)+I(Q_{X\tilde{X}})-\RR\right\} \label{eq: channel expurgated bound}
\end{equation}
where 
\begin{equation}
B(Q_{X\tilde{X}},W)\teq\E_{Q_{X\tilde{X}}}[d_{W}(X,\tilde{X})]\label{eq: Bhat distance}
\end{equation}
 is the \emph{Bhattacharyya distance,} namely 
\[
d_{W}(x,\tilde{x})\teq-\log\sum_{y\in{\cal Y}}\sqrt{W(y|x)W(y|\tilde{x})}.
\]
The sphere packing exponent \cite[Theorem 10.3]{csiszar2011information}
is an upper bound on the supremum fixed-composition reliability function
and given by 
\begin{equation}
\overline{E}_{e}^{*}(\RR,Q{}_{X},W)\leq E_{\s[sp]}(\RR,Q{}_{X},W)\teq\min_{Q_{Y|X}:\: I(Q_{X}\times Q_{Y|X})\leq\RR}D(Q_{Y|X}||W|Q_{X})\label{eq: channel sphere packing bound}
\end{equation}
which is valid for rates except $R_{\infty}(Q_{X},W)$, defined as
the infimum of all rates such that $E_{\s[sp]}(\RR,Q{}_{X},W)<\infty$.
An improved upper bound on the supremum fixed-composition reliability
for low rates, is the straight line bound \cite[Problem 10.30]{csiszar2011information},
\cite[Section 3.8]{viterbi2009principles}. This bound is obtained
by connecting the expurgated bound at $\RR=0$, which is known to
be tight \cite[Problem 10.21]{csiszar2011information}, with the sphere
packing bound. Since specifying our results on the optimal rate function
(Section \ref{sec:Optimal rate functions}) and excess rate exponents
(Section \ref{sec:Excess-Rate-Performance}) of SW codes is fairly
simple and does not contribute to intuition, we will not discuss this
bound henceforth. On the same note, since $E_{\s[rc]}(\RR,Q{}_{X},W)$
and $E_{\s[ex]}(\RR,Q{}_{X},W)$ are not concave in $Q_{X}$, in general,
then the error performance for a given fixed composition of type $Q_{X}$
can be improved by a certain time-sharing structure in the random
coding mechanism. According to this structure, for each randomly selected
codeword, the block length is optimally subdivided into codeword segments
that are randomly drawn from optimally chosen types, whose weighted
average (with weights proportional to the segment lengths) conforms
with the given $Q_{X}$. At zero-rate, the resulting expurgated error
exponent is given by the upper concave envelope (UCE) of $E_{\s[ex]}(0,Q{}_{X},W)$
\cite[Problem 10.22]{csiszar2011information}. Nonetheless, in many
cases (see discussion in \cite{Jelinek_expurgated} and \cite[Section 2]{merhav2014zero_rate}),
$E_{\s[ex]}(0,Q{}_{X},W)$ is already concave, and no improved bound
can be obtained by taking the UCE (e.g., when $|{\cal X}|=2$, $E_{\s[ex]}(0,Q{}_{X},W)$
is concave). In ordinary channel coding (without input constraints)
this improvement is usually not discussed, because the time-sharing
structure does increase the maximum of $E_{\s[rc]}(\RR,Q{}_{X},W)$
and $E_{\s[ex]}(\RR,Q{}_{X},W)$ over $Q_{X}$. However, for the utilization
of channel codes as components of a SW code the value of $E_{\s[rc]}(\RR,Q{}_{X},W)$
and $E_{\s[ex]}(\RR,Q{}_{X},W)$ at any given $Q_{X}$ is of interest.
Nonetheless, for the sake of simplicity of the exposition, throughout
the sequel, we will not include this time-sharing mechanism in our
discussions and derivations, although their inclusion is conceptually
not difficult. 

In \cite[Proposition 4]{ChenUnpublished} these bounds were shown
to hold for any $Q_{X}\in{\cal Q}({\cal X})$ from continuity arguments.
We will use the convention that all the above bounds are formally
infinite for negative rates. It can be deduced from the above bounds
\cite[Corollary 10.4]{csiszar2011information}, that there exists
a \emph{critical rate }$R_{\s[cr]}(Q_{X},W)$ such that for $\RR\in[R_{\s[cr]}(Q_{X},W),I(Q_{X}\times W)]$,
$E_{\s[rb]}(\RR,Q{}_{X},W)=E_{\s[sp]}(\RR,Q{}_{X},W)$, and consequently
$\underline{E}_{e}^{*}(\RR,Q_{X},W)=\overline{E}_{e}^{*}(\RR,Q_{X},W)$. 

In Appendix \ref{sec:The-Fixed-Composition}, we discuss the fixed-composition
reliability functions, and obtain some of their properties, which
are required for the proof of the theorems in Section \ref{sec:Exponents}.

\section{Error and Excess-Rate Exponents\label{sec:Exponents}}

For a SW code, a trade-off exists between the error exponent $\EE$,
the target rate $\RR$, and the excess-rate exponent $\mathsf{E}_{r}$.
In subsection \ref{sub:Previous-Work SW codes}, we discuss informally
some known results regarding error exponents of fixed-rate SW codes
and variable-rate SW codes, under an average rate constraint. We also
discuss the excess-rate exponent function that they achieve.

Then, in subsection \ref{sub:Bounds-on SW codes}, upper bounds (converse
results) will be found on the supremum error and excess-rate exponents,
and lower bounds (achievability results) on the infimum error and
excess-rate exponents will be derived for type-dependent, variable-rate
SW codes. It will be apparent that the gap between the lower and upper
bounds is only due to the gap which exists, in general, between, the
infimum and supremum channel reliability functions. Thus, whenever
the channel reliability functions are equal, type-dependent, variable-rate
SW codes are \emph{optimal}. For this reason, as well as their intuitive
plausibility, we will later analyze optimal (in a sense that will
be made precise) type-dependent, variable-rate SW codes.

\subsection{Previous Work \label{sub:Previous-Work SW codes}}

For a sequence of fixed-rate SW codes ${\cal S}$ at rate $\RR_{0}$,
the excess-rate exponent function is trivially given by 
\begin{equation}
{\cal E}_{r}({\cal S},\RR)=\begin{cases}
0, & \RR\leq\RR_{0}\\
\infty, & \mbox{otherwise}.
\end{cases}\label{eq: excess rate coding fixed rate}
\end{equation}
Evidently, this function bears a strong dichotomy between rates below
and above $\RR_{0}$. Bounds on the error exponents for fixed-rate
SW coding were derived in \cite[Theorems 2 and 3]{csiszar1980towards},
\cite[Theorem 1]{Ahlswede_permutations}, \cite[Theorem 2]{csiszar1981graph}.
The analysis is essentially based on considering each type class of
the source separately. Loosely speaking, for any given $Q_{X}\in{\cal P}_{n}({\cal X})$,
there exists a partition of the type class ${\cal T}_{n}(Q_{X})$
into bins, such that every bin corresponds to a channel code of rate
$H(Q_{X})-\RR_{0}$, which achieves an error exponent function $\underline{E}_{e}^{*}(H(Q_{X})-\RR_{0},Q_{X},P_{Y|X})$.
Since $\P({\cal T}_{n}(Q_{X}))\doteq\exp\left[-nD(Q_{X}||P_{X})\right]$,
and the number of types increases only polynomially, the error exponent
is given by%
\footnote{We will prove \eqref{eq: error exponent fixed rate} rigorously in
Theorem \ref{thm: achieavbility with type dependent}.%
}
\begin{equation}
{\cal E}_{e}^{-}({\cal S})\geq\min_{Q_{X}}\left\{ D(Q_{X}||P_{X})+\underline{E}_{e}^{*}(H(Q_{X})-\RR_{0},Q_{X},P_{Y|X})\right\} .\label{eq: error exponent fixed rate}
\end{equation}
It was observed in \cite{ChenUnpublished,Chen2007reliability} that
sequences of variable-rate SW codes ${\cal S}$ may have better error
exponents than those of fixed-rate SW codes, when an average rate
constraint is imposed, i.e. $\E[r(\mathbf{X})]\leq\RR_{0}$. Intuitively,
since asymptotically, the average rate is only determined by the rate
of types $\{Q_{X}\}$ that are `close' (in a sense that was made precise
in \cite[Theorem 1]{ChenUnpublished} and \cite[Theorems 1 and 2]{Chen2007reliability})
to the source $P_{X}$, one can allocate large rates to a-typical
source blocks, transmit them uncoded using $\log_{2}|{\cal X}|$ bits,
and the decoder will have zero-error for source blocks from these
type classes. The result is that for such variable-rate SW codes,
the supremum error exponent equals the conditional supremum error
exponent of $P_{X}$ at the rate $\RR_{0}$ assigned for source blocks
with type `close' to $P_{X}$, namely 
\begin{equation}
{\cal E}_{e}^{+}({\cal S})=\overline{E}_{e}^{*}(H(P_{X})-\RR_{0},P_{X},P_{Y|X}).\label{eq: error exponent average rate}
\end{equation}
This can be thought of as a generalization of \cite[Theorem 1]{Ahlswede_permutations}
to variable-rate codes under an average-rate constraint. However,
since the probability that $Q_{X}$ would be away from $P_{X}$ decays
with an arbitrary small error exponent, the resulting excess-rate
exponent function is given by ${\cal E}_{r}({\cal S},\RR)=0$ for
$\RR\leq\log|{\cal X}|$, which is inferior to the infinite excess-rate
exponent of fixed-rate coding for $\RR\in(\RR_{0},\log|{\cal X}|)$.
This excess-rate function can be improved, e.g., by coding each of
the source blocks in the `uncoded type classes' with $\log_{2}|{\cal T}_{n}(Q_{X})|\approx nH(Q_{X})$
bits, and obtaining the same error exponent \eqref{eq: error exponent average rate},
and the excess-rate exponent for $\RR\in(\RR_{0},\log|{\cal X}|)$
will be \emph{
\begin{equation}
{\cal E}_{r}({\cal S},\RR)=\begin{cases}
0 & \RR\leq\RR_{0}\\
\min_{H(Q_{X})\geq\RR}D(Q_{X}||P_{X}) & \RR_{0}<\RR\leq\log|{\cal X}|
\end{cases}.\label{eq: excess rate coding average rate improved}
\end{equation}
}While this excess-rate exponent may be positive for $\RR\in(\RR_{0},\log|{\cal X}|)$,
it is nonetheless \emph{finite}, in contrast to fixed-rate coding
\eqref{eq: excess rate coding fixed rate}. In this paper, we will
analyze systematically the trade-off between the error and excess-rate
exponents for variable-rate codes, where the two above cases, i.e.
fixed-rate and variable-rate with average rate constraint, may be
considered as two extremes of this trade-off.

Since in \cite{ChenUnpublished,Chen2007reliability} the focus was
on coding the source type $P_{X}$%
\footnote{If $P_{X}\not\in{\cal P}({\cal X})$ then one can alternatively consider
$P'_{X}\in{\cal P}({\cal X})$ arbitrarily `close' to $P_{X}$. %
}, the essence of \cite[Theorem 1]{ChenUnpublished} and \cite[Theorems 1 and 2]{Chen2007reliability}
is an upper bound and a lower bound for ${\cal E}_{e}^{+}({\cal S},P_{X})$.
Nonetheless, the proofs of these bounds are similar for any given
type $Q_{X}\in{\cal P}({\cal X})$. For the sake of completeness,
and in order to establish this result in the current setting, we include
a proof of the lower bound in Appendix \ref{sec:Proofs}. 
\begin{thm}[{Variation of \cite[Theorem 1]{ChenUnpublished}}]
\label{thm: Error exponent bounds, single type}Let ${\cal S}$ be
an arbitrary sequence of SW codes. Then, for every $Q_{X}\in{\cal P}({\cal X})$
\begin{equation}
{\cal E}_{e}^{+}({\cal S},Q_{X})\leq\overline{E}_{e}^{*}(H(Q_{X})-\overline{R}(Q_{X};{\cal S}),Q_{X},P_{Y|X}).\label{eq: SW conditional error exponent bound sup}
\end{equation}
Also, for any $Q_{X}\in{\cal P}({\cal X})\cap\interior{\cal Q}({\cal X})$
there exists a sequence of type-dependent SW codes ${\cal S}^{*}$
with rates $r^{*}(\mathbf{x})$, such that for any $\delta>0$ and
sufficiently large $n$, we have $r^{*}(\mathbf{x})\leq\overline{R}(Q_{X};{\cal S})+\delta$
for all $\mathbf{x}\in{\cal T}_{n}(Q_{X})$ and 
\begin{equation}
{\cal E}_{e}^{-}({\cal S}^{*},Q_{X})\geq\underline{E}_{e}^{*}(H(Q_{X})-\overline{R}(Q_{X};{\cal S}),Q_{X},P_{Y|X})-\delta.\label{eq: SW conditional error exponent bound inf}
\end{equation}

\end{thm}
In \cite{ChenUnpublished,Chen2007reliability}, the achievability
result actually obtained was 
\[
{\cal E}_{e}^{+}({\cal S},Q_{X})\geq\overline{E}_{e}^{*}(H(Q_{X})-\overline{R}(Q_{X};{\cal S}),Q_{X},P_{Y|X})-\delta,
\]
and \eqref{eq: error exponent average rate} was proved. However,
for the sake of the current setting, the statement of Theorem \ref{thm: Error exponent bounds, single type}
is required, and it based on the additional properties of optimal
channel codes derived in Lemma \ref{lem: exact type codebook} in
Appendix \ref{sec:The-Fixed-Composition}. The reason is that in the
proof of \cite[Theorem 1]{ChenUnpublished} and Theorem \ref{thm: Error exponent bounds, single type}
a \emph{single} channel code is constructed and utilized for SW coding
of a \emph{single} type $Q_{X}$. By contrast, when considering the
more refined notion of excess rate, \emph{all} types ${\cal P}({\cal X})$
of the source should be considered at the same time, and as a result,
many channel codes should be constructed (see the proof of Theorem
\ref{thm: achieavbility with type dependent} henceforth). Now, consider
the simplified case of SW coding for just two different types. In
this case, two channel codes are required for a `good' SW coding of
the two types. However, if the codes are designed to achieve the supremum
reliability function, there is no guarantee that the block lengths
of the codes will match, because the limit superior might not be achieved
by the same sub-sequence of block lengths for both types. Specifically,
for any given block length such that one of the codes has `good' error
probability (i.e., close to the probability guaranteed by the supremum
reliability function), the other code might have `poor' error probability,
and vice versa. Since in order to construct a good SW code, we need
to find a sequence of block lengths such that \emph{both} channel
codes have good error probability, this can only be guaranteed for
the lower error exponent of the infimum reliability function. Indeed,
for the infimum reliability function, good error probability is assured
for all sufficiently large block lengths, and so, when the block length
is sufficiently large, both channel codes, if properly designed, will
have error probability close to the one guaranteed by the infimum
reliability function.

\subsection{Bounds on Exponents for General SW Codes \label{sub:Bounds-on SW codes}}

In this subsection, we derive upper and lower bounds on the error
exponent and excess-rate exponent, which hold for any sequence of
variable-rate SW codes. Unlike the case of Theorem \ref{thm: Error exponent bounds, single type},
the exponent bounds, in this subsection should consider \emph{all}
possible types in ${\cal P}({\cal X})$. 
\begin{thm}
\label{thm: General upper bound on error exponent}Let ${\cal S}$
be an arbitrary sequence of SW codes. Then, 
\[
{\cal E}_{e}^{+}({\cal S})\leq\inf_{Q_{X}\in{\cal P}({\cal X})}\left\{ D(Q_{X}||P_{X})+\overline{E}_{e}^{*}(H(Q_{X})-\overline{R}(Q_{X};{\cal S}),Q_{X},P_{Y|X})\right\} .
\]

\end{thm}

\begin{thm}
\label{thm: General upper bound on excess rate exponent}Let ${\cal S}$
be any arbitrary sequence of SW codes. Then, 
\begin{equation}
{\cal E}_{r}({\cal S},\RR)\leq\inf_{Q_{X}\in{\cal P}({\cal X})}\left\{ D(Q_{X}||P_{X})+{\cal E}_{r}({\cal S},\RR,Q_{X})\right\} .\label{eq: general excess rate exponent upper bound}
\end{equation}

\end{thm}
Next, we derive an achievable error exponent and excess-rate exponent
for type-dependent, variable-rate SW codes. The proof is based on
the achievability result of Theorem \ref{thm: Error exponent bounds, single type},
but when considering the notion of excess-rate exponent, attention
need to be given to all types of the source. 
\begin{thm}
\label{thm: achieavbility with type dependent} For any given rate
function $\rho(Q_{X})$, there exists a sequence of type-dependent,
variable-rate SW codes ${\cal S}$ such that 
\begin{equation}
{\cal E}_{e}^{-}({\cal S})\geq\inf_{Q_{X}\in{\cal P}({\cal X})}\left\{ D(Q_{X}||P_{X})+\underline{E}_{e}^{*}(H(Q_{X})-\rho(Q_{X}),Q_{X},P_{Y|X})\right\} \label{eq: achievable inf error exponent variable SW}
\end{equation}
and

\begin{equation}
{\cal E}_{r}({\cal S},\RR)\geq\inf_{Q_{X}\in{\cal P}({\cal X}):\:\rho(Q_{X})>\RR}D(Q_{X}||P_{X}).\label{eq: lower bound on variable rate excess rate exponent}
\end{equation}

\end{thm}
The proof is deferred to Appendix \ref{sec:Proofs}, but here we provide
an intuitive outline of the SW code constructed. From Theorem \ref{thm: Error exponent bounds, single type},
it is possible to construct a SW code ${\cal S}_{n}^{*}(Q_{X})$ for
any given \emph{$Q_{X}\in\interior{\cal Q}({\cal X})$}, with conditional
error probability converging to about $\exp\left[-n\underline{E}_{e}^{*}(H(Q_{X})-\rho(Q_{X}),Q_{X},P_{Y|X})\right]$\emph{.}
However, to obtain a SW code which satisfies \eqref{eq: achievable inf error exponent variable SW}
for a sufficiently long block lengths, the conditional error probability
should be about $\exp\left[-n\underline{E}_{e}^{*}(H(Q_{X})-\rho(Q_{X})),Q_{X},P_{Y|X})\right]$
\emph{uniformly }over all types. Indeed, if uniform convergence is
not satisfied then, for any given finite block length, there might
be types $Q_{X}$ such that the error probability of ${\cal S}_{n}^{*}(Q_{X})$
is still far from its limit, and the error probability of this code
may be a dominant factor in the total error probability. Thus, we
have to prove uniform convergence of the error probability. Our strategy
is as follows. We choose a large block length $n_{0}$, such that
the types of ${\cal P}_{n_{0}}({\cal X})$ are good approximations
for all types in ${\cal P}({\cal X})$, and construct good SW codes
${\cal S}_{n}^{*}(Q_{X})$ for all $Q_{X}\in{\cal P}_{n_{0}}({\cal X})$.
Since $|{\cal P}_{n_{0}}({\cal X})|$ is finite, uniform convergence
of the error probability of ${\cal S}_{n}^{*}(Q_{X})$ holds. For
any given $n$, upon observing a block from the source, we will modify
it (namely, by truncating it and altering some of its components),
so that the modified source block would have a type within ${\cal P}_{n_{0}}({\cal X})$,
and can then be encoded by one of the `good' SW codes ${\cal S}_{n}^{*}(Q_{X})$.
The encoded modified vector will be sent to the decoder, along with
the modification data. Then, at the decoder, the side information
vector will be modified accordingly, so it appears as resulting from
the memoryless source $P_{Y|X}$, but conditioning on the modified
source block. Thus, the decoder of ${\cal S}_{n}^{*}(Q_{X})$ can
be used to decode the modified vector, and the modification data can
be used to recover the actual source block. 
\begin{rem}
According to Theorem \ref{thm: achieavbility with type dependent}
and the proof of the achievability part of Theorem \ref{thm: Error exponent bounds, single type},
it is implicit that the \emph{random binning exponent, }defined as
\begin{equation}
\min_{Q_{XY}}\left\{ D\left(Q_{XY}||P_{XY}\right)+\left[\RR-H(Q_{X|Y}|Q_{Y})\right]_{+}\right\} .\label{eq: random binning exponent}
\end{equation}
may be achieved by using permutations of a channel code which achieve
the random coding exponent. However, as is well known, for fixed-rate
SW coding \cite{gallager1976source}, one can achieve the error exponent
by simple random binning, i.e. assigning source blocks to bins independently,
with a uniform probability distributions over the bins. As a side
result, in Appendix \ref{sec:Tightness-of-the-Random-Binning}, we
generalize this ensemble to type-dependent, variable-rate random binning
SW codes (defined rigorously therein), and prove that \eqref{eq: random binning exponent}
(with some given rate function $\rho(Q_{X})$ replacing $\RR$) is
the \emph{exact} exponent of this ensemble; a result analogous to
\cite{Gallager1973_tight} for random channel coding. 
\end{rem}

\subsection{Trade-Off Between Exponents}

As common in variable-rate SW coding, a trade-off exists between the
error exponent and excess-rate exponent. In the remaining part of
the paper, we explore this trade-off by requiring the achievability
of a certain target error exponent $\EE$ with maximal excess rate
exponent. Theorem \ref{thm: achieavbility with type dependent} shows
that in order to achieve a target error exponent $\EE$, a type-dependent,
variable-rate SW code may be employed. Then, Theorems \ref{thm: General upper bound on error exponent}
and \ref{thm: General upper bound on excess rate exponent} provide
upper bounds which quantify the gap from optimal performance. Comparing
Theorem \ref{thm: achieavbility with type dependent} with Theorems
\ref{thm: General upper bound on error exponent} and \ref{thm: General upper bound on excess rate exponent},
it is evident that there might be two origins for a gap between the
bounds. The first one lies in the error exponent expression, and the
second is in the excess-rate exponent. We now discuss these differences.

First, in general, it is yet to be known whether the inequality $\underline{E}_{e}^{*}(\RR,Q_{X},W)\leq\overline{E}_{e}^{*}(\RR,Q_{X},W)$
may be strict. Thus, if for the minimizers in \eqref{eq: lower bound on variable rate excess rate exponent}
and \eqref{eq: achievable inf error exponent variable SW} a strict
inequality occurs, then a gap exists between the upper and lower bounds
for the SW code%
\footnote{In \eqref{eq: lower bound on variable rate excess rate exponent}
and \eqref{eq: achievable inf error exponent variable SW}, a minimum
might not be achieved. In this case, the last statement should be
valid for all sequence of distributions which achieves the infimum.%
}. Nonetheless, it is also well known that for $\RR\geq R_{\s[cr]}(Q_{X},W)$,
$\underline{E}_{e}^{*}(\RR,Q_{X},W)=\overline{E}_{e}^{*}(\RR,Q_{X},W)$
is guaranteed, and so there are cases in which the upper and lower
bounds coincide, especially at low target error exponents $\EE$.
Second, on substituting a rate function $\rho(Q_{X})$ of a type-dependent,
variable-rate codes in \eqref{eq: general excess rate exponent upper bound},
the resulting upper bound is different from the lower bound of \eqref{eq: lower bound on variable rate excess rate exponent},
only if the function 
\[
\inf_{Q_{X}:\rho(Q_{X})>\RR}D(Q_{X}||P_{X})
\]
is not left-continuous in $\RR$. As will turn out, for the class
of rate functions of interest, left-continuity is satisfied, and the
upper and lower bounds coincide. Thus, from the above discussion,
we conclude that type-dependent, variable-rate SW codes are optimal
for sufficiently low target error exponents $\EE.$ 

Since from Theorem \ref{thm: achieavbility with type dependent},
any target error exponent can be achieved with type-dependent, variable-rate
SW codes, and because they are provably optimal in some domain, we
henceforth consider only such SW codes. We will define \emph{optimal}
rate functions as follows.
\begin{defn}
\label{def: optimal rate function}A rate function $\underline{\rho}^{*}(Q_{X},\EE)$
is said to be \emph{inf-optimal}, if for any $\delta>0$, there exists
a sequence of type-dependent, variable-rate SW codes ${\cal S}$ with
$\overline{R}(Q_{X};{\cal S})\leq\underline{\rho}^{*}(Q_{X},\EE)+\delta$
and ${\cal E}_{e}^{-}({\cal S})\geq\EE$, and for every other rate
function $\rho(Q_{X})$ with the above property, we have $\underline{\rho}^{*}(Q_{X},\EE)\leq\rho(Q_{X})$,
for all $Q_{X}\in{\cal P}({\cal X})$. The \emph{sup-optimal} rate
function $\overline{\rho}^{*}(Q_{X},\EE)$ is defined analogously. 

Notice that by definition, we have 
\[
\overline{\rho}^{*}(Q_{X},\EE)\leq\underline{\rho}^{*}(Q_{X},\EE).
\]
In Section \ref{sec:Optimal rate functions}, we will obtain bounds
on the optimal rate functions for any given $\EE$, and in Section
\ref{sec:Excess-Rate-Performance}, we will obtain bounds on the excess-rate
performance for these optimal rate functions. 
\end{defn}

\section{Optimal Rate Functions \label{sec:Optimal rate functions}}

In this section, we explore the optimal rate functions, for any given
$\EE$. Before discussing specific bounds, we characterize them using
the inverse of the fixed composition reliability function. 

Theorem \ref{thm: achieavbility with type dependent} implies that
for $\underline{\rho}^{*}(Q_{X},\EE)$ to be inf-optimal, we must
have
\begin{equation}
\EE\leq D(Q_{X}||P_{X})+\underline{E}_{e}^{*}(H(Q_{X})-\underline{\rho}^{*}(Q_{X},\EE),Q_{X},P_{Y|X})\label{eq: achieving EE inf condition}
\end{equation}
for any given $Q_{X}\in{\cal P}({\cal X})$. The following corollary
is immediate from Proposition \ref{prop: property of channel reliability}
in Appendix \ref{sec:The-Fixed-Composition}.
\begin{cor}
\label{cor:reliabiliy function has inverse}As a function of $\RR$,
the function $\underline{E}_{e}^{*}(\RR,Q_{X},W)$ has a continuous
inverse $\underline{R}^{*}(\EE,Q_{X},W)$ across the interval $\EE\in[0,E_{0}^{-}(Q_{X},W))$.
An analogous result holds for $\overline{E}_{e}^{*}(\RR,Q_{X},W)$.
\end{cor}
Now, Corollary \ref{cor:reliabiliy function has inverse} immediately
implies the following:
\begin{multline}
\underline{\rho}^{*}(Q_{X},\EE)\leq\\
\begin{cases}
0, & \EE\leq D(Q_{X}||P_{X})\\
H(Q_{X})-\underline{R}^{*}(\EE-D(Q_{X}||P_{X}),Q_{X},P_{Y|X}), & D(Q_{X}||P_{X})<\EE<D(Q_{X}||P_{X})+E_{0}^{-}(Q_{X},W)\\
H(Q_{X})-C_{0}(Q_{X},P_{Y|X}), & \EE\geq D(Q_{X}||P_{X})+E_{0}^{-}(Q_{X},W),
\end{cases}
\end{multline}
where $E_{0}^{-}(Q_{X},W)$ is as defined in \eqref{eq: maximal inf error exponent channel coding},
and $\underline{R}^{*}(\EE-D(Q_{X}||P_{X}),Q_{X},P_{Y|X})$ is as
defined in Corollary \ref{cor:reliabiliy function has inverse}. 

Similarly, Theorem \ref{thm: General upper bound on error exponent}
implies that $\overline{\rho}^{*}(Q_{X},\EE)$ cannot be sup-optimal
unless 
\begin{equation}
\EE\leq D(Q_{X}||P_{X})+\overline{E}_{e}^{*}(H(Q_{X})-\overline{\rho}^{*}(Q_{X},\EE),Q_{X},P_{Y|X})\label{eq: achieving EE sup condition}
\end{equation}
for any given $Q_{X}\in{\cal P}({\cal X})$. Now Corollary \ref{cor:reliabiliy function has inverse}
implies:
\begin{multline}
\overline{\rho}^{*}(Q_{X},\EE)\geq\\
\begin{cases}
0, & \EE\leq D(Q_{X}||P_{X})\\
H(Q_{X})-\overline{R}^{*}(\EE-D(Q_{X}||P_{X}),Q_{X},P_{Y|X}), & D(Q_{X}||P_{X})<\EE<D(Q_{X}||P_{X})+E_{0}^{+}(Q_{X},W)\\
H(Q_{X})-C_{0}(Q_{X},P_{Y|X}), & \EE\geq D(Q_{X}||P_{X})+E_{0}^{+}(Q_{X},W).
\end{cases}
\end{multline}
In Definition \ref{def: optimal rate function}, $\underline{\rho}^{*}(Q_{X},\EE)$
is only defined for $Q_{X}\in{\cal P}({\cal X})$. This is because
the value of $\underline{\rho}^{*}(Q_{X},\EE)$ for $Q_{X}\in{\cal Q}({\cal X})\backslash{\cal P}({\cal X})$
(any irrational PMF) has no operational meaning, and does not affect
exponents (see Theorems \ref{thm: Error exponent bounds, single type},
\ref{thm: General upper bound on excess rate exponent}, and \ref{thm: achieavbility with type dependent}).
Thus, for $Q_{X}\in{\cal Q}({\cal X})\backslash{\cal P}({\cal X})$,
we may arbitrarily define it as the lower semi-continuous extension
of $\underline{\rho}^{*}(Q_{X},\EE)$. Specifically, for any given
$Q_{X}\in{\cal Q}({\cal X})\backslash{\cal P}({\cal X})$ we henceforth
define 
\[
\underline{\rho}^{*}(Q_{X},\EE)\teq\lim_{\epsilon\downarrow0}\inf_{Q'_{X}\in{\cal P}({\cal X}):||Q'_{X}-Q_{X}||\leq\epsilon\:}\underline{\rho}^{*}(Q_{X},\EE),
\]
and the same convention will be used for $\overline{\rho}^{*}(Q_{X},\EE)$. 
\begin{lem}
The rate function $\underline{\rho}^{*}(Q_{X},\EE)$ is regular and
strictly increasing in the range 
\[
\EE\in\left(D(Q_{X}||P_{X}),E_{0}^{-}(Q_{X},W)\right).
\]
The same properties hold for $\overline{\rho}^{*}(Q_{X},\EE)$.\end{lem}
\begin{IEEEproof}
These properties follow directly from Proposition \ref{prop: property of channel reliability}
and Corollary \ref{cor:reliabiliy function has inverse}. 
\end{IEEEproof}
Next, we provide specific bounds on the optimal rate functions. Generally,
any bound on the reliability function may be used, but we will focus
on the \emph{random binning exponent} and \emph{expurgated exponent}
as lower bounds to the largest achievable exponent, and the \emph{sphere
packing exponent} as an upper bound. In essence, these bounds are
generalizations of the random binning bound \cite[Theorem 2]{csiszar1980towards},
\cite[Theorem 2]{csiszar1981graph}, the expurgated bound, which follows
from \cite[Theorem 2]{csiszar1981graph}, and the sphere packing bound
\cite[Theorem 3]{csiszar1980towards} for type-dependent, variable-rate
SW coding. For the sake of simplicity, we assume that $C_{0}(Q_{X},P_{Y|X})=0$
for all $Q_{X}$, and so the expurgated and sphere packing exponents
are finite for every positive rate. The results are easily generalized
to the case of $C_{0}(Q_{X},P_{Y|X})>0$. 

We first need some definitions. For brevity, the dependency in $Q_{X}$
for the defined quantities is omitted. Let 
\begin{equation}
Q'_{Y|X}\teq\argmin_{Q_{Y|X}}\left\{ I(Q_{X}\times Q_{Y|X})+D(Q_{Y|X}||P_{Y|X}|Q_{X})\right\} ,\label{eq: random binning optimal rate function affine part}
\end{equation}
\[
Q'_{\tilde{X}|X}\teq\argmin_{Q_{\tilde{X}|X}:\: Q_{\tilde{X}}=Q_{X}}\left\{ B(Q_{X}\times Q_{\tilde{X}|X},P_{Y|X})+I(Q_{X}\times Q_{\tilde{X}|X})\right\} ,
\]
where $B(Q_{X\tilde{X}},P_{Y|X})$ is defined in \eqref{eq: Bhat distance}.
Next, define $\EE[,0]=D(Q_{X}||P_{X})$ as well as%
\footnote{The subscript $\textnormal{`a'}$ represents the word `affine'.%
} 
\[
\EE[,\textnormal{\scriptsize a-rb}]\teq D(Q_{X}||P_{X})+D(Q'_{Y|X}||P_{Y|X}|Q_{X}),
\]
\[
\EE[,\textnormal{\scriptsize max-rb}]\teq D(Q_{X}||P_{X})+I(Q_{X}\times Q'_{Y|X})+D(Q'_{Y|X}||P_{Y|X}|Q_{X}),
\]
\[
\EE[,\textnormal{\scriptsize a-ex}]\teq D(Q_{X}||P_{X})+B(Q_{X}\times Q'_{\tilde{X}|X},P_{Y|X}),
\]
\[
\EE[,\textnormal{\scriptsize max-ex}]\teq D(Q_{X}||P_{X})+B(Q_{X}\times Q_{X},P_{Y|X}),
\]
and 
\[
\EE[,\textnormal{\scriptsize max-sp}]\teq D(Q_{X}||P_{X})+D(Q_{X}\times(Q_{X}\times P_{Y|X})_{Y}||P_{XY}).
\]
Also, define the sets 
\[
{\cal A}_{\s[rb]}\teq\left\{ Q_{Y|X}:D(Q_{X}\times Q_{Y|X}||P_{XY})\leq\EE\right\} ,
\]
\[
{\cal A}_{\s[ex]}\teq\left\{ Q_{\tilde{X}|X}:Q_{\tilde{X}}=Q_{X},\:\EE=D(Q_{X}||P_{X})+B(Q_{X\tilde{X}},P_{Y|X})\right\} 
\]
and let ${\cal A}_{\s[sp]}\teq{\cal A}_{\s[rb]}$. 

The random binning rate function is defined as
\begin{equation}
\rho_{\s[rb]}(Q_{X},\EE)\teq\begin{cases}
0, & \EE\leq\EE[,0]\\
\EE+H(Q_{X})-D(Q_{X}||P_{X})\\
-\min_{Q_{Y|X}\in{\cal A}_{\s[rb]}}\left\{ I(Q_{X}\times Q_{Y|X})+D(Q_{Y|X}||P_{Y|X}|Q_{X})\right\} , & \EE[,0]<\EE\leq\EE[,\textnormal{\scriptsize a-rb}]\\
\EE-\EE[,\textnormal{\scriptsize a-rb}]+H(Q_{X})-I(Q_{X}\times Q'_{Y|X}), & \EE[,\textnormal{\scriptsize a-rb}]<\EE\leq\EE[,\textnormal{\scriptsize max-rb}]\\
H(Q_{X}), & \EE[,\textnormal{\scriptsize max-rb}]<\EE
\end{cases}\label{eq: random binning rate function expression}
\end{equation}
and the expurgated rate function is defined as 

\begin{equation}
\rho_{\s[ex]}(Q_{X},\EE)\teq\begin{cases}
0, & \EE\leq\EE[,0]\\
\EE-\EE[,\textnormal{\scriptsize a-ex}]+H(Q_{X})-I(Q_{X}\times Q'_{\tilde{X}|X}), & \EE[,0]<\EE\leq\EE[,\textnormal{\scriptsize a-ex}]\\
H(Q_{X})-\min_{Q_{\tilde{X}|X}\in{\cal A}_{\s[ex]}}I(Q_{X\tilde{X}}), & \EE[,\textnormal{\scriptsize a-ex}]<\EE\leq\EE[,\textnormal{\scriptsize max-ex}]\\
H(Q_{X}), & \EE[,\textnormal{\scriptsize max-ex}]>D(Q_{X}||P_{X})+B(Q_{X}\times Q_{X},P_{Y|X})
\end{cases}\label{eq: expurgated rate function expression}
\end{equation}
and the sphere packing rate function is defined as
\begin{equation}
\rho_{\s[sp]}(Q_{X},\EE)\teq\begin{cases}
0, & \EE\leq\EE[,0]\\
H(Q_{X})-\min_{Q_{Y|X}\in{\cal A}_{\s[sp]}}I(Q_{X}\times Q_{Y|X}), & \EE[,0]<\EE\leq\EE[,\textnormal{\scriptsize max-sp}]\\
H(Q_{X}), & \EE[,\textnormal{\scriptsize max-sp}]<\EE.
\end{cases}\label{eq: sphere packing rate function expression}
\end{equation}

\begin{thm}
\label{thm: Optimal rate function bounds} For any given $\EE$ and
$Q_{X}\in{\cal P}({\cal X})$
\[
\rho_{\s[sp]}(Q_{X},\EE)\leq\overline{\rho}^{*}(Q_{X},\EE)\leq\underline{\rho}^{*}(Q_{X},\EE)\leq\min\{\rho_{\s[rb]}(Q_{X},\EE),\rho_{\s[ex]}(Q_{X},\EE)\}.
\]

\end{thm}
Due to the similarity between the random binning bound and sphere
packing bound, we obtain the known property from channel coding: For
any $Q_{X}$ there exists $\EE[,\textnormal{\scriptsize cr}](Q_{X})$
such that if $\EE\leq\EE[,\textnormal{\scriptsize cr}](Q_{X})$ we
get $\rho_{\s[rb]}(Q_{X},\EE)=\rho_{\s[sp]}(Q_{X},\EE).$ Thus, for
any required $\EE$, if $\EE\leq\EE[,\textnormal{\scriptsize cr}](Q_{X})$
then the optimal rate function is exactly known. Specifically, the
right limit of the optimal rate function $\rho_{\s[rb]}(Q_{X},\EE)$
at its discontinuity point $\EE[,0]$ can be easily evaluated from
\eqref{eq: random binning rate function expression} to be 
\[
\lim_{\EE\downarrow\EE[,0]}\rho_{\s[rb]}(Q_{X},\EE)=H(Q_{X})-D(P_{Y|X}||Q_{Y}^{*}|Q_{X})
\]
where $Q_{Y}^{*}(y)=\sum_{x\in{\cal X}}Q_{X}(x)P_{Y|X}(y|x)$. Namely,
the resulting rate is the conditional entropy $H(Q_{X|Y}|Q_{Y})$
of the distribution $Q_{XY}=Q_{X}\times P_{Y|X}$. Especially, for
$Q_{X}=P_{X}$ we have that $\rho_{\s[rb]}(Q_{X},\epsilon)\geq H(P_{X|Y}|P_{Y})$,
for all $\epsilon>0$, as expected. The following lemma provides several
simple properties of the rate functions $\rho_{\s[rb]}(Q_{X},\EE)$,
$\rho_{\s[ex]}(Q_{X},\EE)$ and $\rho_{\s[sp]}(Q_{X},\EE)$. 
\begin{lem}
\label{lem: rate function bounds properties}The rate functions $\rho_{\s[rb]}(Q_{X},\EE)$,
$\rho_{\s[ex]}(Q_{X},\EE)$ and $\rho_{\s[sp]}(Q_{X},\EE)$ have the
following properties:
\begin{itemize}
\item Strictly positive for \textup{\emph{$\EE>\EE[,0]$.}}
\item Strictly increasing as a function of $\EE\geq\EE[,0]$ and $\EE\leq\EE[,\textnormal{\scriptsize max-rb}]$
(for random binning) or $\EE\leq\EE[,\textnormal{\scriptsize max-ex}]$
(for expurgated) and $\EE\leq\EE[,\textnormal{\scriptsize max-sp}]$
(for sphere packing).
\item Concave in $\EE\in(\EE[,0],\infty)$.
\item Regular rate functions. 
\end{itemize}
\end{lem}
As for capacity and error exponents in channel coding, the computation
of the bounds on the optimal rate function requires the solution of
a non-trivial optimization problem. We defer the discussion on this
matter to Section \ref{sec:Computational algorithms}, where we discuss
iterative algorithms for the computation of the bounds on the optimal
rate functions, as well as their excess-rate performance. Nonetheless,
in Appendix \ref{sec: wakly correlated sources}, we provide analytic
approximations for $\rho_{\s[rb]}(Q_{X},\EE)$ and $\rho_{\s[sp]}(Q_{X},\EE)$
in the case of weakly correlated sources%
\footnote{The expurgated bound is not very useful in this regime \cite[Section 3.4]{viterbi2009principles}.%
}.

\section{Excess-Rate Performance\label{sec:Excess-Rate-Performance}}

In this section, we evaluate the excess-rate exponent of the optimal
rate functions bounds, as defined in Section \ref{sec:Optimal rate functions}.
This results in lower and upper bounds on the maximal achievable excess
rate exponent for a given error exponent, and thus the characterization
of the optimal trade-off between error exponent and excess-rate exponent. 

Notice that for a general rate function $\rho(\cdot)$, and target
rates $\RR\in(\E[r(\mathbf{x})],\max_{Q_{X}}\rho(Q_{X}))$, the excess-rate
exponent is strictly positive and finite. The next lemma shows that
the upper bound of \eqref{eq: general excess rate exponent upper bound}
and the lower bound of \eqref{eq: lower bound on variable rate excess rate exponent}
coincide for regular rate functions. Since in Lemma \ref{lem: rate function bounds properties}
it was shown that inf/sup optimal rate functions as well as the random
binning, expurgated and sphere packing rate functions are all regular
rate functions, this means that we have the exact expression for their
excess-rate performance. 
\begin{lem}
\label{lem: excess rate exponent expression}For a regular rate function
$\rho(Q_{X})$ 
\begin{equation}
\inf_{Q_{X}:\rho(Q_{X})>\RR}D(Q_{X}||P_{X})=\min_{Q_{X}:\rho(Q_{X})\geq\RR}D(Q_{X}||P_{X}).\label{eq: left continuity for regular rate functions}
\end{equation}

\end{lem}
We now mention a few general properties of excess-rate exponents functions.
\begin{lem}
\label{lem: Excess Rate Exponent Properties} Let $\rho(Q_{X})$ be
a rate function, and $\RR_{\max}\teq\sup_{Q_{X}}\rho(Q_{X})$. If
$\rho(Q_{X})$ is regular, let $\RR_{\max}'\teq\sup_{Q_{X}\in{\cal V}}\rho(Q_{X})$.
The excess-rate exponent $E_{r}(\RR)$ for the rate function $\rho(Q_{X})$
has the following properties:
\begin{itemize}
\item $E_{r}(\RR)=0$ for $\RR\in[0,\rho(P_{X})]$. 
\item $E_{r}(\RR)=\infty$ for $\RR\in(\RR_{\max},\infty)$. 
\item $E_{r}(\RR)$ is increasing in $[\rho(P_{X}),\RR_{\max}]$. If $\rho(Q_{X})$
is regular, then $E_{r}(\RR)$ is strictly increasing in $[\rho(P_{X}),\RR_{\max}']$.
\item $E_{r}(\RR)$ is continuous in $[\rho(P_{X}),\RR_{\max}]$ except
for a countable number of points. If $\rho(Q_{X})$ is regular, then
$E_{r}(\RR)$ is left-continuous in $[\rho(P_{X}),\RR_{\max}']$.
\end{itemize}
\end{lem}
In the rest of the section, we assume that a target error exponent
$\EE$ is given and fixed. Thus, for brevity, we omit the notation
of the dependence of various quantities on it. We define the excess-rate
exponent of the inf-optimal rate function as 
\[
\underline{E}_{r}^{*}(\RR)\teq\min_{Q_{X}:\underline{\rho}^{*}(Q_{X},\EE)\geq\RR}D(Q_{X}||P_{X}),
\]
and analogously, define $\overline{E}_{r}^{*}(\RR)$. Similarly, we
define the random-binning excess-rate exponent as
\begin{equation}
E_{r,\s[rb]}(\RR)\teq\min_{Q_{X}:\rho_{\st[rb]}(Q_{X},\EE)\geq\RR}D(Q_{X}||P_{X}),\label{eq:excess rate exponent random binning}
\end{equation}
and analogously define $E_{r,\s[ex]}(\RR)$ and $E_{r,\s[sp]}(\RR)$.
For a given $\RR$, we evidently have 
\begin{equation}
\max\{E_{r,\s[rb]}(\RR),E_{r,\s[ex]}(\RR)\}\leq\underline{E}_{r}^{*}(\RR)\leq\overline{E}_{r}^{*}(\RR)\leq E_{r,\s[sp]}(\RR).\label{eq: excess rate bounds}
\end{equation}
For some $\RR$, let the minimizer in \eqref{eq:excess rate exponent random binning}
be $Q_{X}^{*}$. Then, if $\rho_{\s[rb]}(Q_{X}^{*},\EE)=\rho_{\s[sp]}(Q_{X}^{*},\EE)$,
it is easy to verify that the bounds in \eqref{eq: excess rate bounds}
are tight, and $\underline{E}_{r}^{*}(\RR)=\overline{E}_{r}^{*}(\RR)=E_{r,\s[rb]}(\RR)$.
In other cases, one can use the upper bound at the rate function $\rho_{\s[ub]}^{*}=\min\{\rho_{\s[rb]}(Q_{X},\EE),\rho_{\s[ex]}(Q_{X},\EE)\}$
to obtain an excess-rate exponent $E_{r,\s[ub]}(\RR)$, defined similarly
to \eqref{eq:excess rate exponent random binning}. In this case,
an improvement over the random-binning and expurgated excess-rate
exponents is guaranteed, as 
\[
\max\{E_{r,\s[rb]}(\RR),E_{r,\s[ex]}(\RR)\}\leq E_{r,\s[ub]}(\RR).
\]

Next, we evaluate the bounds on the optimal excess-rate exponent,
e.g., as in \eqref{eq:excess rate exponent random binning}. However,
as we have seen, $\rho_{\s[rb]}(Q_{X},\EE)$, as well as the other
rate functions, are not given analytically, and performing the maximization
in \eqref{eq:excess rate exponent random binning} directly may be
prohibitively complex, especially when ${\cal \left|X\right|}$ is
large. Thus, we describe an indirect method to evaluate the excess-rate
bounds. For a given $\RR$, any curve $\mathsf{E}_{r}=E_{r}(\RR)$
may be characterized by a condition that verifies whether the rate
and excess-rate pair $(\RR,\mathsf{E}_{r})$ is either below or above
the curve. The proof is based on the following lemma, that introduces
a rate function which is designed to achieve pointwise $(\RR,\mathsf{E}_{r})$,
but not necessarily $\EE$. 
\begin{lem}
\label{lem: Dominating and simple rate function equivalence}Let \textup{
\[
\hat{\rho}(Q_{X};\RR,\mathsf{E}_{r})\teq\begin{cases}
\RR, & D(Q_{X}||P_{X})<\mathsf{E}_{r}\\
\RR_{0}, & \mathrm{otherwise}
\end{cases}.
\]
}Then, if there exists $\RR_{0}$ such that $\hat{\rho}(Q_{X};\RR,\mathsf{E}_{r})$
achieves infimum error exponent $\EE$ then $\underline{\rho}^{*}(Q_{X},\EE)$
achieves infimum error exponent $\EE$ with rate $\RR$ and excess-rate
exponent $\mathsf{E}_{r}$. If $\hat{\rho}(Q_{X};\RR,\mathsf{E}_{r})$
does not achieve supremum error exponent $\EE$ then $\overline{\rho}^{*}(Q_{X},\EE)$
does not achieve supremum error exponent $\EE$ with rate $\RR$ and
excess-rate exponent $\mathsf{E}_{r}$.\end{lem}
\begin{IEEEproof}
~

$(\Leftarrow)$ Assume that $\hat{\rho}(Q_{X};\RR,\mathsf{E}_{r})$
achieves $(\RR,\mathsf{E}_{r})$ with an infimum error exponent $\EE$.
Clearly the definition of an\emph{ optimal} rate function imply that
$\underline{\rho}^{*}(Q_{X},\EE)$ also achieves $(\RR,\mathsf{E}_{r})$.

$(\Rightarrow)$ Assume that $\overline{\rho}^{*}(Q_{X},\EE)$ achieves
$(\RR,\mathsf{E}_{r})$. If $Q_{X}$ satisfies $D(Q_{X}||P_{X})\geq\mathsf{E}_{r}$
then for 
\[
\RR_{0}\geq\max_{Q_{X}:\: D(Q_{X}||P_{X})\geq\mathsf{E}_{r}}\overline{\rho}^{*}(Q_{X},\EE)
\]
we get $\hat{\rho}(Q_{X};\RR,\mathsf{E}_{r})\geq\overline{\rho}^{*}(Q_{X},\EE)$.
Else, if $\hat{\rho}(Q_{X};\RR,\mathsf{E}_{r})>\RR$ for some $Q_{X}$
that satisfies $D(Q_{X}||P_{X})<\mathsf{E}_{r}$, then $\overline{\rho}^{*}(Q_{X},\EE)$
does not achieve $(\RR,\mathsf{E}_{r})$ using Lemma \ref{lem: excess rate exponent expression}.
Thus, we must have $\hat{\rho}(Q_{X};\RR,\mathsf{E}_{r})\geq\overline{\rho}^{*}(Q_{X},\EE)$
for all $Q_{X}$ and this implies that $\hat{\rho}(Q_{X};\RR,\mathsf{E}_{r})$
also achieves supremum error exponent $\EE$. It is easy to see that
$\hat{\rho}(Q_{X};\RR,\mathsf{E}_{r})$ has excess-rate exponent $\mathsf{E}_{r}$
at rate $\RR$ directly from its construction and Lemma \ref{lem: excess rate exponent expression}.
\end{IEEEproof}
Notice that the rate function $\hat{\rho}(Q_{X};\RR,\mathsf{E}_{r})$,
introduced in the previous lemma, has only \emph{pointwise} optimal
excess-rate exponent, in the sense that for the given $(\RR,\mathsf{E}_{r})$
it achieves the optimal trade-off between the error exponent and excess-rate
exponent. By contrast, the optimal rate functions $\underline{\rho}^{*}(Q_{X},\EE)$
and $\overline{\rho}^{*}(Q_{X},\EE)$ achieve the optimal excess-rate
exponent, at any given rate. 

Define for a given $(\RR,\mathsf{E}_{r})$ 
\begin{eqnarray}
\Gamma_{\s[rb]}(t,Q_{X},Q_{Y|X}) & \triangleq & D(Q_{X}||P_{X})+D(Q_{Y|X}||P_{Y|X}|Q_{X})\nonumber \\
 &  & +t\cdot\left[\RR-H(Q_{X|Y}|Q_{Y})\right]\label{eq: defintion of Gamma for excess rate random binning}
\end{eqnarray}
\begin{equation}
e_{\s[rb]}(t)\teq\min_{Q_{X}:D(Q_{X}||P_{X})\leq\mathsf{E}_{r}}\min_{Q_{Y|X}}\Gamma_{\s[rb]}(t,Q_{X},Q_{Y|X}).\label{eq: minimization problem excess rate random binning}
\end{equation}
and
\begin{eqnarray}
\Gamma_{\s[ex]}(t,Q_{X},Q_{\tilde{X}|X}) & \triangleq & D(Q_{X}||P_{X})+B(Q_{X\tilde{X}},P_{Y|X})\nonumber \\
 &  & +t\left[\RR-H(Q_{X|\tilde{X}}|Q_{\tilde{X}})\right]\label{eq: defintion of Gamma for excess rate expurgated}
\end{eqnarray}
\begin{equation}
e_{\s[ex]}(t)\teq\min_{Q_{X}:D(Q_{X}||P_{X})\leq\mathsf{E}_{r}}\min_{Q_{\tilde{X}|X:\:}Q_{\tilde{X}}=Q_{X}}\Gamma_{\s[ex]}(t,Q_{X},Q_{\tilde{X}|X}).\label{eq: minimization problem excess rate expurgated}
\end{equation}
Also, define 
\begin{equation}
\Gamma_{\s[sp]}(t,Q_{X},Q_{Y|X})\teq\Gamma_{\s[rb]}(t,Q_{X},Q_{Y|X})\label{eq: defintion of Gamma for excess rate sphere packing}
\end{equation}
\begin{equation}
e_{\s[sp]}(t)\teq e_{\s[rb]}(t).\label{eq: minimization problem excess rate sphere packing}
\end{equation}

\begin{thm}
\label{thm: excess rate and rate curve condition}If 

\emph{
\[
\max\left\{ \max_{0\leq t\leq1}e_{\s[rb]}(t),\max_{t\geq1}e_{\s[ex]}(t)\right\} \geq\EE
\]
}\textup{\emph{then there exists a sequence of SW codes with infimum
error exponent $\EE$, and }}excess-rate exponent \textup{$\mathsf{E}_{r}$}\emph{
}\textup{\emph{at rate }}$\RR$. Conversely, \textup{\emph{if }}

\emph{
\[
\max_{t\geq0}e_{\s[sp]}(t)<\EE
\]
}\textup{\emph{then there is no sequence of SW codes with supremum
error exponent $\EE$, and }}excess-rate exponent \textup{$\mathsf{E}_{r}$}\emph{
}\textup{\emph{at rate }}$\RR$. 
\end{thm}
Notice that the functions $e_{\s[rb]}(t)$, $e_{\s[ex]}(t)$ and $e_{\s[sp]}(t)$
are concave functions of $t$ (as pointwise minimum of linear functions
in $t$), and thus the maximization over $t$ is relatively simple
to perform. In addition, $\max_{0\leq t\leq1}e_{\s[rb]}(t)$, $\max_{t\geq1}e_{\s[ex]}(t)$
and $\max_{t\geq0}e_{\s[sp]}(t)$ are non-increasing functions of
$\mathsf{E}_{r}$ and so for any given constraint on $\EE$ and target
rate $\RR$, a simple line search algorithm will find $E_{r}(\RR)=\min\left\{ \mathsf{E}_{r}:\left(\RR,\mathsf{E}_{r}\right)\mbox{is achievable for }\EE\right\} $.
Thus, the computational problem is to compute $e_{\s[rb]}(t)$, $e_{\s[ex]}(t)$
and $e_{\s[sp]}(t)$, for any given $t$. We address this matter in
Section \ref{sec:Computational algorithms}. 

For the sake of comparison, we mention fixed-rate coding and coding
under average rate constraint. In the case of fixed-rate coding, to
ensure an infimum error exponent of $\EE$ one must use $\rho(Q_{X})=\RR_{0}=\max_{Q'_{X}}\underline{\rho}^{*}(Q'_{X},\EE)$
for all $Q_{X}$, and the excess-rate exponent is as in \eqref{eq: excess rate coding fixed rate}.
For coding under average rate constraint, to ensure an infimum error
exponent of $\EE$ one can choose $\rho(P_{X})=\underline{\rho}^{*}(P_{X},\EE)$
and $\rho(Q_{X})=H(Q_{X})$ otherwise, and the excess-rate exponent
is as in \eqref{eq: excess rate coding average rate improved}. It
is also evident that if $\max_{Q'_{X}}\underline{\rho}^{*}(Q'_{X},\EE)=\underline{\rho}^{*}(P_{X},\EE)$
then fixed-rate coding is optimal and the excess rate exponent cannot
be improved beyond that of fixed-rate.

\section{Computational Algorithms\label{sec:Computational algorithms}}

As we have seen in Sections \ref{sec:Optimal rate functions} and
\ref{sec:Excess-Rate-Performance}, in order to compute the bounds
on the optimal rate functions and the resulting excess-rate performance,
some optimization problems need to be solved. In essence, since the
bounds on the optimal rate functions stem from the bounds on channel
coding error exponents, any computational algorithm for channel coding
error exponents may be used, e.g. \cite{arimoto_exponents_computation,Lesh}.
However, these classical algorithms are given for Gallager-style bounds
\cite{gallager1968information}, not the form of Csisz{\'a}r and K{\"o}rner
\cite{csiszar2011information}, used in this paper. In addition, they
form the basis for the computational algorithm of the excess-rate
performance for the random binning and sphere packing bounds. 

For the random binning and sphere packing rate functions \eqref{eq: random binning rate function expression},\eqref{eq: sphere packing rate function expression},
it is required to compute%
\footnote{Notice that the affine (third) term in \eqref{eq: random binning rate function expression}
can simply be obtained by setting $\EE=\infty$. The solution in this
case is $Q'_{Y|X}$.%
} 

\begin{equation}
v_{\s[rb]}(P_{XY},Q_{X},\EE,\eta)\teq\min_{Q_{Y|X}:D(Q_{X}\times Q_{Y|X}||P_{XY})\leq\EE}\left\{ D(Q_{Y|X}||Q_{Y}|Q_{X})+\eta\cdot D(Q_{Y|X}||P_{Y|X}|Q_{X})\right\} ,\label{eq: value definition random binning}
\end{equation}
where $\eta$ is $1$ for \eqref{eq: random binning rate function expression}
and $0$ for \eqref{eq: sphere packing rate function expression}.
For expurgated rate function \eqref{eq: expurgated rate function expression}
it is required to compute%
\footnote{The affine (second) term in \eqref{eq: expurgated rate function expression}
can be handled by similar methods. The solution in this case is simply
$Q'_{Z|X}$.%
} 
\begin{equation}
v_{\s[ex]}(P_{XY},Q_{X},\EE)\teq\min_{Q_{X\tilde{X}}:B(Q_{X\tilde{X}})+D(Q_{X}||P_{X})=\EE,\: Q_{X}=Q_{\tilde{X}}}D(Q_{\tilde{X}|X}||Q_{\tilde{X}}|Q_{X}).\label{eq: value definition expurgated}
\end{equation}
Moreover, to compute the bounds on the excess-rate performance in
\eqref{eq: minimization problem excess rate random binning} and \eqref{eq: minimization problem excess rate expurgated},
the values of $e_{\s[rb]}(P_{XY},\RR,\mathsf{E}_{r},t)$ and $e_{\s[ex]}(P_{XY},\RR,\mathsf{E}_{r},t)$
need to be computed%
\footnote{For concreteness, we have made explicit the dependence of $e_{\s[rb]}(t)$
and $e_{\s[ex]}(t)$ on $(P_{XY},\RR,\mathsf{E}_{r})$.%
}. In this section, we provide explicit iterative algorithms to compute
$v_{\s[rb]}(P_{XY},Q_{X},\EE,\eta)$, $v_{\s[ex]}(P_{XY},Q_{X},\EE)$
and $e_{\s[rb]}(P_{XY},\RR,\mathsf{E}_{r},t)$, and prove their correctness.
The merit of these algorithms is that they require at most a one-dimensional
optimization, regardless of the alphabet sizes $|{\cal X}|$ and $|{\cal Y}|$.
The optimization problem of $e_{\s[ex]}(P_{XY},\RR,\mathsf{E}_{r},t)$
is briefly discussed, and shown to be convex, rendering it feasible
to compute using generic algorithms. \textbf{}

Throughout, we will utilize an auxiliary PMF $\tilde{Q}_{Y}$. For
$0\leq\alpha\leq1$, define the \emph{geometric combination mapping}
$\mathbb{M}_{\s[g]}(P_{Y|X},\tilde{Q}_{Y},\alpha)$ whose output $\acute{Q}_{Y|X}$
satisfies 

\begin{equation}
\acute{Q}_{Y|X}(y|x)\teq\psi_{x}P_{Y|X}^{\alpha}(y|x)\tilde{Q}_{Y}^{1-\alpha}(y),\label{eq: definition of alpha conditional probability}
\end{equation}
for all $x\in{\cal X},y\in{\cal Y}$, where $\psi_{x}$ is a normalization
factor, chosen such that $\sum_{y\in{\cal Y}}\acute{Q}_{Y|X}(y|x)=1$
for all $x\in{\cal X}$. The following algorithm provides a method
to compute $v_{\s[rb]}(P_{XY},Q_{X},\EE,\eta)$. 

\begin{algorithm}
\begin{onehalfspace}
\textbf{Input}: A source $P_{XY}$, a type $Q_{X}$, a target error
exponent $\EE$ and $\eta\geq0$. 
\end{onehalfspace}

\textbf{Output}: The value of $v_{\s[rb]}(P_{XY},Q_{X},\EE,\eta)$
and the optimal solution $Q_{Y|X}^{*}$.
\begin{enumerate}
\item Initialize $\tilde{Q}_{Y}$ randomly such that $\supp(\tilde{Q}_{Y})=\supp(\sum_{x\in{\cal X}}Q_{X}P_{Y|X})$. 
\item Iterate over the following steps until convergence:

\begin{enumerate}
\item \label{enu: Interative minimization algorithm, bisection step}Set
$Q_{Y|X}=\mathbb{M}_{\s[g]}(P_{Y|X},\tilde{Q}_{Y},\frac{\eta}{1+\eta})$.
If $D(Q_{Y|X}||P_{Y|X}|Q_{X})<\EE$ then set $\alpha=\frac{\eta}{1+\eta}$.
Else, find $\alpha^{*}\in[\frac{\eta}{1+\eta},1]$ that satisfies\emph{
\[
D\left(\mathbb{M}_{\s[g]}(P_{Y|X},\tilde{Q}_{Y},\alpha^{*})||P_{Y|X}|Q_{X}\right)=\EE-D(Q_{X}||P_{X})
\]
}and set $Q_{Y|X}=\mathbb{M}_{\s[g]}(P_{Y|X},\tilde{Q}_{Y},\alpha^{*})$.
\item Set\emph{ $\tilde{Q}_{Y}(y)=\sum_{x\in{\cal X}}Q_{X}(x)Q_{Y|X}(y|x)$.}
\end{enumerate}
\item Let the converged variable be $\alpha^{*}$ and $\tilde{Q}_{Y}^{*}$.
Then, set $Q_{Y|X}=\mathbb{M}_{\s[g]}(P_{Y|X},\tilde{Q}_{Y}^{*},\alpha^{*})$
in \eqref{eq: value definition random binning}. Return. 
\end{enumerate}
\protect\caption{Alternating minimization algorithm for the computation of $v_{\s[rb]}(P_{XY},Q_{X},\EE,\eta)$
\label{alg:Alternating minimization algorithm v rb}}
\end{algorithm}

\begin{lem}
\label{lem:Proof of alternating minimization algorithm v rb}Algorithm
\ref{alg:Alternating minimization algorithm v rb} outputs $v_{\s[rb]}(P_{XY},Q_{X},\EE,\eta)$.
\end{lem}
Algorithm \ref{alg:Alternating minimization algorithm v rb} is presented
for a specific $\EE$, but it is also useful if one is interested
in the full curve $\rho_{\s[rb]}(Q_{X},\EE)$. To compute the second
term in the random binning rate function \eqref{eq: random binning rate function expression}
one needs to compute
\begin{eqnarray}
 &  & \min_{Q_{Y|X}\in{\cal A}_{\s[rb]}}I(Q_{X}\times Q_{Y|X})+D(Q_{Y|X}||P_{Y|X}|Q_{X})\label{eq: iterative max min rb first}\\
 &  & =\min_{Q_{Y|X}}\max_{\lambda\geq0}D(Q_{Y|X}||Q_{Y}|Q_{X})+D(Q_{Y|X}||P_{Y|X}|Q_{X})\nonumber \\
 &  & \,\,\,\,\,\,\,\,\,\,\,\,\,\,\,\,\,\,\,\,\,\,\,\,\,\,\,\,+\lambda\left(D(Q_{X}||P_{X})+D(Q_{Y|X}||P_{Y|X}|Q_{X})-\EE\right)\\
 &  & \trre[=,a]\max_{\lambda\geq0}\left\{ \lambda\left(D(Q_{X}||P_{X})-\EE\right)+\min_{Q_{Y|X}}D(Q_{Y|X}||Q_{Y}|Q_{X})+(1+\lambda)D(Q_{Y|X}||P_{Y|X}|Q_{X})\right\} \label{eq: iterative max min rb last}
\end{eqnarray}
where $(a)$ is because the minimization problem is convex. The KKT
optimality conditions \cite[Section 5.5.3]{boyd2004convex}\emph{
}imply that for any given $\lambda\in[0,\infty)$ the inner minimizer
$Q_{Y|X}^{*}(\lambda)$ of last line in \eqref{eq: iterative max min rb last}
is also the optimal solution for \eqref{eq: iterative max min rb last},
whenever the error exponent constraint in ${\cal A}_{\s[rb]}$ is
given by

\[
\EE(\lambda)=D(Q_{X}||P_{X})+D(Q_{Y|X}^{*}(\lambda)||P_{Y|X}|Q_{X}).
\]
Clearly, Algorithm \ref{alg:Alternating minimization algorithm v rb}
is suitable for the inner minimization in \eqref{eq: iterative max min rb last},
simply by setting $\eta=\lambda+1$ and $\EE=\infty$. Equivalently,
this means that in step \ref{enu: Interative minimization algorithm, bisection step}
of the algorithm, we always set $\alpha^{*}=\frac{\eta}{1+\eta}=\frac{\lambda+1}{\lambda+2}$.
Otherwise stated, when $\alpha^{*}$ varies from $1$ to $\frac{1}{2}$,
the curved part of $\rho_{\s[rb]}(Q_{X},\EE)$ is exhausted.

Next, Algorithm \ref{alg:Alternating minimization algorithm V rb}
provides a method to compute $e_{\s[rb]}(P_{XY},\RR,\mathsf{E}_{r},t)$.
The technique is somewhat similar to Algorithm \ref{alg:Alternating minimization algorithm v rb},
but here an additional optimization is carried out over $Q_{X}$.
For this algorithm, we define
\[
h_{1,t}(x)\teq D(\overline{Q}_{Y|X}(\cdot|x)||P_{Y|X}(\cdot|x)),
\]
\[
h_{2,t}(x)\teq D(\overline{Q}_{Y|X}(\cdot|x)||\tilde{Q}_{Y}),
\]
where $\overline{Q}_{Y|X}=\mathbb{M}_{\s[g]}(P_{Y|X},\tilde{Q}_{Y},\frac{1}{1+t})$,
as well as the mapping $\mathbb{M}_{\s[h]}(P_{X},h_{1},h_{2},\lambda,t)$
whose output $\overline{Q}_{X}$ satisfies 
\begin{equation}
\overline{Q}_{X}(x)=\psi\cdot\left[P_{X}(x)\right]^{\frac{1+\lambda}{1+\lambda+t}}\cdot\exp\left(-\frac{1}{1+t+\lambda}\cdot h_{1,t}(x)-\frac{t}{1+t+\lambda}\cdot h_{2,t}(x)\right),\label{eq: optimal prior excess rate}
\end{equation}
for all $x\in{\cal X}$, where $\psi$ is a normalization factor\emph{,
}such that $\sum_{x\in{\cal X}}\overline{Q}_{X}(x)=1$\emph{. }

\begin{algorithm}
\textbf{Input}: A source $P_{XY}$, a target rate $\RR$, a target
excess-rate $\mathsf{E}_{r}$ and $t\geq0$.

\textbf{Output}: The value of $e_{\s[rb]}(P_{XY},\RR,\mathsf{E}_{r},t)$.
\begin{enumerate}
\item Initialize $\tilde{Q}_{Y}$ randomly such that $\supp(\tilde{Q}_{Y})={\cal Y}$,
and set $\overline{Q}_{Y|X}=\mathbb{M}_{\s[g]}(P_{Y|X},\tilde{Q}_{Y},\frac{1}{1+t})$
and compute $h_{1}$ and $h_{2}$. 
\item Iterate over the following steps until convergence:

\begin{enumerate}
\item \label{enu: excess rate algorithm, bisection step}Set $\overline{Q}_{X}=\mathbb{M}_{\s[h]}(P_{X},h_{1},h_{2},0,t)$.
If $D(\overline{Q}_{X}||P_{X})<\mathsf{E}_{r}$ then set $\lambda=0$.
Else, find $\lambda^{*}>0$ that satisfies 
\[
D(\mathbb{M}_{\s[h]}(P_{X},h_{1},h_{2},\lambda,t)||P_{X})=\mathsf{E}_{r}
\]
and set $\overline{Q}_{X}=\mathbb{M}_{\s[h]}(P_{X},h_{1},h_{2},\lambda^{*},t)$.\emph{ }
\item Set $\tilde{Q}_{Y}(y)=\sum_{x\in{\cal X}}\overline{Q}_{X}(x)\overline{Q}_{Y|X}(y|x)$
for all $y\in{\cal Y}$, set $\overline{Q}_{Y|X}=\mathbb{M}_{\s[g]}(P_{Y|X},\tilde{Q}_{Y},\frac{1}{1+t})$
and compute $h_{1}$ and $h_{2}$. 
\end{enumerate}
\item Let the converged variables be $\lambda^{*}$ and $\tilde{Q}_{Y}^{*}$\emph{.
}Set $\overline{Q}_{X},\overline{Q}_{Y|X}$ in \eqref{eq: minimization problem excess rate random binning}.
Return\emph{. }
\end{enumerate}
\protect\caption{Alternating minimization algorithm for the computation of $e_{\s[rb]}(P_{XY},\RR,\mathsf{E}_{r},t)$
\label{alg:Alternating minimization algorithm V rb}}
\end{algorithm}

\begin{lem}
\label{lem:Proof of alternating minimization algorithm Value rb}Algorithm
\ref{alg:Alternating minimization algorithm V rb} outputs $e_{\s[rb]}(P_{XY},\RR,\mathsf{E}_{r},t)$.
\end{lem}
Next, Algorithm \ref{alg:Iterative scaling algorithm value ex} provides
a method to compute $v_{\s[ex]}(P_{XY},Q_{X},\EE)$. We define the
\emph{Bhattacharyya mapping} $\mathbb{M}_{\s[B]}(Q_{\tilde{X}|X},P_{Y|X},\lambda)$
whose output $\acute{Q}_{\tilde{X}|X}$ satisfies 
\[
\acute{Q}_{\tilde{X}|X}(\tilde{x}|x)\teq\psi_{x}Q_{\tilde{X}|X}(\tilde{x}|x)\exp\left[-\lambda\cdot d_{P_{Y|X}}(x,\tilde{x})\right]
\]
for all $x,\tilde{x}\in{\cal X}$, and $\psi_{x}$ is a normalization
constant, such that $\sum_{\tilde{x}\in{\cal X}}\acute{Q}_{\tilde{X}|X}(\tilde{x}|x)=1$,
for any $x\in\supp(Q_{X})$. Similarly, define the \emph{lumping mapping}
$\mathbb{M}_{\s[l]}(Q_{X\tilde{X}})$ whose output $\acute{Q}_{X\tilde{X}}$
satisfies 
\[
\acute{Q}_{X\tilde{X}}(x,\tilde{x})=Q_{X}(\tilde{x})\cdot\frac{Q_{X\tilde{X}}(x,\tilde{x})}{\sum_{x'\in{\cal X}}Q_{X\tilde{X}}(x',\tilde{x})}
\]
for all $x,\tilde{x}\in{\cal X}$.

\begin{algorithm}
\begin{onehalfspace}
\textbf{Input}: A source $P_{XY}$, a type $Q_{X}$ and a target error
exponent $\EE$.
\end{onehalfspace}

\textbf{Output}: The value of $v_{\s[ex]}(P_{XY},Q_{X},\EE)$ and
the optimal solution $Q_{\tilde{X}|X}^{*}$.
\begin{enumerate}
\item Initialize $Q_{X\tilde{X}}^{(0)}(x,\tilde{x})=Q_{X}(x)Q_{X}(\tilde{x})$
for all $x,\tilde{x}\in{\cal X}$. 
\item For $k=1,\ldots$, iterate over the following steps until convergence: 

\begin{enumerate}
\item Find $\lambda^{*}\in\mathbb{R}$ that satisfies $B(Q_{X}\times\mathbb{M}_{\s[B]}(Q_{\tilde{X}|X}^{(2k-2)},P_{Y|X},\lambda))+D(Q_{X}||P_{X})=\EE$,
and\emph{ }set $Q_{\tilde{X}|X}^{(2k-1)}=\mathbb{M}_{\s[B]}(Q_{\tilde{X}|X}^{(2k-2)},P_{Y|X},\lambda^{*})$.
\item Set $Q_{X\tilde{X}}^{(2k)}=\mathbb{M}_{\s[l]}(Q_{X}\times Q_{\tilde{X}|X}^{(2k-1)})$.
\end{enumerate}
\item Let the converged variables be $Q_{X\tilde{X}}^{*}(x,\tilde{x})$.
Then, set $Q_{X\tilde{X}}^{*}(x,\tilde{x})$ in \eqref{eq: value definition expurgated}.
Return. 
\end{enumerate}
\protect\caption{Iterative scaling algorithm for the computation of $v_{\s[ex]}(P_{XY},Q_{X},\EE)$
\label{alg:Iterative scaling algorithm value ex}}
\end{algorithm}

\begin{lem}
\label{lem:Proof of alternating minimization algorithm value ex}Algorithm
\ref{alg:Iterative scaling algorithm value ex} outputs $v_{\s[ex]}(P_{XY},Q_{X},\EE)$.
\end{lem}
Finally, we discuss the computation of $e_{\s[ex]}(P_{XY},\RR,\mathsf{E}_{r},t)$.
We have 
\begin{eqnarray}
e_{\s[ex]}(P_{XY},\RR,\mathsf{E}_{r},t) & = & \min_{Q_{X}:D(Q_{X}||P_{X})\leq\mathsf{E}_{r}}\min_{Q_{\tilde{X}|X}:(Q_{X}\times Q_{\tilde{X}|X})=Q_{X}}\biggl\{ D(Q_{X}||P_{X})+B(Q_{X\tilde{X}})\nonumber \\
 &  & +t\cdot\left[\RR-H(Q_{X})+D(Q_{\tilde{X}|X}||Q_{\tilde{X}}|Q_{X})\right]\biggr\}\\
 & = & \min_{Q_{X}:D(Q_{X}||P_{X})\leq\mathsf{E}_{r}}\min_{Q_{\tilde{X}|X}:(Q_{X}\times Q_{\tilde{X}|X})=Q_{X}}\min_{\tilde{Q}_{\tilde{X}}}\biggl\{ D(Q_{X}||P_{X})+B(Q_{X\tilde{X}})\nonumber \\
 &  & +t\cdot\left[\RR-H(Q_{X})+D(Q_{\tilde{X}|X}||\tilde{Q}_{\tilde{X}}|Q_{X})\right]\biggr\}\\
 & = & \min_{Q_{X\tilde{X}}:D(Q_{X}||P_{X})\leq\mathsf{E}_{r},\: Q_{X}=Q_{\tilde{X}}}\min_{\tilde{Q}_{\tilde{X}}}\Bigl\{ D(Q_{X}||P_{X})+B(Q_{X\tilde{X}})\nonumber \\
 &  & +t\cdot\left[\RR-H(Q_{X})+D(Q_{X\tilde{X}}||Q_{X}\times\tilde{Q}_{\tilde{X}})\right]\Bigr\}.
\end{eqnarray}
It can be easily seen that the resulting optimization problem is convex
in the variables $(\tilde{Q}_{\tilde{X}},Q_{X\tilde{X}})$, and can
be solved by any general solver. Unfortunately, we have not been able
to prove that alternating minimization algorithm converges (and even
in this event, there is no explicit solution for the optimal $Q_{X\tilde{X}}$
given some $\tilde{Q}_{\tilde{X}}$).

\section{\label{sec:A-Numerical-Example}A Numerical Example}

In this section, we provide a simple numerical example to illustrate
the bounds obtained in previous sections, utilizing the computational
algorithms of Section \ref{sec:Computational algorithms}. Let the
alphabets be ${\cal X}=\{0,1\}$ and ${\cal Y}=\{0,\chi,1\}$, $P_{X}$
be given by $P_{X}(0)=1-P_{X}(1)=0.2$, and $P_{Y|X}$ be given by
the following transition probability matrix 
\[
P_{Y|X}=\left[\begin{array}{ccc}
0.8 & 0.15 & 0.05\\
0.05 & 0.15 & 0.8
\end{array}\right].
\]

\begin{figure}
\begin{centering}
\includegraphics[scale=0.5]{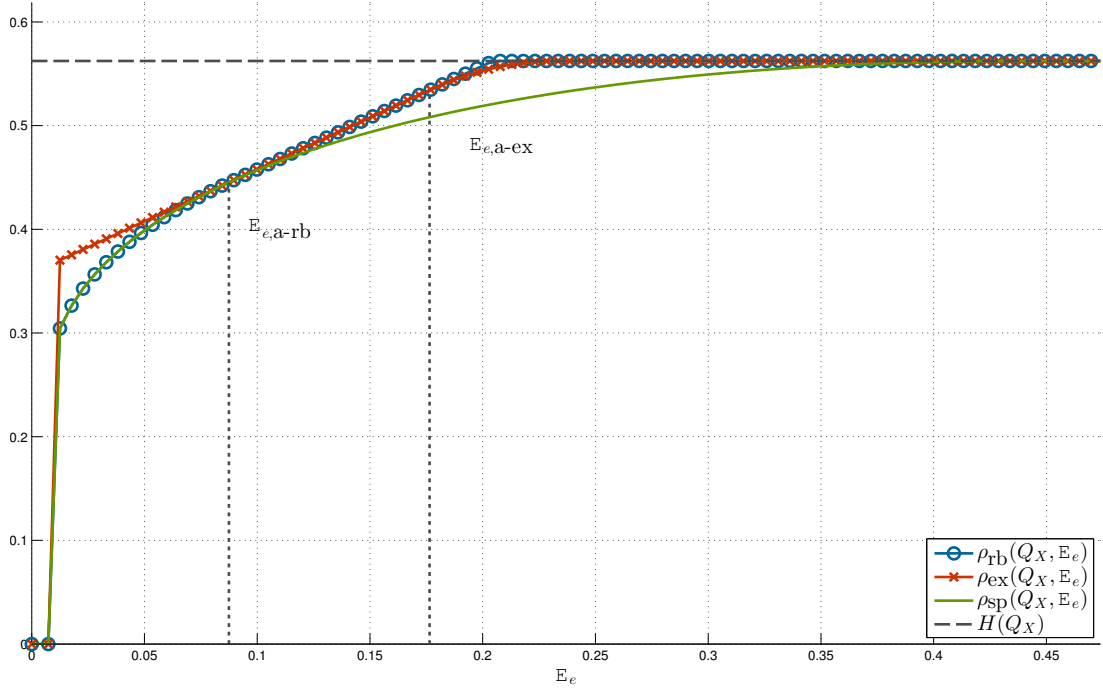}
\par\end{centering}

\protect\caption{Bounds on the optimal rate functions for the type $Q_{X}(0.25)=0.25$.\label{fig: dominating rate function for given type}}
\end{figure}

Figure \ref{fig: dominating rate function for given type} shows the
bounds on the optimal rate functions (in nats) for $Q_{X}$ given
by $Q_{X}(0)=1-Q_{X}(1)=0.25$ as a function of $\EE$. The points
at which $\rho_{\s[rb]}(Q_{X},\EE)$ ($\rho_{\s[ex]}(Q_{X},\EE)$)
becomes (respectively, ceases to be) affine with a unity slope are
indicated by a vertical lines. For small $\EE$, the random binning
and sphere packing bounds coincide, and so, $\overline{\rho}^{*}(Q_{X},\EE)=\rho_{\s[rb]}(Q_{X},\EE)=\rho_{\s[sp]}(Q_{X},\EE)=\underline{\rho}^{*}(Q_{X},\EE)$.

\begin{figure}
\begin{centering}
\includegraphics[scale=0.5]{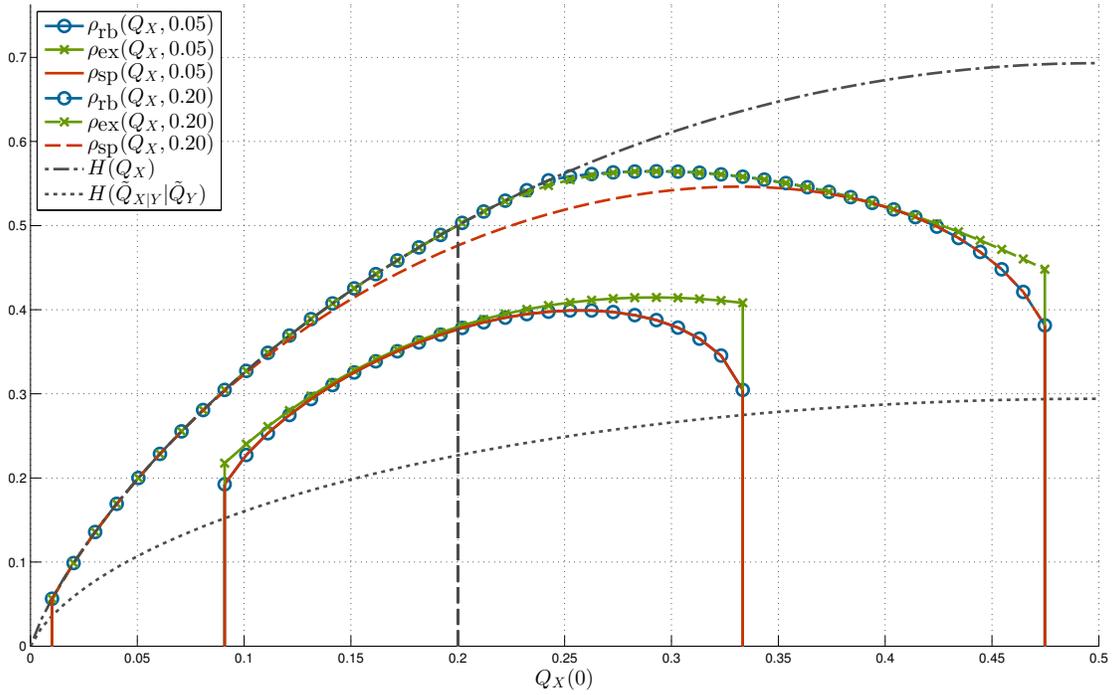}
\par\end{centering}

\protect\caption{Bounds on the optimal rate functions for a given $\EE$.\label{fig: dominating rate function for given error}}
\end{figure}

\begin{figure}
\centering{}\includegraphics[scale=0.5]{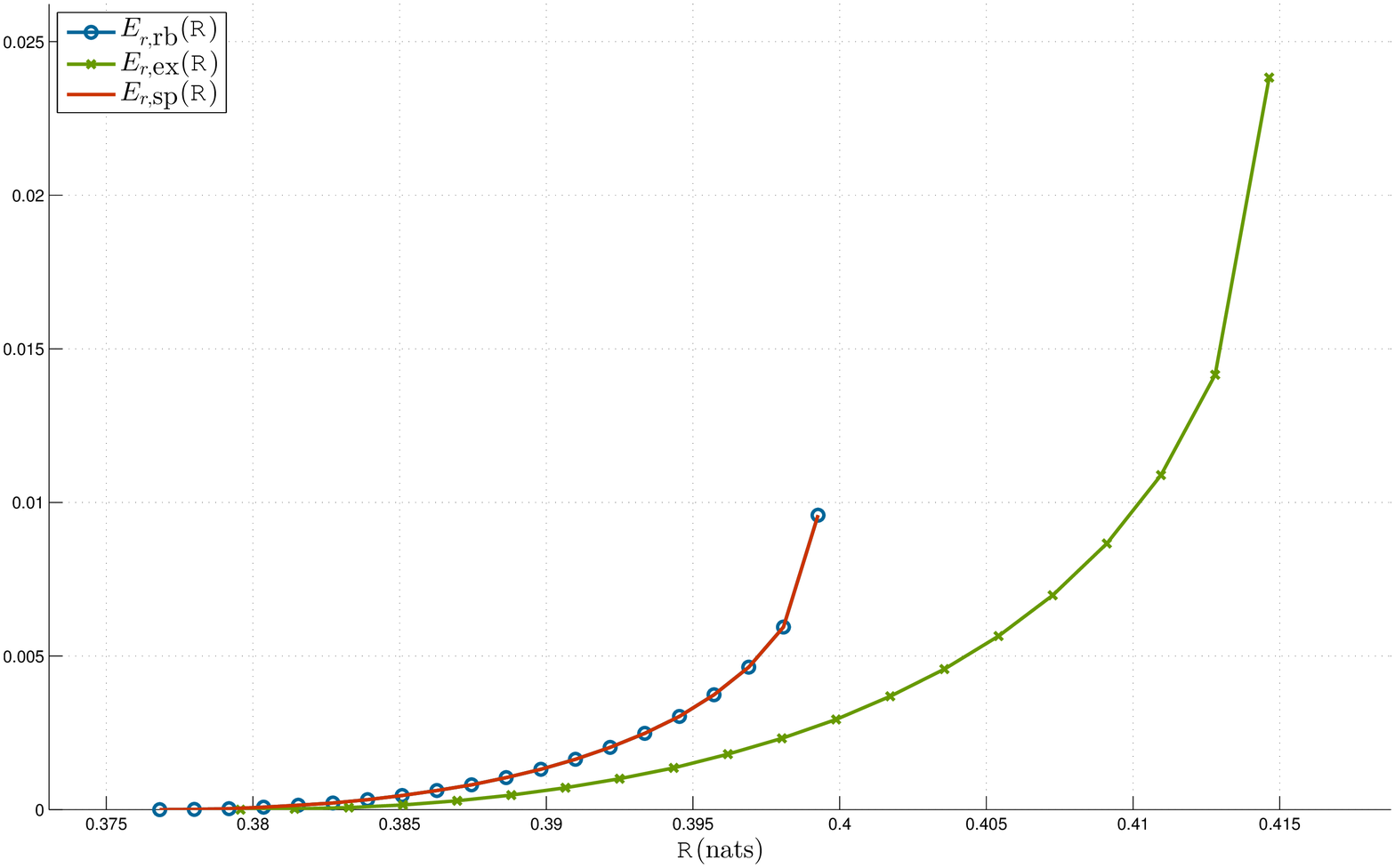}\protect\caption{Bounds on the optimal excess-rate exponent for $\EE=0.05$. \label{fig: Excess rate exponent}}
\end{figure}

\begin{figure}
\centering{}\includegraphics[scale=0.5]{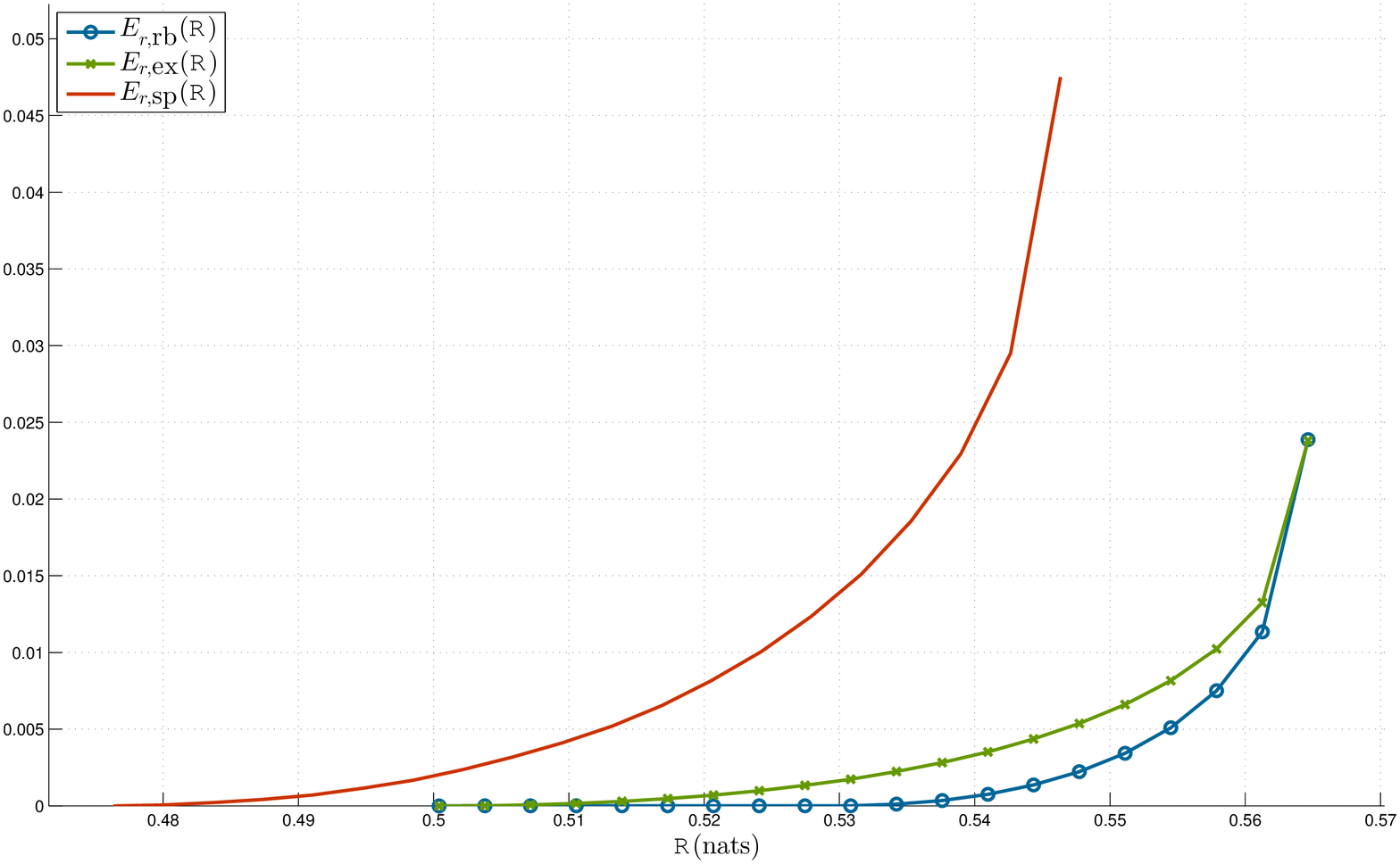}\protect\caption{Bounds on the optimal excess-rate exponent for $\EE=0.2$. \label{fig: Excess rate exponent 2}}
\end{figure}

Figure \ref{fig: dominating rate function for given error} shows
the bounds on the optimal rate functions (in nats), for all possible
types (indexed by $Q_{X}(0)$) for $\EE=0.05$ and $\EE=0.2$. It
can be seen that indeed this optimal function is in the form of a
regular rate function, and that for $\EE=0.05$ the optimal rate function
is exactly known, for all types of the source. For comparison, the
entropies $H(Q_{X})$ and $H(\tilde{Q}_{X|Y}|\tilde{Q}_{Y})$ where
$\tilde{Q}_{XY}=Q_{X}\times P_{Y|X}$ are also plotted, and the rates
for $P_{X}$ are marked. The bounds on the optimal excess-rate exponent
are computed and plotted in Figure \ref{fig: Excess rate exponent}
for $\EE=0.05$ and in Figure \ref{fig: Excess rate exponent 2} for
 $\EE=0.2$. As before, for the smaller $\EE$ the optimal excess-rate
exponent is obtained exactly, while a gap exists for the larger $\EE$.
It can be verified that Figure \ref{fig: dominating rate function for given error}
and Figure \ref{fig: Excess rate exponent} are consistent. For example,
for $\EE=0.05$ it can be seen in Figure \ref{fig: dominating rate function for given error}
that when the type is $Q_{X}=P{}_{X}$, the rate is $\underline{\rho}^{*}(P_{X},0.05)=\overline{\rho}^{*}(P_{X},0.05)\approx0.377$
nats so the excess-rate exponent is $\underline{E}_{r}^{*}(0.377)=\overline{E}_{r}^{*}(0.377)=0$.
Then, as $Q_{X}(0)$ increases, the rate also increases, up to its
maximal value of $\underline{\rho}^{*}(Q_{X}^{*},0.05)=\overline{\rho}^{*}(Q_{X}^{*},0.05)\approx0.4$,
for $Q_{X}^{*}(0)\approx0.2574$. The excess-rate exponent is determined
by the divergence of this type from the true source $P_{X}$, and
given by $\underline{E}_{r}^{*}(0.4)=\overline{E}_{r}^{*}(0.4)\approx D(Q_{X}^{*}||P_{X})\approx10^{-2}$.
This is the maximal value of $\underline{E}_{r}^{*}(\RR)$ shown in
Figure \ref{fig: Excess rate exponent}, and for larger rates, clearly
$\underline{E}_{r}^{*}(\RR)=\infty$. 

For comparison, we also consider fixed-rate coding. From Figure \ref{fig: Excess rate exponent},
for $\EE=0.05$ we have $\underline{E}_{r}^{*}(0.3921)=2\cdot10^{-3}$.
It can be found that if one uses fixed-rate coding, at rate $\RR_{0}=0.3921$,
for all $Q_{X}$ then the error exponent achieved is only $\EE\approx0.045$.
Therefore, if the \emph{finite} excess-rate exponent of variable-rate
coding is tolerated, then this provides an improvement in the error
exponent over fixed-rate coding.

\section{Summary\label{sec:Summary}}

In this paper, we have considered the trade-off between error and
excess-rate exponents for variable-rate SW coding. The cases of fixed-rate
coding and variable-rate coding under average constraints may be considered
as two extreme points in this trade-off. In fixed-rate coding the
same rate is assigned to all possible types, and so, the maximal excess-rate
exponent is achieved, but at the price of minimal error exponent.
In average-rate coding, the main concern is the coding of the true
type of the source, and all other types are sent uncoded. The resulting
error exponent is maximal, but at the price of minimal excess-rate
exponent. Thus, for a coding system with more stringent instantaneous
rate demands, it is necessary to lose some of the gains in error exponent
of variable-rate coding, and improve the excess-rate exponent. In
this work, we have derived bounds on rate functions which achieve
the optimal trade-off, and analyzed their excess-rate performance,
for a given requirement on the error exponent. 

Before we conclude, we briefly outline two possible extensions. In
many practical cases, there is some uncertainty regarding the source
$P_{XY}=P_{X}\times P_{Y|X}$. Clearly, if independence between $X$
and $Y$ is a possible scenario, then in this worst case, the side
information $\mathbf{y}$ is useless (when no feedback link exists).
In other cases, it may be known that $P_{XY}\in{\cal F}\subset{\cal Q}({\cal X}\times{\cal Y})$
for some family of distributions ${\cal F}$. In this case, a possible
requirement is that the rate function $\rho(Q_{X})$ will be chosen
to achieve error exponent of $\EE$ uniformly for all sources in ${\cal F}$.
With a slight change and abuse of notation, we define, e.g. the infimum
optimal rate function for the source $P_{XY}$ as $\underline{\rho}^{*}(Q_{X},\EE;P_{XY})$
and the optimal rate function for the family ${\cal F}$ as 

\begin{equation}
\underline{\rho}^{*}(Q_{X},\EE;{\cal F})\teq\max_{P_{XY}\in{\cal F}}\underline{\rho}^{*}(Q_{X},\EE;P_{XY}).\label{eq: Maximizinig unknown source}
\end{equation}

This maximization is (relatively) easy to perform if, e.g., the conditional
probability $P_{Y|X}$ is known exactly, and in addition, a \emph{nominal
$\tilde{P}_{X}$ }is known such that the actual $P_{X}$ satisfies
$D(\tilde{P}_{X}||P_{X})\leq\mathsf{U}$, for some given uncertainty
level $\mathsf{U}>0$ (recall Pinsker's inequality \cite[Lemma 11.6.1]{Cover:2006:EIT:1146355}
and see also the discussion in \cite{rezaei2005robust}). A direction
for future research is to derive bounds on optimal rate functions
and their excess-rate performance which are robust for source uncertainty
of various kinds. 

In this paper, we have focused on the SW scenario in which the side
information vector $\mathbf{y}$ is known exactly to the source. Similar
techniques can also be applied to the more general case of SW coding,
where the side information is also encoded. In this case, there are
two encoders, $s_{n}$ for encoding $\mathbf{x}$ and $s_{n}'$ for
encoding $\mathbf{y}$, while the central decoder $\sigma_{n}$ now
uses both codewords $s_{n}(\mathbf{x})$ and $s'_{n}(\mathbf{y})$.
For type-dependent, variable-rate codes, two rate functions $\rho_{X}(Q_{X})$
and $\rho_{Y}(Q_{Y})$ may be defined accordingly. While bounds on
the resulting error exponent may be derived, the trade-off in this
case is more complicated. First, there are two excess-rate exponents,
one for each of the decoders. Second, a trickle of coordination might
be required between the two encoders in order to ensure a required
error exponent.Specifically, at least one of the encoders needs to
know the current rate (or equivalently, the type class of the current
source block) of the other encoder.

\section*{Acknowledgments}

The authors would like to thank the Associate Editor, Jun Chen, for
providing them the unpublished manuscript \cite{ChenUnpublished},
and for his useful comments. Specifically, the proof of Theorem \ref{thm: achieavbility with type dependent}
follows from the proof of Theorem \ref{thm: Error exponent bounds, single type}
which is a reproduction of \cite[Theorem 1]{ChenUnpublished}. Useful
comments made by the anonymous referees are also acknowledged with
thanks.

\appendices{\numberwithin{equation}{section}}

\section{\label{sec:Proofs}}
\begin{IEEEproof}[Proof of Theorem \ref{thm: Error exponent bounds, single type}]

\emph{\uline{Upper bound }}\eqref{eq: SW conditional error exponent bound sup}\emph{:}
Follows exactly as in the proof of \cite[Theorem 1]{ChenUnpublished}.

\emph{\uline{Lower bound }}\uline{\mbox{\eqref{eq: SW conditional error exponent bound inf}}}\emph{:}
The proof of the achievable bound is also very similar to the proof
of \cite[Theorem 1]{ChenUnpublished}, with a slight modification.
For completeness, we provide a proof here. 

For brevity, we will omit the notation of the dependence of $\overline{R}(Q_{X};{\cal S})$
in ${\cal S}$ and denote it by $\overline{R}(Q_{X})$. Assume that
$Q_{X}\in\interior{\cal Q}({\cal X}),$ and $Q_{X}\in{\cal P}_{n_{0}}({\cal X})$
for some minimal $n_{0}\in\mathbb{N}$. Since the statement in \eqref{eq: SW conditional error exponent bound inf}
is only about the conditional error exponent of the type $Q_{X}$,
it is clear that the SW codes constructed, may only encode $\mathbf{x}\in{\cal T}_{n}(Q_{X})$,
and so only block lengths $n\mod n_{0}\neq0$ should be considered,
as otherwise ${\cal T}_{n}(Q_{X})$ is empty, and the conditional
error probability is $0$, by definition. 

Let $\delta>0$ be given, and let ${\cal C}$ be a sequence of constant
composition channel codes of type $Q_{X}^{(n)}\to Q_{X}$, asymptotic
rate $\liminf_{n\to\infty}\frac{\log|{\cal C}_{n}|}{n}\geq H(Q_{X})-\overline{R}(Q_{X})-\frac{\delta}{2}$,
which also achieves the infimum reliability function for the channel
$P_{Y|X}$, i.e.
\begin{equation}
\liminf_{n\to\infty}-\frac{1}{n}\log p_{e}({\cal C}_{n})\geq\underline{E}_{e}^{*}(H(Q_{X})-\overline{R}(Q_{X}),Q_{X},P_{Y|X})-\delta.\label{eq: beginning with good channel code}
\end{equation}
From Lemma \ref{lem: exact type codebook}, it can be assumed w.l.o.g.
that for $n$ sufficiently large, whenever, $n\mod n_{0}=0$, the
codebook satisfies ${\cal C}_{n}\in{\cal T}_{n}(Q_{X})$. Now, assume
that $n$ is sufficiently large and that $n\mod n_{0}\neq0$. From
the covering lemma \cite[Section 6, Covering Lemma 2]{ahlswede1979coloring},
one can find 
\[
T_{n}=\exp\left[n\left(\overline{R}(Q_{X})+\delta\right)\right]
\]
permutations $\{\pi_{n,t}\}_{t=1}^{T_{n}}$, such that ${\cal T}_{n}(Q_{X})=\bigcup_{t=1}^{T_{n}}\pi_{n,t}({\cal C}_{n})$,
where $\pi_{n,t}({\cal C}_{n})$ means that the same permutation $\pi_{n,t}(\cdot)$
operates on codewords in the codebook. Since the channel $P_{Y|X}$
is memoryless then clearly $p_{e}(\pi({\cal C}_{n}))=p_{e}({\cal C}_{n})$
for any permutation $\pi$, since the decoder can always apply the
inverse permutation on $\mathbf{y}$ and decode as if the codebook
is ${\cal C}_{n}$. Let us define the following sequence of SW codes
${\cal S}^{*}=\{{\cal S}{}_{n}^{*}=(s_{n}^{*},\sigma_{n}^{*})\}$
from the channel codes ${\cal C}=\{{\cal C}_{n}=(f_{n},\varphi_{n})\}$. 
\begin{itemize}
\item \emph{Codebook Construction:} Generate the codebook ${\cal C}_{n}$
and enumerate the permutations $\{\pi_{n,t}\}_{t=1}^{T_{n}}$ such
that ${\cal T}_{n}(Q_{X})=\bigcup_{l=1}^{T_{n}}\pi_{n,t}({\cal C}_{n})$.
The above information is revealed to both the encoder and the decoder
off-line. 
\item \emph{Encoding:} Upon observing $\mathbf{x}$, determine its empirical
distribution $\hat{Q}_{\mathbf{x}}$. If $\hat{Q}_{\mathbf{x}}\not=Q_{X}$
the codeword is $s_{n}^{*}(\mathbf{x})=0$. Else, find $t^{*}(\mathbf{x})\teq\min\{t':\mathbf{x}\in\pi_{n,t'}({\cal C}_{n}(Q_{X}))\}$.
The codeword is $s_{n}^{*}(\mathbf{x})=(1,\tau(t^{*}(\mathbf{x})))$
where $\tau(t)$ is the binary representation of $t$ in $\left\lceil \log_{2}T_{n}\right\rceil $
bits. 
\item \emph{Decoding:} If $s{}_{n}^{*}(\mathbf{x})=0$ then declare an error.
Else, recover from $s_{n}^{*}(\mathbf{x})$ the permutation $\pi=\pi_{n,t(\mathbf{x})}$.
Find $\hat{t}(\mathbf{y})\teq t^{*}(\pi(\varphi_{n}(\pi^{-1}(\mathbf{y})))$,
and if $\hat{t}(\mathbf{y})=t^{*}(\mathbf{x})$ then decode $\hat{\mathbf{x}}=\pi(\varphi_{n}(\pi^{-1}(\mathbf{y})))$,
and otherwise declare an error. 
\end{itemize}

The conditional average rate of ${\cal S}^{*}$ satisfies $\limsup_{n\to\infty}\E[r^{*}(\mathbf{X})|\mathbf{X}\in{\cal T}_{n}(Q_{X})]=\overline{R}(Q_{X})+\delta$.
Since all source blocks in ${\cal T}_{n}(Q_{X})$ are equiprobable,
the conditional error probability satisfies 
\begin{alignat}{1}
\liminf_{n\to\infty}-\frac{1}{n}\log\P(\sigma{}_{n}^{*}(s{}_{n}^{*}(\mathbf{X}),\mathbf{Y})\neq\mathbf{X}|\mathbf{X}\in{\cal T}_{n}(Q_{X})) & \geq\liminf_{n\to\infty}-\frac{1}{n}\log\P(\varphi{}_{n}(\mathbf{Y})\neq\mathbf{X}|\mathbf{X}\in{\cal T}_{n}(Q_{X}))\nonumber \\
 & =\underline{E}_{e}^{*}(H(Q_{X})-\overline{R}(Q_{X}),Q_{X},P_{Y|X})-\delta,
\end{alignat}
where it should be emphasized that whenever $n\mod n_{0}\neq0$ then
$\P(\sigma{}_{n}^{*}(s{}_{n}^{*}(\mathbf{X}),\mathbf{Y})\neq\mathbf{X}|\mathbf{X}\in{\cal T}_{n}(Q_{X}))=0$
by convention. The result follows since $\delta>0$ was arbitrary.
Before concluding the proof, we make the following remark.
\begin{rem}
In the proof, the actual choice of the decoder was implicit since
the SW codes are constructed from channel codes. However, as is well
known, the optimal decoder in terms of minimum error probability is
to decode $\tilde{\mathbf{x}}\in{\cal T}_{n}(Q_{X})\cap s_{n}^{-1}(s_{n}(\mathbf{x}))$
that maximizes $P_{\mathbf{X}|\mathbf{Y}}(\tilde{\mathbf{x}}|\mathbf{y})$.
Since all $\tilde{\mathbf{x}}\in s_{n}^{-1}(s_{n}(\mathbf{x}))$ are
in the type class $Q_{X}$, they have the same probability $P_{\mathbf{X}}(\tilde{\mathbf{x}})$,
so this decoding rule is equivalent to maximizing $P_{\mathbf{Y}|\mathbf{X}}(\mathbf{y}|\tilde{\mathbf{x}})$,
which is a \emph{maximum likelihood} (ML) decoding rule. Nonetheless,
there are cases in which other decoders, such as the minimum conditional
entropy decoder, also achieve the same error exponent (see Appendix
\ref{sec:Tightness-of-the-Random-Binning} for a precise definition).
This decoder has the merit of not depending on $P_{XY}$ and is therefore
a \emph{universal} decoder.
\end{rem}
\end{IEEEproof}

\begin{IEEEproof}[Proof of Theorem \ref{thm: General upper bound on error exponent}]
Since $|{\cal P}_{n}({\cal X})|\leq(n+1)^{|{\cal X}|}$, the error
probability satisfies 
\begin{alignat}{1}
p_{e}({\cal S}_{n}) & =\sum_{Q_{X}\in{\cal P}_{n}({\cal X})}\P({\cal T}_{n}(Q_{X}))\P(\hat{\mathbf{X}}\neq\mathbf{X}|\mathbf{X}\in{\cal T}_{n}(Q_{X}))\label{eq: conditional error probability method of types derivation begin}\\
 & \doteq\max_{Q_{X}\in{\cal P}_{n}({\cal X})}e^{-nD\left(Q_{X}||P_{X}\right)}\cdot\P(\hat{\mathbf{X}}\neq\mathbf{X}|\mathbf{X}\in{\cal T}_{n}(Q_{X}))\\
 & =\exp\left(-n\cdot\min_{Q_{X}\in{\cal P}_{n}({\cal X})}\left\{ D\left(Q_{X}||P_{X}\right)-\frac{1}{n}\log\P(\hat{\mathbf{X}}\neq\mathbf{X}|\mathbf{X}\in{\cal T}_{n}(Q_{X}))\right\} \right).\label{eq: conditional error probability method of types derivation end}
\end{alignat}
Now, for every $\epsilon>0$, let $Q_{X}^{*}\in{\cal P}({\cal X})$
be such that 
\begin{multline}
D\left(Q_{X}^{*}||P_{X}\right)+\limsup_{n\to\infty}\left\{ -\frac{1}{n}\log\P(\hat{\mathbf{X}}\neq\mathbf{X}|\mathbf{X}\in{\cal T}_{n}(Q_{X}^{*}))\right\} \leq\\
\inf_{Q_{X}\in{\cal P}({\cal X})}\left\{ D\left(Q_{X}||P_{X}\right)+\limsup_{n\to\infty}\left\{ -\frac{1}{n}\log\P(\hat{\mathbf{X}}\neq\mathbf{X}|\mathbf{X}\in{\cal T}_{n}(Q_{X}))\right\} \right\} +\epsilon\label{eq: ineq1}
\end{multline}
and let $m_{0}$ be sufficiently large so that
\begin{equation}
\sup_{n>m_{0}}\left\{ -\frac{1}{n}\log\P(\hat{\mathbf{X}}\neq\mathbf{X}|\mathbf{X}\in{\cal T}_{n}(Q_{X}^{*}))\right\} \leq\limsup_{n\to\infty}\left\{ -\frac{1}{n}\log\P(\hat{\mathbf{X}}\neq\mathbf{X}|\mathbf{X}\in{\cal T}_{n}(Q_{X}^{*}))\right\} +\epsilon.\label{eq:ineq2}
\end{equation}
Then, 
\begin{alignat}{1}
{\cal E}_{e}^{+}({\cal S}) & =\limsup_{n\to\infty}\min_{Q_{X}\in{\cal P}_{n}({\cal X})}\left\{ D\left(Q_{X}||P_{X}\right)-\frac{1}{n}\log\P(\hat{\mathbf{X}}\neq\mathbf{X}|\mathbf{X}\in{\cal T}_{n}(Q_{X}))\right\} \\
 & =\lim_{m\to\infty}\sup_{n\geq m}\min_{Q_{X}\in{\cal P}_{n}({\cal X})}\left\{ D\left(Q_{X}||P_{X}\right)-\frac{1}{n}\log\P(\hat{\mathbf{X}}\neq\mathbf{X}|\mathbf{X}\in{\cal T}_{n}(Q_{X}))\right\} \\
 & \trre[=,a]\lim_{m\to\infty}\sup_{n\geq m}\inf_{Q_{X}\in{\cal P}({\cal X})}\left\{ D\left(Q_{X}||P_{X}\right)-\frac{1}{n}\log\P(\hat{\mathbf{X}}\neq\mathbf{X}|\mathbf{X}\in{\cal T}_{n}(Q_{X}))\right\} \\
 & \leq\sup_{n\geq m_{0}}\inf_{Q_{X}\in{\cal P}({\cal X})}\left\{ D\left(Q_{X}||P_{X}\right)-\frac{1}{n}\log\P(\hat{\mathbf{X}}\neq\mathbf{X}|\mathbf{X}\in{\cal T}_{n}(Q_{X}))\right\} \\
 & \leq\inf_{Q_{X}\in{\cal P}({\cal X})}\left\{ D\left(Q_{X}||P_{X}\right)+\sup_{n\geq m_{0}}\left\{ -\frac{1}{n}\log\P(\hat{\mathbf{X}}\neq\mathbf{X}|\mathbf{X}\in{\cal T}_{n}(Q_{X}))\right\} \right\} \\
 & \leq D\left(Q_{X}^{*}||P_{X}\right)+\sup_{n>m_{0}}\left\{ -\frac{1}{n}\log\P(\hat{\mathbf{X}}\neq\mathbf{X}|\mathbf{X}\in{\cal T}_{n}(Q_{X}^{*}))\right\} \\
 & \trre[\leq,b]\inf_{Q_{X}\in{\cal P}({\cal X})}\left\{ D\left(Q_{X}||P_{X}\right)+\limsup_{n\to\infty}\left\{ -\frac{1}{n}\log\P(\hat{\mathbf{X}}\neq\mathbf{X}|\mathbf{X}\in{\cal T}_{n}(Q_{X}))\right\} \right\} +2\epsilon\\
 & \trre[\leq,c]\inf_{Q_{X}\in{\cal P}({\cal X})}\left\{ D\left(Q_{X}||P_{X}\right)+\overline{E}_{e}^{*}(H(Q_{X})-\overline{R}(Q_{X};{\cal S}),Q_{X},P_{Y|X})\right\} +2\epsilon
\end{alignat}
where $(a)$ is because, by assumption, if ${\cal T}_{n}(Q_{X})$
is empty then $\P(\hat{\mathbf{X}}\neq\mathbf{X}|\mathbf{X}\in{\cal T}_{n}(Q_{X}))=0$
, and $(b)$ is from \eqref{eq: ineq1} and \eqref{eq:ineq2}. The
inequality $(c)$ is due to the upper bound of Theorem \ref{thm: Error exponent bounds, single type}. 
\end{IEEEproof}

\begin{IEEEproof}[Proof of Theorem \ref{thm: General upper bound on excess rate exponent}]
The excess-rate exponent at the target rate $\RR$ is
\begin{alignat}{1}
p_{r}({\cal S}_{n},\RR) & =\sum_{Q_{X}\in{\cal P}_{n}({\cal X})}\P({\cal T}_{n}(Q_{X}))\P(r(\mathbf{X})\geq\RR|\mathbf{X}\in{\cal T}_{n}(Q_{X}))\label{eq: conditional excess rate probability method of types derivation begin}\\
 & \doteq\exp\left(-n\cdot\min_{Q_{X}\in{\cal P}_{n}({\cal X})}\left\{ D(Q_{X}||P_{X})-\frac{1}{n}\log\P(r(\mathbf{X})\geq\RR|\mathbf{X}\in{\cal T}_{n}(Q_{X}))\right\} \right).\label{eq: conditional excess rate probability method of types derivation end}
\end{alignat}
Now, let $\epsilon>0$ be given and let $m_{0}$ be sufficiently large
such that
\begin{multline}
\liminf_{n\to\infty}\left\{ D(Q_{X}||P_{X})-\frac{1}{n}\log\P(r(\mathbf{X})\geq\RR|\mathbf{X}\in{\cal T}_{n}(Q_{X}))\right\} \\
\leq\inf_{n\geq m_{0}}\min_{Q_{X}\in{\cal P}_{n}({\cal X})}\left\{ D(Q_{X}||P_{X})+\liminf_{n\to\infty}\left\{ -\frac{1}{n}\log\P(r(\mathbf{X})\geq\RR|\mathbf{X}\in{\cal T}_{n}(Q_{X}))\right\} \right\} +\epsilon.\label{eq: ineq2 excess rate}
\end{multline}
Also, choose $Q_{X}^{*}\in{\cal P}_{n_{0}}({\cal X})$ such that 
\begin{multline}
D(Q_{X}^{*}||P_{X})+\liminf_{n\to\infty}\left\{ -\frac{1}{n}\log\P(r(\mathbf{X})\geq\RR|\mathbf{X}\in{\cal T}_{n}(Q_{X}^{*}))\right\} \\
\leq\inf_{Q_{X}\in{\cal P}({\cal X})}\left\{ D(Q_{X}||P_{X})+\liminf_{n\to\infty}\left\{ -\frac{1}{n}\log\P(r(\mathbf{X})\geq\RR|\mathbf{X}\in{\cal T}_{n}(Q_{X}))\right\} \right\} +\epsilon.\label{eq: ineq1 excess rate}
\end{multline}
Then, 
\begin{alignat}{1}
{\cal E}_{r}({\cal S},\RR) & =\liminf_{n\to\infty}\min_{Q_{X}\in{\cal P}_{n}({\cal X})}\left\{ D(Q_{X}||P_{X})-\frac{1}{n}\log\P(r(\mathbf{X})\geq\RR|\mathbf{X}\in{\cal T}_{n}(Q_{X}))\right\} \\
 & \trre[\leq,a]\inf_{n\geq m_{0}}\min_{Q_{X}\in{\cal P}_{n}({\cal X})}\left\{ D(Q_{X}||P_{X})-\frac{1}{n}\log\P(r(\mathbf{X})\geq\RR|\mathbf{X}\in{\cal T}_{n}(Q_{X}))\right\} +\epsilon\\
 & \trre[\leq,b]\inf_{n\geq m_{0}}\left\{ D(Q_{X}^{*}||P_{X})-\frac{1}{n}\log\P(r(\mathbf{X})\geq\RR|\mathbf{X}\in{\cal T}_{n}(Q_{X}^{*}))\right\} +\epsilon\\
 & =D(Q_{X}^{*}||P_{X})+\inf_{n\geq m_{0}}\left\{ -\frac{1}{n}\log\P(r(\mathbf{X})\geq\RR|\mathbf{X}\in{\cal T}_{n}(Q_{X}^{*}))\right\} +\epsilon\\
 & \leq\left\{ D(Q_{X}^{*}||P_{X})+\lim_{m\to\infty}\inf_{n\geq m}-\frac{1}{n}\log\P(r(\mathbf{X})\geq\RR|\mathbf{X}\in{\cal T}_{n}(Q_{X}^{*}))\right\} +\epsilon\\
 & \trre[\leq,c]\inf_{Q_{X}\in{\cal P}({\cal X})}\left\{ D(Q_{X}||P_{X})+\liminf_{n\to\infty}\left\{ -\frac{1}{n}\log\P(r(\mathbf{X})\geq\RR|\mathbf{X}\in{\cal T}_{n}(Q_{X}))\right\} \right\} +2\epsilon\nonumber 
\end{alignat}
where $(a)$ is due to \eqref{eq: ineq2 excess rate}, $(b)$ is because
there exists $l\in\mathbb{N}$ so that $l\cdot n_{0}>m_{0}$ and then
$Q_{X}^{*}\in{\cal P}_{n_{0}}({\cal X})\subset{\cal P}_{l\cdot n_{0}}({\cal X})$,
and $(c)$ is due to \eqref{eq: ineq1 excess rate}. As $\epsilon>0$
is arbitrary we get the desired result. 
\end{IEEEproof}

\begin{IEEEproof}[Proof of Theorem \ref{thm: achieavbility with type dependent}]
 We will use the following two lemmas:
\begin{lem}
\label{lem: hamming distance close types}Let $Q_{X},Q'_{X}\in{\cal P}_{n}({\cal X})$
and assume that%
\footnote{For two different types in ${\cal P}_{n}({\cal X})$, the minimal
variation distance is $\frac{2}{n}$.%
} $||Q_{X}-Q'_{X}||=\frac{2d^{*}}{n}$ where $d^{*}>0$. If $\mathbf{x}\in{\cal T}_{n}(Q_{X})$
then 
\[
\min_{\mathbf{z}\in{\cal T}_{n}(Q'_{X})}d_{\s[H]}(\mathbf{x},\mathbf{z})\leq d^{*}.
\]
\end{lem}
\begin{IEEEproof}[Proof of Lemma \ref{lem: hamming distance close types}]
We prove this Lemma by modifying the vector $\mathbf{x}\in{\cal T}_{n}(Q_{X})$
into a vector $\mathbf{x}'\in{\cal T}_{n}(Q'_{X})$ by less than $d^{*}$
letter substitutions. Clearly, for some letters $x_{1}^{-},x_{1}^{+}\in{\cal X}$
we have $Q_{X}(x_{1}^{-})<Q'_{X}(x_{1}^{-})$ and $Q_{X}(x_{1}^{+})>Q'_{X}(x_{1}^{+})$.
Find an index $i_{1}$ such that the $i$th entry of $\mathbf{x}$
is $x_{1}^{+}$ and change it to $x_{1}^{-}$. Denote the resulting
vector by $\mathbf{x}_{1}$, and let its type by $Q_{X}^{(1)}$. If,
$Q_{X}^{(1)}=Q'_{X}$ then we have found a vector $\mathbf{x}'\in{\cal T}_{n}(Q'_{X})$
such that $d_{\s[H]}(\mathbf{x},\mathbf{x}')\leq1\leq d^{*}$ and
thus we are done. Otherwise, we have 
\[
||Q'_{X}-Q_{X}^{(1)}||=\frac{2(d^{*}-1)}{n}.
\]
In this case, repeat the same steps for $\mathbf{x}_{1}$, and at
each step, the variation distance between $Q'_{X}$ and $Q_{X}^{(k)}$
decreases by $\frac{2}{n}$. Thus, after at most $d^{*}$ stages,
a vector $\mathbf{x}^{(d^{*})}\in{\cal T}_{n}(Q'_{X})$ is found,
such that $d_{\s[H]}(\mathbf{x},\mathbf{x}')\leq d^{*}$.\end{IEEEproof}
\begin{lem}
\label{lem: truncation of types}Let $Q_{X}\in{\cal P}_{n}({\cal X})$
and $\mathbf{x}\in{\cal T}_{n}(Q_{X})$. For $1\leq k<n$ we have
\[
||\hat{Q}_{\mathbf{x}}-\hat{Q}_{\mathbf{x}(1:n-k)}||<|{\cal X}|\cdot\frac{k}{n-k}.
\]
\end{lem}
\begin{IEEEproof}[Proof of Lemma \ref{lem: truncation of types}]
For any given letter $x\in{\cal X}$, we denote $\hat{Q}_{\mathbf{x}}(x)=\frac{m}{n}$,
and analyze 
\[
|\hat{Q}_{\mathbf{x}}(x)-\hat{Q}_{\mathbf{x}(1:n-k)}(x)|.
\]
The largest difference possible is either when $\mathbf{x}(i)=x$
for $\min\{k,m\}$ letters out of $n-k+1\leq i\leq n$ or $\mathbf{x}(i)\neq x$
for all $n-k+1\leq i\leq n$. In the former case, when $m\geq k$,
then 
\begin{alignat}{1}
|\hat{Q}_{\mathbf{x}}(x)-\hat{Q}_{\mathbf{x}(1:n-k)}(x)| & =\left|\frac{m}{n}-\frac{m-k}{n-k}\right|=\frac{m}{n}-\frac{m-k}{n-k}\\
 & <\frac{m}{n}-\frac{m-k}{n}<\frac{k}{n-k}
\end{alignat}
and when $m<k$ then 
\begin{alignat}{1}
|\hat{Q}_{\mathbf{x}}(x)-\hat{Q}_{\mathbf{x}(1:n-k)}(x)| & =\left|\frac{m}{n}-\frac{0}{n-k}\right|=\frac{m}{n}\\
 & <\frac{k}{n}<\frac{k}{n-k}.
\end{alignat}
In the later case
\begin{alignat}{1}
|\hat{Q}_{\mathbf{x}}(x)-\hat{Q}_{\mathbf{x}(1:n-k)}(x)| & =\left|\frac{m}{n}-\frac{m}{n-k}\right|=\frac{m}{n-k}-\frac{m}{n}\\
 & \leq\frac{m}{n-k}-\frac{m-k}{n-k}=\frac{k}{n-k}.
\end{alignat}
The result follows by summing over $x\in{\cal X}$. 
\end{IEEEproof}
We can now prove Theorem \ref{thm: achieavbility with type dependent}.
Let $\epsilon>0$ be given, and find $n_{0}$ sufficiently large such
that for any $Q'_{X}\in{\cal P}({\cal X})$ there exists $Q_{X}\in{\cal P}_{n_{0}}({\cal X})\cap\interior{\cal Q}({\cal X})$
such that $||Q_{X}-Q'_{X}||\leq\frac{\epsilon}{2}$. For a given pair
of vectors $\mathbf{x},\mathbf{x}'\in{\cal X}^{n}$, define the binary
vector $\Delta_{\mathbf{x}\mathbf{x}'}\in\{0,1\}^{n}$ where 
\[
\Delta_{\mathbf{x}\mathbf{x}'}(i)=\begin{cases}
1, & \mathbf{x}(i)\neq\mathbf{x}'(i)\\
0, & \mathbf{x}(i)=\mathbf{x}'(i)
\end{cases}
\]
and also define ${\cal H}_{\mathbf{x}\mathbf{x}'}=\{i:\mathbf{x}(i)\neq\mathbf{x}'(i)\}$.
Also let $n_{1}=n_{0}\epsilon+2n_{0}|{\cal X}|$. We construct the
following SW codes ${\cal S}$ for all $n>\max\{n_{0},n_{1}\}$: 
\begin{itemize}
\item \emph{Codebook Construction:} 

\begin{itemize}
\item Compute $k^{*}(n)\teq\left\lfloor \frac{n}{n_{0}}\right\rfloor $.
\item Assign a binary string $\tau_{1}(Q_{X})$ for each type in $Q_{X}\in{\cal P}_{n}({\cal X})$. 
\item Assign a binary string $\tau_{2}(a)$ for each letter $x\in{\cal X}$.
For any vector $\mathbf{x}\in{\cal X}^{m}$, define 
\[
\tau_{2}(\mathbf{x})\teq(\tau_{2}(\mathbf{x}(1)),\ldots,\tau_{2}(\mathbf{x}(m))).
\]

\item Assign a binary string $\tau_{3}(\mathbf{b})$ for each binary vector
$\mathbf{b}\in\{0,1\}^{n}$ such that $d_{\s[H]}(\mathbf{0},\mathbf{b})\leq\lceil\frac{n\epsilon}{2}\rceil$,
where $\mathbf{0}$ is the all-zero vector of length $n$. 
\item Construct the SW codes ${\cal S}_{k^{*}(n)\cdot n_{0}}^{*}(Q_{X})=(s_{k^{*}(n)\cdot n_{0},Q_{X}}^{*},\sigma_{k^{*}(n)\cdot n_{0},Q_{X}}^{*})$
of rate $\rho(Q_{X})$ as in Theorem \ref{thm: Error exponent bounds, single type},
for all $Q{}_{X}\in{\cal P}_{n_{0}}({\cal X})\cap\interior{\cal Q}({\cal X})$. 
\item For any given $Q_{X}\in{\cal P}_{n}({\cal X})$ find 
\[
\Phi_{\epsilon}(Q_{X})\teq\argmin_{Q'_{X}\in{\cal P}_{n_{0}}({\cal X})\cap\interior{\cal Q}({\cal X})}||Q_{X}-Q'_{X}||.
\]

\end{itemize}

The above information is revealed to both the encoder and the decoder
off-line.

\item \emph{Encoding:} Upon observing $\mathbf{x}$, determine its empirical
distribution $\hat{Q}_{\mathbf{x}}$ and find 
\begin{equation}
\mathbf{w}=\argmin_{\overline{\mathbf{w}}\in{\cal T}_{k^{*}(n)\cdot n_{0}}(\Phi_{\epsilon}(\hat{Q}_{\mathbf{x}}))}d_{\s[H]}(\mathbf{x}(1:k^{*}(n)\cdot n_{0}),\overline{\mathbf{w}}).\label{eq: modified source block to encode}
\end{equation}
Let $\mathbf{x}'=\mathbf{x}(1:k^{*}(n)\cdot n_{0})$, and encode the
source block $\mathbf{x}$ as: 
\[
s_{n}(\mathbf{x})=(\tau_{1}(\hat{Q}_{\mathbf{x}}),\tau_{2}(\mathbf{w}({\cal H}_{\mathbf{x}'\mathbf{w}}),\tau_{2}(\mathbf{\mathbf{x}}({\cal H}_{\mathbf{x}'\mathbf{w}})),\tau_{3}(\Delta_{\mathbf{x}'\mathbf{w}}),\tau_{2}(\mathbf{x}(k^{*}(n)\cdot n_{0}+1:n)),s_{k^{*}(n)\cdot n_{0},Q_{X}}^{*}(\mathbf{w})).
\]

\item \emph{Decoding:} Upon observing $\mathbf{y}$ and $s_{n}(\mathbf{x})$:

\begin{itemize}
\item From $s_{n}(\mathbf{x})$, recover $\hat{Q}_{\mathbf{x}}$ and determine
$\Phi_{\epsilon}(\hat{Q}_{\mathbf{x}})$. Recover $\Delta_{\mathbf{x}'\mathbf{w}}$,
$\mathbf{w}({\cal H}_{\mathbf{x}'\mathbf{w}})$,$\mathbf{x}({\cal H}_{\mathbf{x}'\mathbf{w}})$,
and $\mathbf{x}(k^{*}(n)\cdot n_{0}+1:n)$. 
\item Generate a vector $\mathbf{y}'\in{\cal Y}^{k^{*}(n)\cdot n_{0}}$
as follows: For any index $1\leq i\leq k^{*}(n)\cdot n_{0}$. If $\Delta_{\mathbf{x}'\mathbf{w}}(i)=0$
then set $\mathbf{y}'(i)=\mathbf{y}(i)$. Otherwise, draw $\mathbf{y}'(i)$
according to the conditional distribution $P_{Y|X}(\cdot|\mathbf{w}(i))$.
\item Decode 
\[
\hat{\mathbf{w}}=\sigma_{k^{*}(n)\cdot n_{0},\Phi_{\epsilon}(\hat{Q}_{\mathbf{x}})}^{*}\left(s_{k^{*}(n)\cdot n_{0},\Phi_{\epsilon}(\hat{Q}_{\mathbf{x}})}^{*}(\mathbf{w}),\mathbf{y}'\right).
\]

\item The decoded source block is 
\[
\hat{\mathbf{x}}(i)=\begin{cases}
\hat{\mathbf{w}}(i), & 1\leq i\leq k^{*}(n)\cdot n_{0},\:\Delta_{\mathbf{x}'\mathbf{w}}(i)=0\\
\mathbf{x}(i), & 1\leq i\leq k^{*}(n)\cdot n_{0},\:\Delta_{\mathbf{x}'\mathbf{w}}(i)=1\\
\mathbf{x}(i), & k^{*}(n)\cdot n_{0}+1\leq i\leq n.
\end{cases}
\]

\end{itemize}
\end{itemize}
To prove that such coding is possible, notice that from Lemma \ref{lem: truncation of types}
and the fact that $n>n_{1}$, we have
\[
||\hat{Q}_{\mathbf{x}'}-\hat{Q}_{\mathbf{x}}||\leq\frac{\epsilon}{2}
\]
and by the triangle inequality 
\[
||\hat{Q}_{\mathbf{x}'}-\hat{Q}_{\mathbf{w}}||\leq||\hat{Q}_{\mathbf{x}'}-\hat{Q}_{\mathbf{x}}||+||\hat{Q}_{\mathbf{x}}-\hat{Q}_{\mathbf{w}}||\leq\frac{\epsilon}{2}+\frac{\epsilon}{2}=\epsilon.
\]
Thus, Lemma \ref{lem: hamming distance close types} implies that
\[
d_{\s[H]}(\Delta_{\mathbf{x}'\mathbf{w}},\mathbf{0})=d_{\s[H]}(\mathbf{x}',\mathbf{w})\leq\left\lceil \frac{n\epsilon}{2}\right\rceil .
\]
Let us now analyze the resulting asymptotic error probability of ${\cal S}$.
For any given $\delta>0$ 
\begin{eqnarray}
{\cal E}_{e}^{-}({\cal S}) & \trre[=,a] & \liminf_{n\to\infty}\min_{Q_{X}\in{\cal P}_{n}({\cal X})}\left\{ D\left(Q_{X}||P_{X}\right)-\frac{1}{n}\log\P(\hat{\mathbf{X}}\neq\mathbf{X}|\mathbf{X}\in{\cal T}_{n}(Q_{X}))\right\} \\
 & \trre[=,b] & \liminf_{n\to\infty}\min_{Q_{X}\in{\cal P}_{n}({\cal X})}\Biggl\{ D\left(Q_{X}||P_{X}\right)+\nonumber \\
 &  & -\frac{1}{n}\log\P(\hat{\mathbf{W}}\neq\mathbf{W}|\mathbf{W}\in{\cal T}_{k^{*}(n)\cdot n_{0}}(\Phi_{\epsilon}(Q_{X})))\Biggr\}\label{eq: achievabiliy erorr exponent modification}\\
 & = & \liminf_{n\to\infty}\min_{Q_{X}\in{\cal P}_{n}({\cal X})}\Biggr\{ D\left(Q_{X}||P_{X}\right)+\nonumber \\
 &  & -\frac{k^{*}(n)\cdot n_{0}}{n}\frac{1}{k^{*}(n)\cdot n_{0}}\log\P(\hat{\mathbf{W}}\neq\mathbf{W}|\mathbf{W}\in{\cal T}_{k^{*}(n)\cdot n_{0}}(\Phi_{\epsilon}(Q_{X})))\Biggr\}\\
 & \trre[\geq,c] & \liminf_{n\to\infty}\min_{Q_{X}\in{\cal P}_{n}({\cal X})}\Biggr\{ D\left(Q_{X}||P_{X}\right)+\nonumber \\
 &  & \underline{E}_{e}^{*}(H(\Phi_{\epsilon}(Q_{X}))-\rho(\Phi_{\epsilon}(Q_{X})),\Phi_{\epsilon}(Q_{X}),P_{Y|X})-\delta\Biggr\}\\
 & \trre[\geq,d] & \liminf_{n\to\infty}\min_{Q_{X}\in{\cal P}_{n}({\cal X})}\Biggr\{ D\left(\Phi_{\epsilon}(Q_{X})||P_{X}\right)+\nonumber \\
 &  & \underline{E}_{e}^{*}(H(\Phi_{\epsilon}(Q_{X}))-\rho(\Phi_{\epsilon}(Q_{X})),\Phi_{\epsilon}(Q_{X}),P_{Y|X})-\delta-\delta_{1}\Biggr\}\\
 & = & \liminf_{n\to\infty}\min_{Q_{X}\in{\cal P}_{n_{0}}({\cal X})}\Biggr\{ D\left(\Phi_{\epsilon}(Q_{X})||P_{X}\right)+\nonumber \\
 &  & \underline{E}_{e}^{*}(H(\Phi_{\epsilon}(Q_{X}))-\rho(\Phi_{\epsilon}(Q_{X})),\Phi_{\epsilon}(Q_{X}),P_{Y|X})-\delta-\delta_{1}\Biggr\}\\
 & = & \liminf_{n\to\infty}\min_{Q_{X}\in{\cal P}_{n_{0}}({\cal X})}\Biggr\{ D\left(Q_{X}||P_{X}\right)+\underline{E}_{e}^{*}(H(Q_{X})-\rho(Q_{X}),Q_{X},P_{Y|X})-\delta-\delta_{1}\Biggr\}\\
 & = & \min_{Q_{X}\in{\cal P}_{n_{0}}({\cal X})}\Biggr\{ D\left(Q_{X}||P_{X}\right)+\underline{E}_{e}^{*}(H(Q_{X})-\rho(Q_{X}),Q_{X},P_{Y|X})-\delta-\delta_{1}\Biggr\}\\
 & \geq & \inf_{Q_{X}\in{\cal P}({\cal X})}\Biggr\{ D\left(Q_{X}||P_{X}\right)+\underline{E}_{e}^{*}(H(Q_{X})-\rho(Q_{X}),Q_{X},P_{Y|X})-\delta-\delta_{1}\Biggr\}
\end{eqnarray}
where the passages are explained as follows:
\begin{itemize}
\item Equality $(a)$ is as in \eqref{eq: conditional error probability method of types derivation begin}-\eqref{eq: conditional error probability method of types derivation end}.
Notice that the error event $\{\hat{\mathbf{X}}\neq\mathbf{X}\}$
in this equation is for the code ${\cal S}_{n}.$
\item Equality $(b)$ is because an error $\hat{\mathbf{x}}\neq\mathbf{x}$
occurs only when the decoder \emph{$\sigma_{k^{*}(n)\cdot n_{0},\Phi_{\epsilon}(Q_{X})}^{*}$
}makes an error, since the vector $\mathbf{v}$ is generated memorylessly
according to $P_{Y|X}$, conditioned on $\mathbf{w}$. Notice that
the error event $\{\hat{\mathbf{W}}\neq\mathbf{W}\}$ in this equation
and the following is for the code ${\cal S}_{k^{*}(n)\cdot n_{0}}^{*}(\Phi_{\epsilon}(Q_{X}))$. 
\item Inequality $(c)$ is because there exists $n_{2}$ sufficiently large,
such that for all $n>n_{2}$ the error probability of the decoder
$\sigma_{k^{*}(n)\cdot n_{0},\Phi_{\epsilon}(Q_{X})}^{*}$ satisfies
\begin{multline*}
-\frac{k^{*}(n)\cdot n_{0}}{n}\frac{1}{k^{*}(n)\cdot n_{0}}\log\P(\hat{\mathbf{W}}\neq\mathbf{W}|\mathbf{W}\in{\cal T}_{k^{*}(n)\cdot n_{0}}(\Phi_{\epsilon}(Q_{X})))\\
\geq\underline{E}_{e}^{*}(H(Q_{X})-\rho(Q_{X}),Q_{X},P_{Y|X})-\delta
\end{multline*}
\emph{uniformly }for all \emph{$Q_{X}\in{\cal P}_{n_{0}}(Q_{X})$
}(notice also that $\frac{k^{*}(n)\cdot n_{0}}{n}\to1$ as $n\to\infty$). 
\item Inequality $(d)$ is because $D(Q_{X}||P_{X})$ is a continuous function
of $Q_{X}$ in ${\cal Q}({\cal X})$ (as $\supp(P_{X})={\cal X}$),
and thus uniformly continuous, and where $\delta_{1}>0$ and $\delta_{1}\downarrow0$
as $\epsilon\downarrow0$.
\end{itemize}
Regarding the rate, observe that the resulting codes of ${\cal S}$
are type-dependent, variable-rate SW codes, since ${\cal S}_{k^{*}(n)\cdot n_{0}}^{*}(Q_{X})$
are such. Let us analyze the total rate required to encode $\mathbf{x}\in Q_{X}$:
\begin{itemize}
\item Since $|{\cal P}_{n}({\cal X})|\leq(n+1)^{|{\cal X}|}$ then for $n$
sufficiently large. 
\[
\frac{1}{n}|\tau_{1}(\hat{Q}_{\mathbf{x}})|\leq\frac{|{\cal X}|}{n}\cdot\log(n+1)\leq\delta
\]

\item Encoding of all possible binary vectors $\mathbf{b}\in\{0,1\}^{n-k^{*}(n)n_{0}}$
such that $d_{\s[H]}(\mathbf{0},\mathbf{b})\leq\lceil\frac{n\epsilon}{2}\rceil$
requires a rate of \cite[Chapter 13.2]{Cover:2006:EIT:1146355} 
\[
\frac{1}{n}|\tau_{3}(\Delta_{\mathbf{x}'\mathbf{w}})|\leq h_{B}\left(\frac{\epsilon}{2}\right)+\delta
\]
for $n$ sufficiently large. 
\item Encoding the components of $\mathbf{\mathbf{x}}({\cal H}_{\mathbf{x}'\mathbf{w}})$
and $\mathbf{w}({\cal H}_{\mathbf{x}'\mathbf{w}})$ letter-wise with
zero error, requires a rate 
\[
\frac{1}{n}|\tau_{2}(\mathbf{w}({\cal H}_{\mathbf{x}'\mathbf{w}}))|+\frac{1}{n}|\tau_{2}(\mathbf{x}'({\cal H}_{\mathbf{x}'\mathbf{w}}))|\leq2\cdot\frac{\epsilon}{2}\log|{\cal X}|+\frac{2}{n}\leq\epsilon\log|{\cal X}|+\delta
\]
for $n$ sufficiently large. 
\item Encoding the components of $\mathbf{x}(k^{*}(n)\cdot n_{0}+1:n)$
letter-wise with zero error, requires a rate of
\[
\frac{1}{n}|\tau_{2}(\mathbf{x}(k^{*}(n)\cdot n_{0}+1:n))|\leq\left(\frac{n-k^{*}(n)n_{0}}{n}\right)\cdot\log|{\cal X}|\leq\delta
\]
for $n$ sufficiently large. 
\item By construction, ${\cal S}_{k^{*}(n)\cdot n_{0},\Phi_{\epsilon}(Q_{X})}^{*}$
is a type-dependent, variable-rate SW code of rate $\rho(\Phi_{\epsilon}(Q_{X}))$
and thus for sufficiently large $n$
\[
\frac{1}{n}\left|s_{k^{*}(n)\cdot n_{0},\Phi_{\epsilon}(Q_{X})}^{*}(\mathbf{w})\right|\leq\frac{k^{*}(n)\cdot n_{0}}{n}\rho(\Phi_{\epsilon}(Q_{X}))+\delta
\]
uniformly over $Q_{X}$.
\end{itemize}
Thus, for sufficiently large $n$, the resulting total rate for coding
$\mathbf{x}\in Q_{X}$ is less than $\rho(\Phi_{\epsilon}(Q_{X}))+\varrho$
where 
\[
\varrho=h_{B}\left(\frac{\epsilon}{2}\right)+\epsilon\log|{\cal X}|+5\delta.
\]
The resulting excess-rate exponent is 
\begin{eqnarray}
{\cal E}_{r}({\cal S},\RR+\varrho) & \trre[=,a] & \liminf_{n\to\infty}\min_{Q_{X}\in{\cal P}_{n}({\cal X})}\left\{ D(Q_{X}||P_{X})-\frac{1}{n}\log\P(r(\mathbf{X})\geq\RR+\varrho|\mathbf{X}\in{\cal T}_{n}(Q_{X}))\right\} \nonumber \\
 & \trre[=,b] & \liminf_{n\to\infty}\min_{Q_{X}\in{\cal P}_{n}({\cal X}):\:\rho(\Phi_{\epsilon}(Q_{X}))\geq\RR}D(Q_{X}||P_{X})\\
 & \trre[\geq,c] & \liminf_{n\to\infty}\min_{Q_{X}\in{\cal P}_{n}({\cal X}):\:\rho(\Phi_{\epsilon}(Q_{X}))\geq\RR}D(\Phi_{\epsilon}(Q_{X})||P_{X})-\delta_{1}\\
 & = & \liminf_{n\to\infty}\min_{Q_{X}\in{\cal P}_{n_{0}}({\cal X}):\:\rho(\Phi_{\epsilon}(Q_{X}))\geq\RR}D(\Phi_{\epsilon}(Q_{X})||P_{X})-\delta_{1}\\
 & = & \min_{Q_{X}\in{\cal P}_{n_{0}}({\cal X}):\:\rho(Q_{X})\geq\RR}D(Q_{X}||P_{X})-\delta_{1}\\
 & \geq & \inf_{Q_{X}\in{\cal P}({\cal X}):\:\rho(Q_{X})\geq\RR}D(Q_{X}||P_{X})-\delta_{1}
\end{eqnarray}
where $(a)$ is as in \eqref{eq: conditional excess rate probability method of types derivation begin}-\eqref{eq: conditional excess rate probability method of types derivation end},
$(b)$ is because the codes ${\cal S}_{n}$ are type-dependent, variable-rate
codes which assign rate $\rho(\Phi_{\epsilon}(Q_{X}))+\varrho$ to
the type $Q_{X}$, and $(c)$ is again by the uniform continuity of
$D(Q_{X}||P_{X})$ in ${\cal Q}({\cal X})$. We obtain the desired
result by taking $\delta\downarrow0$ and then $\epsilon\downarrow0$.

Before completing the proof, we make the following two remarks.
\begin{rem}
The vector actually coded is $\mathbf{w}$ \eqref{eq: modified source block to encode},
not the original source block $\mathbf{x}$. Thus, after modifying
$\mathbf{x}$ to $\mathbf{w}$, the distribution of $\mathbf{w}$
may not be uniform within its type class (even when conditioned on
the event that $\mathbf{x}$ belongs to some type class), which might
affect \eqref{eq: achievabiliy erorr exponent modification}. There
are two possibilities to circumvent this%
\footnote{This matter was not addressed in the body of the proof in order not
to over-complicate it.%
}. The first is to use common randomness at the encoder and decoder,
and to generate a uniformly random permutation. Prior to encoding,
the source block $\mathbf{x}$ is permuted, and the decoder simply
applies the inverse permutation after decoding. In this case, the
uniform distribution of $\mathbf{w}$ is assured. The second possibility
is to construct the SW codes from channel codes (as was done in Theorem
\ref{thm: Error exponent bounds, single type}) which have maximal
error probability according to the reliability function (see \eqref{eq: beginning with good channel code}),
and not just the average error probability. As is well known, such
a channel code can be generated from a good average error probability
codebook, by simply expurgating the worst half of the codebook. The
rate loss is negligible, and here too, good error probability is assured
uniformly over $\mathbf{w}$ in the type class. 
\end{rem}

\begin{rem}
In the proof above, the actual decoders $\sigma_{n,Q_{X}}^{*}$ of
${\cal S}_{n}^{*}(Q_{X})$ were not specified, and any decoder which
achieves the error exponent for the underlying channel code can be
used. Thus, in the proof of \ref{thm: achieavbility with type dependent},
a randomized decoder was required, in order to mimic the channel operation
$P_{Y|X}$ for the vector $\mathbf{w}$. However, this might not be
required if $\sigma_{n,Q_{X}}^{*}$ is more specific. For example,
if the decoder $\sigma_{n,Q_{X}}^{*}$ is the ML decoder, then instead
of drawing $\mathbf{y}'(i)$ according to the conditional distribution
$P_{Y|X}(\cdot|\mathbf{w}(i))$, it can be simply set to the letter
with maximal likelihood, i.e., $\mathbf{y}'(i)=\argmax_{y\in{\cal Y}}P_{Y|X}(y|\mathbf{w}(i))$.
This only improves the error probability, and thus the results of
Theorem \ref{thm: achieavbility with type dependent} remain valid.
\end{rem}
\end{IEEEproof}

\begin{IEEEproof}[Proof of Theorem \ref{thm: Optimal rate function bounds}]
The proof is divided into three parts, one for each of the bounds.

\emph{\uline{Random binning bound:}} From Theorem \ref{thm: achieavbility with type dependent},
we may clearly assume that $\rho_{\s[rb]}(Q_{X},\EE)\leq H(Q_{X})$,
as otherwise the random coding bound in \eqref{eq: channel random coding bound}
is infinite, and $\EE$ is trivially achieved. Now, from the random
coding bound in \eqref{eq: channel random coding bound}, the condition
in \eqref{eq: achieving EE inf condition} will be satisfied for a
rate function $\rho_{\s[rb]}(Q_{X},\EE)$ which satisfies 
\begin{equation}
\EE\leq D(Q_{X}||P_{X})+\min_{Q_{Y|X}}D(Q_{Y|X}||P_{Y|X}|Q_{X})+\left[\rho_{\s[rb]}(Q_{X},\EE)-H(Q_{X|Y}|Q_{Y})\right]_{+}.\label{eq: EE requirement random coding}
\end{equation}
Clearly, if $\EE\leq D(Q_{X}||P_{X})$, no actual constraint is imposed
on the rate, and \eqref{eq: EE requirement random coding} is satisfied
even for $\rho_{\s[rb]}(Q_{X},\EE)=0$. Otherwise, \eqref{eq: EE requirement random coding}
is equivalent to 

\[
\EE\leq D(Q_{X}||P_{X})+\min_{Q_{Y|X}}\max_{0\leq\lambda\leq1}D(Q_{Y|X}||P_{Y|X}|Q_{X})+\lambda\left[\rho_{\s[rb]}(Q_{X},\EE)-H(Q_{X|Y}|Q_{Y})\right]
\]
or 
\begin{eqnarray}
\rho_{\s[rb]}(Q_{X},\EE) & \geq & \max_{Q_{Y|X}}\min_{0\leq\lambda\leq1}\left[\frac{\EE-D(Q_{XY}||P_{XY})}{\lambda}+H(Q_{X|Y}|Q_{Y})\right]\\
 & = & \max_{Q_{Y|X}\in{\cal A}}\left[\EE-D(Q_{X}||P_{X})-D(Q_{Y|X}||P_{Y|X}|Q_{X})+H(Q_{X|Y}|Q_{Y})\right]
\end{eqnarray}
which directly leads to the second term in \eqref{eq: random binning rate function expression}.
For the third term in \eqref{eq: random binning rate function expression},
let us notice that for $\EE\geq D(Q_{X}||P_{X})+D(Q'_{Y|X}||P_{Y|X}|Q_{X})$
we have that $\rho_{\s[rb]}(Q_{X},\EE)$ is affine with slope $1$.
Indeed, using \eqref{eq: random binning rate function expression}
we get for $\EE>D(Q_{X}||P_{X})$ 
\begin{eqnarray}
\rho_{\s[rb]}(Q_{X},\EE) & \geq & \EE+H(Q_{X})\\
 &  & -D(Q_{X}||P_{X})-\left\{ I(Q_{X}\times Q'_{Y|X})+D(Q'_{Y|X}||P_{Y|X}|Q_{X})\right\} \nonumber 
\end{eqnarray}
and for $\EE\geq D(Q_{X}||P_{X})+D(Q'_{Y|X}||P_{Y|X}|Q_{X})$ equality
is achieved since $Q'_{Y|X}\in{\cal A}_{\s[rb]}$. For the fourth
term in \eqref{eq: random binning rate function expression}, notice
that the minimal $\EE$ such that $\rho_{\s[rb]}(Q_{X},\EE)=H(Q_{X})$
is given by 
\[
\EE=D(Q_{X}||P_{X})+I(Q_{X}\times Q'_{Y|X})+D(Q'_{Y|X}||P_{Y|X}|Q_{X}).
\]

\emph{\uline{Expurgated bound:}}\emph{ }From Theorem \ref{thm: achieavbility with type dependent},
we may clearly assume that $\rho_{\s[ex]}(Q_{X},\EE)\leq H(Q_{X})$,
as otherwise the expurgated bound in \eqref{eq: channel expurgated bound}
is infinite, and $\EE$ is trivially achieved. Now, from the expurgated
bound in \eqref{eq: channel expurgated bound}, the condition in \eqref{eq: achieving EE inf condition}
will be satisfied for a rate function $\rho_{\s[ex]}(Q_{X},\EE)$
which satisfies 
\begin{equation}
\EE\leq D(Q_{X}||P_{X})+\min_{Q_{\tilde{X}|X}:\: Q_{\tilde{X}}=Q_{X},\rho_{\s[ex]}(Q_{X},\EE)\leq H(Q_{X|\tilde{X}}|Q_{\tilde{X}})}\left\{ B(Q_{X\tilde{X}},P_{Y|X})+\rho_{\s[ex]}(Q_{X},\EE)-H(Q_{X|\tilde{X}}|Q_{\tilde{X}})\right\} .\label{eq: EE requirement expurgated}
\end{equation}
Clearly, if $\EE\leq D(Q_{X}||P_{X})$ then $\rho_{\s[ex]}(Q_{X},\EE)=0$.
Now, \eqref{eq: EE requirement expurgated} is equivalent to
\begin{eqnarray*}
\EE & \leq & D(Q_{X}||P_{X})+\min_{Q_{\tilde{X}|X}:\: Q_{\tilde{X}}=Q_{X}}\max_{\lambda\geq0}\biggl\{ B(Q_{X\tilde{X}},P_{Y|X})+(1+\lambda)\left[\rho_{\s[ex]}(Q_{X},\EE)-H(Q_{X|\tilde{X}}|Q_{\tilde{X}})\right]\biggr\}
\end{eqnarray*}
or equivalently,
\[
\rho_{\s[ex]}(Q_{X},\EE)\geq\max_{Q_{\tilde{X}|X}:\: Q_{\tilde{X}}=Q_{X}}\min_{\lambda\geq0}\frac{\EE-D(Q_{X}||P_{X})-B(Q_{X\tilde{X}},P_{Y|X})}{1+\lambda}+H(Q_{X|\tilde{X}}|Q_{\tilde{X}})=\max\{v_{1},v_{2}\}
\]
where
\begin{align}
v_{1} & \teq\max_{Q_{\tilde{X}|X}:\:\EE-D(Q_{X}||P_{X})\leq B(Q_{X\tilde{X}},P_{Y|X})}\min_{\lambda\geq0}\frac{\EE-D(Q_{X}||P_{X})-B(Q_{X\tilde{X}},P_{Y|X})}{1+\lambda}+H(Q_{X|\tilde{X}}|Q_{\tilde{X}})\\
 & =\max_{Q_{\tilde{X}|X}:\:\EE-D(Q_{X}||P_{X})\leq B(Q_{X\tilde{X}},P_{Y|X})}\EE-D(Q_{X}||P_{X})-B(Q_{X\tilde{X}},P_{Y|X})+H(Q_{X|\tilde{X}}|Q_{\tilde{X}}),
\end{align}
and 
\begin{align}
v_{2} & \teq\max_{Q_{\tilde{X}|X}:\:\EE-D(Q_{X}||P_{X})\geq B(Q_{X\tilde{X}},P_{Y|X})}\min_{\lambda\geq0}\frac{\EE-D(Q_{X}||P_{X})-B(Q_{X\tilde{X}},P_{Y|X})}{1+\lambda}+H(Q_{X|\tilde{X}}|Q_{\tilde{X}})\\
 & =\max_{Q_{\tilde{X}|X}:\:\EE-D(Q_{X}||P_{X})\geq B(Q_{X\tilde{X}},P_{Y|X})}H(Q_{X|\tilde{X}}|Q_{\tilde{X}})
\end{align}
and in both the maximization problems of $v_{1}$ and $v_{2}$, the
constraint $Q_{X}=(Q_{X}\times Q_{\tilde{X}|X})_{\tilde{X}}$ is also
imposed. Notice that the maximizer of $H(Q_{X|\tilde{X}}|Q_{\tilde{X}})$
under the constraint $Q_{\tilde{X}}=Q_{X}$, is given by $Q_{X\tilde{X}}=Q_{X}\times Q_{\tilde{X}}=Q_{X}\times Q_{X}$.
We now have two cases, depending whether $\EE-D(Q_{X}||P_{X})-B(Q_{X}\times Q_{X},P_{Y|X})\leq0$
or $\EE-D(Q_{X}||P_{X})-B(Q_{X}\times Q_{X},P_{Y|X})>0$. In the first
case, $\EE-D(Q_{X}||P_{X})-B(Q_{X}\times Q_{X},P_{Y|X})\leq0$ and
then the solution of $v_{2}$ must be on the boundary of the constraint
set (as this optimization problem is concave), i.e.
\[
v_{2}=\max_{Q_{\tilde{X}|X}:\: Q_{\tilde{X}}=Q_{X},\:\EE-D(Q_{X}||P_{X})-B(Q_{X\tilde{X}},P_{Y|X})=0}H(Q_{X|\tilde{X}}|Q_{\tilde{X}})\leq v_{1}
\]
 So, for this case
\begin{multline}
\rho_{\s[ex]}(Q_{X},\EE)=\EE+H(Q_{X})-D(Q_{X}||P_{X})\\
-\min_{Q_{\tilde{X}|X}:\: Q_{\tilde{X}}=Q_{X},\:\EE-D(Q_{X}||P_{X})-B(Q_{X\tilde{X}},P_{Y|X})\leq0}\left\{ B(Q_{X\tilde{X}},P_{Y|X})+I(Q_{X\tilde{X}})\right\} .\label{eq: expurgated rate function proof 1st expression}
\end{multline}
Now, for $\EE\leq\EE[,\textnormal{\scriptsize a-ex}]$ we have
\[
\rho_{\s[ex]}(Q_{X},\EE)\geq\EE+H(Q_{X})-D(Q_{X}||P_{X})-\min_{Q_{\tilde{X}|X}:\: Q_{\tilde{X}}=Q_{X}}\left\{ B(Q_{X\tilde{X}},P_{Y|X})+I(Q_{X\tilde{X}})\right\} 
\]
and for $\EE\leq D(Q_{X}||P_{X})+B(Q_{X}\times Q'_{\tilde{X}|X},P_{Y|X})$
equality is achieved since $Q'_{\tilde{X}|X}\in\in{\cal A}_{\s[ex]}$.
Thus, the second term in \eqref{eq: expurgated rate function expression}
follows. For the third term, we must have that the constraint in \eqref{eq: expurgated rate function proof 1st expression}
is satisfied with an equality. In the second case, $\EE-D(Q_{X}||P_{X})-B(Q_{X}\times Q_{X},P_{Y|X})\geq0$
and then $v_{2}\geq v_{1}$ and\emph{ }the fourth term in \eqref{eq: expurgated rate function expression}
is obtained.

\emph{\uline{Sphere packing bound:}} From Theorem \ref{thm: General upper bound on error exponent},
we may clearly assume that $\rho_{\s[sp]}(Q_{X},\EE)\leq H(Q_{X})$,
as otherwise the sphere packing bound in \eqref{eq: channel sphere packing bound}
is infinite, and the upper bound on the error exponent is trivial.
From Theorem \ref{thm: General upper bound on error exponent} and
the sphere packing bound in \eqref{eq: channel sphere packing bound},
the condition in \eqref{eq: achieving EE inf condition} will be not
be satisfied unless that rate function $\rho_{\s[sp]}(Q_{X},\EE)$
satisfies 
\begin{equation}
\EE\leq D(Q_{X}||P_{X})+\min_{Q_{Y|X}:\:\rho_{\s[sp]}(Q_{X},\EE)\leq H(Q_{X|Y}|Q_{Y})}D(Q_{Y|X}||P_{Y|X}|Q_{X}).\label{eq: EE requirement sphere packing}
\end{equation}
Clearly, if $\EE\leq D(Q_{X}||P_{X})$ then $\rho_{\s[ex]}(Q_{X},\EE)=0$.
Otherwise, 
\[
\EE\leq D(Q_{X}||P_{X})+\min_{Q_{Y|X}}\max_{\lambda\geq0}D(Q_{Y|X}||P_{Y|X}|Q_{X})+\lambda\left(\rho_{\s[sp]}(Q_{X},\EE)-H(Q_{X|Y}|Q_{Y})\right)
\]
which is equivalent to 
\begin{eqnarray*}
\rho_{\s[sp]}(Q_{X},\EE) & \geq & \max_{Q_{Y|X}}\min_{\lambda\geq0}\frac{\EE-D(Q_{X}||P_{X})-D(Q_{Y|X}||P_{Y|X}|Q_{X})}{\lambda}+H(Q_{X|Y}|Q_{Y})\\
 & = & \max_{Q_{Y|X}:\: D(Q_{X}||P_{X})+D(Q_{Y|X}||P_{Y|X}|Q_{X})\leq\EE}H(Q_{X|Y}|Q_{Y})
\end{eqnarray*}
which directly leads to the second term in \eqref{eq: sphere packing rate function expression}.
For the third term in \eqref{eq: sphere packing rate function expression},
let us find the minimal $\EE$ such that $\rho_{\s[sp]}(Q_{X},\EE)=H(Q_{X})$,
or equivalently
\[
\min_{Q_{Y|X}\in{\cal {\cal A}}}I(Q_{X}\times Q_{Y|X})=0.
\]
Obviously, for minimal $\EE$ with this property, the inequality in
${\cal A}$ must be achieved with an equality, and so 
\[
\min_{D(Q_{X}\times Q_{Y|X}||P_{XY})=\EE}I(Q_{X}\times Q_{Y|X})=0.
\]
Thus, using Lemma \ref{lem: Divergence output marginal minimization backwards},
the minimal $\EE$ is given by
\[
\min_{Q_{Y}}D(Q_{X}\times Q_{Y}||P_{XY})=D(Q_{X}\times(Q_{X}\times P_{Y|X})_{Y}||P_{XY}).
\]

\end{IEEEproof}

\begin{IEEEproof}[Proof of Lemma \ref{lem: rate function bounds properties}]
Most of the properties can be immediately obtained, and so we only
provide the less trivial proofs. 
\begin{itemize}
\item \emph{\uline{Positivity:}} For $\rho_{\s[rb]}(Q_{X},\EE)$, observe
that if $\EE>D(Q_{X}||P_{X})$ then to satisfy \eqref{eq: EE requirement random coding}
for $Q_{Y|X}=P_{Y|X}$, we must have $\rho_{\s[rb]}(Q_{X},\EE)>H(Q_{X|Y}|Q_{Y})$,
where here $Q_{X|Y}$ is induced from $Q_{X}\times P_{Y|X}$. If $H(Q_{X|Y}|Q_{Y})>0$
then we are done. Else, slightly alter $Q_{Y|X}$ from $P_{Y|X}$
such that $D(Q_{X}||P_{X})+D(Q_{Y|X}||P_{Y|X}|Q_{X})<\EE$ but $H(Q_{X|Y}|Q_{Y})>0$.
For $\rho_{\s[ex]}(Q_{X},\EE)$, observe that if $\EE>D(Q_{X}||P_{X})$
then 
\[
\rho_{\s[ex]}(Q_{X},\EE)=H(Q_{X})-\min_{Q_{\tilde{X}|X}\in{\cal A}_{\s[ex]}}I(Q_{X\tilde{X}})
\]
and $\rho_{\s[ex]}(Q_{X},\EE)=0$ iff $I(Q_{X\tilde{X}})=H(Q_{X})$,
namely, the channel $Q_{\tilde{X}|X}$ is noiseless, $X=\tilde{X}$
with probability $1$. However, for this channel $B(Q_{X\tilde{X}},P_{Y|X})=0<\EE-D(Q_{X}||P_{X})$,
and so the constraint $Q_{\tilde{X}|X}\in{\cal A}$ is not satisfied.
Thus, $\rho_{\s[ex]}(Q_{X},\EE)>0$. 
\item \emph{\uline{Monotonicity:}} For $\rho_{\s[ex]}(Q_{X},\EE)$, notice
that from\textbf{ \eqref{eq: expurgated rate function expression}},
we have
\[
\rho_{\s[ex]}(Q_{X},\EE)=H(Q_{X})-\min_{Q_{\tilde{X}|X}\in{\cal A}}I(Q_{X\tilde{X}}).
\]
Now, since $I(Q_{X\tilde{X}})$ is a convex function of $Q_{\tilde{X}|X}$
and its minimizer $Q_{X}\times Q_{X}$ is outside the set 
\[
{\cal A}'\teq\left\{ Q_{\tilde{X}|X}:Q_{\tilde{X}}=Q_{X},\:\EE\geq D(Q_{X}||P_{X})+B(Q_{X\tilde{X}},P_{Y|X})\right\} 
\]
then we also have 
\[
\rho_{\s[ex]}(Q_{X},\EE)=H(Q_{X})-\min_{Q_{\tilde{X}|X}\in{\cal A}'}I(Q_{X\tilde{X}})
\]
and the solution is always on the boundary. Since the set ${\cal A}'$
is strictly increasing as a function of $\EE$, the result follows. 
\item \emph{\uline{Concavity:}} Can be verified using Lemma \ref{lem: concave of min problem}
(Appendix \ref{sec: Useful Lemmas}). 
\item \emph{\uline{Regularity:}} Obtained by letting ${\cal V}=\{Q_{X}:D(Q_{X}||P_{X})<\EE\}$. 
\end{itemize}
\end{IEEEproof}

\begin{IEEEproof}[Proof of Lemma \ref{lem: excess rate exponent expression}]
This can be proved if we show that the infimum of $\inf_{Q_{X}:\rho(Q_{X})\geq\RR}D(Q_{X}||P_{X})$
is attained, and that the function $\min_{Q_{X}:\rho(Q_{X})\geq\RR}D(Q_{X}||P_{X})$
is left-continuous in $\RR$. We begin by showing that the infimum
of $\inf_{Q_{X}:\rho(Q_{X})\geq\RR}D(Q_{X}||P_{X})$ is attained.
Recall that $\rho(Q_{X})$ is regular, and so there exists a $d>0$
such that $\rho(Q_{X})$ is continuous in ${\cal V}=\{Q_{X}\in{\cal Q}({\cal X}):D(Q_{X}||P_{X})<d\}$,
and equals a constant $\rho(Q_{X})=\RR_{0}$, for $Q_{X}\in{\cal V}^{c}$.
Thus, 
\[
\inf_{Q_{X}:\rho(Q_{X})\geq\RR}D(Q_{X}||P_{X})=\min\left\{ \inf_{Q_{X}\in{\cal V}:\rho(Q_{X})\geq\RR}D(Q_{X}||P_{X}),\min_{Q_{X}\in{\cal V}^{c}:\rho(Q_{X})\geq\RR}D(Q_{X}||P_{X})\right\} 
\]
and so, if $\inf_{Q_{X}:\rho(Q_{X})\geq\RR}D(Q_{X}||P_{X})$ is not
attained, then the infimum of $\inf_{Q_{X}\in{\cal V}:\rho(Q_{X})\geq\RR}D(Q_{X}||P_{X})$
is not attained for some $Q_{X}\in{\cal V}$, and so 
\[
\inf_{Q_{X}\in{\cal V}:\rho(Q_{X})\geq\RR}D(Q_{X}||P_{X})=d.
\]
However, in this case, there also must exist a sequence $\overline{Q}_{X}^{(n)}\in{\cal V}$
such that $\rho(\overline{Q}_{X}^{(n)})\to\RR$ and $\rho(\overline{Q}_{X}^{(n)})>\RR$.
But since $D(\overline{Q}_{X}^{(n)}||P_{X})<d$ this is a contradiction
that $d=\inf_{Q_{X}\in{\cal V}:\rho(Q_{X})\geq\RR}D(Q_{X}||P_{X})$. 

Now, to show left continuity of $\min_{Q_{X}:\rho(Q_{X})\geq\RR}D(Q_{X}||P_{X})$
as a function of $\RR$, let $\delta>0$ be given. For any $\epsilon>0$
we clearly have 
\[
\min_{Q_{X}:\rho(Q_{X})\geq\RR-\epsilon}D(Q_{X}||P_{X})\leq\min_{Q_{X}:\rho(Q_{X})\geq\RR}D(Q_{X}||P_{X})=D(Q_{X}^{*}||P_{X}).
\]
To obtain the reversed inequality, we divide the proof into two cases,
depending on whether $Q_{X}^{*}\in{\cal V}$ or not.

\emph{Case 1:} $Q_{X}^{*}\in{\cal V}$. Recall that $\rho(Q_{X})$
is continuous and finite inside the interior of ${\cal V}$, and $D(Q_{X}||P_{X})$
is a continuous function of $Q_{X}$. Now, we may define for any $Q_{X}\in{\cal V}$
such that $\rho(Q_{X})\geq\RR$, the closed neighborhood 
\[
{\cal D}(Q_{X},\RR,\delta)\teq\left\{ \tilde{Q}_{X}:D(\tilde{Q}_{X}||P_{X})\geq D(Q_{X}||P_{X})-\delta\right\} \cap\overline{{\cal V}}.
\]
Also, we may define the set 
\[
{\cal V}'(\RR)\teq\left\{ Q_{X}\in\partial{\cal V}:\lim_{\tilde{Q}_{X}\to Q_{X}}\rho(\tilde{Q}_{X})=\RR\right\} 
\]
where $\partial{\cal V}={\cal \overline{V}}\backslash{\cal V}$ is
the boundary of ${\cal V}$, and for any $Q_{X}\in{\cal V}'(\RR)$
\[
{\cal D}'(Q_{X},\RR,\delta)\teq\{Q_{X}\}\cup{\cal D}(Q_{X},\RR,\delta).
\]
Now, consider the set
\[
{\cal U}\teq{\cal V}\backslash\left\{ \left\{ \bigcup_{\left\{ Q_{X}\in{\cal V}'(\RR)\right\} }{\cal D}'(Q_{X},\RR,\delta)\right\} \cup\left\{ \bigcup_{\left\{ Q_{X}\in{\cal V}:\rho(Q_{X})\geq\RR\right\} }{\cal D}(Q_{X},\RR,\delta)\right\} \right\} 
\]
and let $\RR'\teq\sup_{Q_{X}\in\overline{{\cal U}}}\rho(Q_{X})$.
Then we must have $\RR'<\RR$. To see this, assume conversely, that
$\RR'=\RR$ and let $\overline{Q}_{X}$ achieve the maximum, namely,
$\rho(\overline{Q}_{X})=\RR$. Now, either $\overline{Q}_{X}\in{\cal V}$
or the supremum is not attained, but both cases lead to contradiction.
Indeed, if the supremum is attained at some $\overline{Q}_{X}\in{\cal V}$
then ${\cal D}(\overline{Q}_{X},\RR,\delta)\notin{\cal U}$ and so
$\overline{Q}_{X}\not\in{\cal U}$ which is a contradiction. Otherwise,
there exists a sequence $\overline{Q}_{X}^{(n)}\in\overline{{\cal U}}$
such that $\rho(\overline{Q}_{X}^{(n)})\to\RR$. Assume that an arbitrary
convergent sub-sequence of $\overline{Q}_{X}^{(n)}$ converges to
$\overline{Q}_{X}\in\overline{{\cal V}}$. But, the definition of
${\cal D}'(Q_{X},\RR,\delta)$ and the continuity of $D(Q_{X}||P_{X})$
in $\overline{{\cal V}}$ imply that for any sufficiently large $n$
we must have $\overline{Q}_{X}^{(n)}\not\in\overline{{\cal U}}$,
which is a contradiction. Now, consider two sub-cases:
\begin{enumerate}
\item $\RR>\RR_{0}$. If we choose $\epsilon\leq\min\{\RR-\RR',\RR-\RR_{0}\}$
we have 
\[
\min_{Q_{X}:\rho(Q_{X})\geq\RR-\epsilon}D(Q_{X}||P_{X})\geq\min_{{\cal V}\backslash{\cal U}}D(\tilde{Q}_{X}||P_{X})\geq D(Q_{X}^{*}||P_{X})-\delta
\]
since the left most minimization is over a smaller set.
\item $\RR\leq\RR_{0}$. Since $Q_{X}^{*}$ is the minimizer for the right
hand side of \eqref{eq: left continuity for regular rate functions}
then 
\[
\min_{\tilde{Q}_{X}\in{\cal V}^{c}}D(\tilde{Q}_{X}||P_{X})\geq D(Q_{X}^{*}||P_{X})
\]
and if we choose $\epsilon\leq\RR-\RR'$ we also have 
\[
\min_{Q_{X}:\rho(Q_{X})\geq\RR-\epsilon}D(Q_{X}||P_{X})\geq\min_{\tilde{Q}_{X}\in{\cal V}^{c}\cup{\cal V}\backslash{\cal U}}D(\tilde{Q}_{X}||P_{X})\geq D(Q_{X}^{*}||P_{X})-\delta.
\]

\end{enumerate}

\emph{Case 2:} $Q_{X}^{*}\in{\cal V}^{c}$. In this case we clearly
have $\RR_{0}\geq\RR$ and
\[
\inf_{Q_{X}:\rho(Q_{X})\geq\RR}D(Q_{X}||P_{X})=d.
\]
Now, if we let $\overline{\RR}\teq\sup_{Q_{X}\in{\cal V}}\rho(Q_{X})$,
then either this supremum is not attained or $\overline{\RR}<\RR$.
To see this, assume conversely, that the supremum is attained by some
$\overline{Q}_{X}\in{\cal V}$ and also $\overline{\RR}\geq\RR$.
Then this implies
\[
\inf_{Q_{X}:\rho(Q_{X})\geq\RR}D(Q_{X}||P_{X})\leq D(\overline{Q}_{X}||P_{X})\leq d
\]
which is a contradiction. Now, we have two sub-cases:
\begin{enumerate}
\item If $\overline{\RR}<\RR$, we can choose $\epsilon=\RR-\overline{\RR}>0$
and obtain
\[
\min_{Q_{X}:\rho(Q_{X})\geq\RR-\epsilon}D(Q_{X}||P_{X})\geq D(Q_{X}^{*}||P_{X}).
\]

\item Otherwise, suppose that $\sup_{Q_{X}\in{\cal V}}\rho(Q_{X})$ is not
attained and $\overline{\RR}\geq\RR$. If $\overline{\RR}>\RR$ then
there exists a sequence $\overline{Q}_{X}^{(n)}\in{\cal V}$ such
that $\rho(\overline{Q}_{X}^{(n)})\to\overline{\RR}$, and so there
exists $n_{0}$ such that $\rho(\overline{Q}_{X}^{(n)})>\overline{\RR}$
which contradicts the optimality of $Q_{X}^{*}$, and so we must have
$\overline{\RR}=\RR$. In this case, $\rho(Q_{X})<\overline{\RR}$
for all $Q_{X}\in{\cal V}$, so define 
\[
{\cal W}\teq\{Q_{X}\in{\cal Q}({\cal X}):D(Q_{X}||P_{X})\leq d-\delta\}
\]
and let $\RR'\teq\max_{Q_{X}\in{\cal W}}\rho(Q_{X})$, where clearly
$\RR'<\RR$. Then, for $\epsilon=\RR-\RR'>0$
\[
\min_{Q_{X}:\rho(Q_{X})\geq\RR-\epsilon}D(Q_{X}||P_{X})\geq d-\delta=D(Q_{X}^{*}||P_{X})-\delta.
\]

\end{enumerate}

To conclude, in both cases, for any given $\delta>0$ we can find
$\epsilon>0$ such that 
\[
\min_{Q_{X}:\rho(Q_{X})\geq\RR-\epsilon}D(Q_{X}||P_{X})\geq D(Q_{X}^{*}||P_{X})-\delta.
\]
This means that $\min_{Q_{X}:\rho(Q_{X})\geq\RR}D(Q_{X}||P_{X})$
is left-continuous as a function of $\RR$, and the desired result
is obtained.

\end{IEEEproof}

\begin{IEEEproof}[Proof of Lemma \ref{lem: Excess Rate Exponent Properties}]
~
\begin{itemize}
\item \emph{\uline{Zero value domain:}} This follows directly from Theorem
\ref{thm: General upper bound on excess rate exponent}.
\item \emph{\uline{Infinite value domain:}} This follows directly from
the excess-rate exponent bound of Theorem \ref{thm: achieavbility with type dependent}.
\item \emph{\uline{Monotonicity:}} The first statement follows directly
from the definition \eqref{eq: excess rate exponent function definition}.
When $\rho(Q_{X})$ is regular, we may use \eqref{eq: left continuity for regular rate functions}.
Now, let $Q_{X}^{*}$ be any minimizer of \eqref{eq: left continuity for regular rate functions},
for a given $\RR<\RR_{\max}'$. We begin by showing that $\rho(Q_{X}^{*})=\RR$.
Assume conversely, that $\rho(Q_{X}^{*})>\RR$. Since $\RR<\RR_{\max}'$
then the same arguments that were used in the proof of Lemma \ref{lem: excess rate exponent expression}
show that $Q_{X}^{*}\in{\cal V}$. Now, consider 
\[
Q_{\alpha,X}=(1-\alpha)P_{X}+\alpha Q_{X}^{*}.
\]
Since $\rho(Q_{X})$ is continuous in ${\cal V}$, then the intermediate
value theorem implies that $\alpha<1$ must exist such that $\rho(Q_{\alpha,X})=\RR$.
Using Lemma \ref{lem:divergence increasing for a ray}, we have that
$D(Q_{\alpha,X}||P_{X})<D(Q_{X}^{*}||P_{X})$ which contradicts the
fact that $Q_{X}^{*}$ is a minimizer of \eqref{eq: left continuity for regular rate functions}.
Now, let ${\cal M}(\RR)$ be the collection of all minimizers of \eqref{eq: left continuity for regular rate functions},
such that for all $Q_{X}\notin{\cal M}(\RR)$ we have either $D(Q_{X}||P_{X})>D(Q_{X}^{*}||P_{X})$
or $\rho(Q_{X})<\RR$. Thus, for any $\RR_{1}>\RR$ we have 
\[
\min_{Q_{X}:\:\rho(Q_{X})\geq\RR_{1}}D(Q_{X}||P_{X})>D(Q_{X}^{*}||P_{X}).
\]

\item \emph{\uline{Continuity:}} The first statement follows from the
fact that monotonic functions are continuous except for a countable
number of points (Froda's theorem). The proof of the second is a part
of the proof of Lemma \ref{lem: excess rate exponent expression}. 
\end{itemize}
\end{IEEEproof}

\begin{IEEEproof}[Proof of Theorem \ref{thm: excess rate and rate curve condition}]
For any given $(\RR,\mathsf{E}_{r})$ we may use the condition of
Lemma \ref{lem: Dominating and simple rate function equivalence}.
Notice that $\hat{\rho}(Q_{X};\RR,\mathsf{E}_{r})$ is a regular rate
function, and so the excess-rate exponent in Lemma \ref{lem: excess rate exponent expression}
is applicable. The proof is divided into three parts, one for each
of the bounds.

\emph{\uline{Random binning bound:}} From Theorem \ref{thm: achieavbility with type dependent}
and the random coding bound in \eqref{eq: channel random coding bound},
the rate function $\rho(Q_{X};\RR,\mathsf{E}_{r})$ will achieve infimum
error exponent $\EE$ if 
\[
\EE\leq D(Q_{X}||P_{X})+\min_{Q_{Y|X}}\left\{ D(Q_{Y|X}||P_{Y|X}|Q_{X})+\left[\rho(Q_{X};\RR,\mathsf{E}_{r})-H(Q_{X|Y}|Q_{Y})\right]_{+}\right\} 
\]
for all $Q_{X}$. Now, choosing \textbf{$\RR_{0}$ }sufficiently large,
this condition will be satisfied for any $Q_{X}$ which satisfies
$D(Q_{X}||P_{X})>\mathsf{E}_{r}$, and then the resulting condition
is 
\begin{alignat}{1}
\EE & \leq\min_{Q_{X}:\: D(Q_{X}||P_{X})\leq\mathsf{E}_{r}}\min_{Q_{Y|X}}\left\{ D(Q_{X}||P_{X})+D(Q_{Y|X}||P_{Y|X}|Q_{X})+\left[\RR-H(Q_{X|Y}|Q_{Y})\right]_{+}\right\} \\
 & =\min_{Q_{X}:\: D(Q_{X}||P_{X})\leq\mathsf{E}_{r}}\min_{Q_{Y|X}}\max_{0\leq t\leq1}\Bigl\{ D(Q_{X}||P_{X})+D(Q_{Y|X}||P_{Y|X}|Q_{X})\nonumber \\
 & +t\left[\RR-H(Q_{X|Y}|Q_{Y})\right]\Bigr\}\label{eq: minmax optimization for error exponent verification random binning}\\
 & \trre[=,a]\max_{0\leq t\leq1}\min_{Q_{X}:\: D(Q_{X}||P_{X})\leq\mathsf{E}_{r}}\min_{Q_{Y|X}}\Bigl\{ D(Q_{X}||P_{X})+D(Q_{Y|X}||P_{Y|X}|Q_{X})\nonumber \\
 & +t\left[\RR-H(Q_{X|Y}|Q_{Y})\right]\Bigr\}
\end{alignat}
where $(a)$ is because the minimization problem in \eqref{eq: minmax optimization for error exponent verification random binning}
is convex in $Q_{X}$ (over the convex set $\{Q_{X}\in{\cal Q}({\cal X}):D(Q_{X}||P_{X})\leq\mathsf{E}_{r}\}$)
and $\{Q_{Y|X}\}$, and the maximization problem is linear in $t$
(over the convex set $[0,1]$), and thus also concave. Therefore,
we can interchange the maximization and minimization \cite{Sion1958minimax}
order, and obtain the condition $\max_{0\leq t\leq1}e_{\s[rb]}(t)\geq\EE$. 

\emph{\uline{Expurgated bound:}} From Theorem \ref{thm: achieavbility with type dependent}
and the expurgated bound in \eqref{eq: channel expurgated bound},
the rate function $\rho(Q_{X};\RR,\mathsf{E}_{r})$ will achieve infimum
error exponent $\EE$ if
\[
\EE\leq D(Q_{X}||P_{X})+\min_{Q_{\tilde{X}|X}:\: Q_{\tilde{X}}=Q_{X},\rho(Q_{X};\RR,\mathsf{E}_{r})\leq H(Q_{X|\tilde{X}}|Q_{\tilde{X}})}\left\{ B(Q_{X\tilde{X}},P_{Y|X})+\rho(Q_{X};\RR,\mathsf{E}_{r})-H(Q_{X|\tilde{X}}|Q_{\tilde{X}})\right\} 
\]
for all $Q_{X}$. Now, choosing \textbf{$\RR_{0}$ }sufficiently large,
this condition will be satisfied for any $Q_{X}$ which satisfies
$D(Q_{X}||P_{X})>\mathsf{E}_{r}$, and then the resulting condition
is 
\begin{alignat}{1}
\EE & \leq\min_{Q_{X}:\: D(Q_{X}||P_{X})\leq\mathsf{E}_{r}}\min_{Q_{\tilde{X}|X}:\:\RR\leq H(Q_{X|\tilde{X}}|Q_{\tilde{X}})}\Bigl\{ D(Q_{X}||P_{X})+B(Q_{X\tilde{X}},P_{Y|X})\nonumber \\
 & +\RR-H(Q_{X|\tilde{X}}|Q_{\tilde{X}})\Bigr\}\\
 & =\min_{Q_{X}:\: D(Q_{X}||P_{X})\leq\mathsf{E}_{r}}\min_{Q_{\tilde{X}|X}}\max_{t\geq0}\Bigl\{ D(Q_{X}||P_{X})+B(Q_{X\tilde{X}},P_{Y|X})\nonumber \\
 & +(t+1)\left[\RR-H(Q_{X|\tilde{X}}|Q_{\tilde{X}})\right]\Bigr\}\label{eq: minmax optimization for error exponent verification expurgated}\\
 & \trre[=,a]\max_{t\geq1}\min_{Q_{X}:\: D(Q_{X}||P_{X})\leq\mathsf{E}_{r}}\min_{Q_{\tilde{X}|X}}\left\{ D(Q_{X}||P_{X})+B(Q_{X\tilde{X}},P_{Y|X})+t\left[\RR-H(Q_{X|\tilde{X}}|Q_{\tilde{X}})\right]\right\} 
\end{alignat}
where in the maximization problems above, the constraint $Q_{X}=(Q_{X}\times Q_{\tilde{X}|X})_{\tilde{X}}$
is also imposed. The passage $(a)$ is because the minimization problem
in \eqref{eq: minmax optimization for error exponent verification expurgated}
is jointly convex in $Q_{X}$ (over the convex set $\{Q_{X}\in{\cal Q}({\cal X}):D(Q_{X}||P_{X})\leq\mathsf{E}_{r}\}$),
$\{Q_{\tilde{X}|X}:\: Q_{\tilde{X}}=Q_{X}\}$, and the maximization
problem is linear in $t$ (over the convex set $[1,\infty)$), and
thus also concave. Therefore, we can interchange the maximization
and minimization \cite{Sion1958minimax} order, and obtain the condition
$\max_{0\leq t\leq1}e_{\s[ex]}(t)\geq\EE$. 

\emph{\uline{Sphere packing bound:}} From Theorem \ref{thm: General upper bound on error exponent}
and the sphere packing bound in \eqref{eq: channel sphere packing bound},
the rate function $\rho(Q_{X};\RR,\mathsf{E}_{r})$ will not achieve
supremum error exponent $\EE$ unless
\[
\EE\leq D(Q_{X}||P_{X})+\min_{Q_{Y|X}:\:\rho(Q_{X};\RR,\mathsf{E}_{r})\leq H(Q_{X|Y}|Q_{Y})}D(Q_{Y|X}||P_{Y|X}|Q_{X})
\]
for all $Q_{X}$. Now, choosing \textbf{$\RR_{0}$ }sufficiently large
this condition will be satisfied for any $Q_{X}$ which satisfies
$D(Q_{X}||P_{X})>\mathsf{E}_{r}$, and then the resulting condition
is 
\begin{eqnarray}
\EE & \leq & \min_{Q_{X}:\: D(Q_{X}||P_{X})\leq\mathsf{E}_{r}}\min_{Q_{Y|X}:\:\RR\leq H(Q_{X|Y}|Q_{Y})}\left\{ D(Q_{X}||P_{X})+D(Q_{Y|X}||P_{Y|X}|Q_{X})\right\} \\
 & = & \min_{Q_{X}:\: D(Q_{X}||P_{X})\leq\mathsf{E}_{r}}\min_{Q_{Y|X}}\max_{t\geq0}\Bigl\{ D(Q_{X}||P_{X})+D(Q_{Y|X}||P_{Y|X}|Q_{X})\nonumber \\
 &  & +t\left[\RR-H(Q_{X|Y}|Q_{Y})\right]\Bigr\}\label{eq: minmax optimization for error exponent verification sphere packing}\\
 & \trre[=,a] & \max_{t\geq0}\min_{Q_{X}:\: D(Q_{X}||P_{X})\leq\mathsf{E}_{r}}\min_{Q_{Y|X}}\Bigl\{ D(Q_{X}||P_{X})+D(Q_{Y|X}||P_{Y|X}|Q_{X})\nonumber \\
 &  & +t\left[\RR-H(Q_{X|Y}|Q_{Y})\right]\Bigr\}
\end{eqnarray}
where $(a)$ is because the minimization problem in \eqref{eq: minmax optimization for error exponent verification sphere packing}
is convex in $Q_{X}$ (over the convex set $\{Q_{X}\in{\cal Q}({\cal X}):D(Q_{X}||P_{X})\leq\mathsf{E}_{r}\}$),
$\{Q_{Y|X}\}$, and the maximization problem is linear in $t$ (over
the convex set $[1,\infty)$), and thus also concave. Therefore, we
can interchange the maximization and minimization \cite{Sion1958minimax}
order, and obtain the condition $\max_{t\geq0}e_{\s[sp]}(t)\geq\EE$. 
\end{IEEEproof}

\begin{IEEEproof}[Proof of Lemma \ref{lem:Proof of alternating minimization algorithm v rb}]
Introducing an auxiliary PMF $\tilde{Q}_{Y}$ and using Lemma \ref{lem: Divergence output marginal minimization}
(Appendix \ref{sec: Useful Lemmas}) we get that
\begin{eqnarray}
v_{\s[rb]}(P_{XY},Q_{X},\EE,\eta) & = & \min_{Q_{Y|X}:D(Q_{X}\times Q_{Y|X}||P_{XY})\leq\EE}\min_{\tilde{Q}_{Y}}\bigl\{ D(Q_{Y|X}||\tilde{Q}_{Y}|Q_{X})\nonumber \\
 &  & +\eta\cdot D(Q_{Y|X}||P_{Y|X}|Q_{X})\bigr\}\\
 & = & \min_{\tilde{Q}_{Y}}\min_{Q_{Y|X}:D(Q_{X}\times Q_{Y|X}||P_{XY})\leq\EE}\bigl\{ D(Q_{Y|X}||\tilde{Q}_{Y}|Q_{X})\nonumber \\
 &  & +\eta\cdot D(Q_{Y|X}||P_{Y|X}|Q_{X})\bigr\}\label{eq: v rb optimization Q_Y_tilde}
\end{eqnarray}
Notice that \eqref{eq: v rb optimization Q_Y_tilde} is an optimization
problem over $(Q_{Y|X},\tilde{Q}_{Y})$ and consider utilizing an
alternating minimization algorithm, where for a given $\tilde{Q}_{Y}$,
the minimizer $Q_{Y|X}$ is found, and vice versa. We divide the rest
of the proof into two main parts. In the first part, we prove that
the alternating minimization algorithm indeed converges to the optimal
solution, and in the second part, we solve the two individual optimization
problems (resulting from keeping one of the optimization variables
fixed).

\emph{\uline{Part 1:}} In \cite[Section 5.2]{csiszar2004information},
\cite{csiszar_information_1984} sufficient conditions were derived
for the convergence of an alternating minimization algorithm. Specifically,
these conditions are met for a minimization problem of the form 
\begin{equation}
\inf_{Q_{1}\in{\cal Q}_{1}}\inf_{Q_{2}\in{\cal Q}_{2}}D(Q_{1}||Q_{2})\label{eq: alternating minimization general form}
\end{equation}
where $Q_{1}$ and $Q_{2}$ are two positive measures (which may not
necessarily sum to $1$) over a finite alphabet ${\cal Z}$, and ${\cal Q}_{1},{\cal Q}_{2}$
are two convex sets. To prove that alternating minimization algorithm
converges for the optimization problem \eqref{eq: v rb optimization Q_Y_tilde},
we now show that it can be written in the form of \eqref{eq: alternating minimization general form}.
The objective function of \eqref{eq: v rb optimization Q_Y_tilde}
is given by 
\begin{alignat}{1}
 & D(Q_{Y|X}||\tilde{Q}_{Y}|Q_{X})+\eta\cdot D(Q_{Y|X}||P_{Y|X}|Q_{X})\nonumber \\
= & \sum_{x,y}Q_{X}(x)Q_{Y|X}(y|x)\log\frac{\left[Q_{Y|X}(y|x)\right]^{1+\eta}}{\tilde{Q}_{Y}(y)P_{Y|X}^{\eta}(y|x)}\\
= & (1+\eta)\sum_{x,y}Q_{XY}(x,y)\log\frac{Q_{XY}(x,y)}{\left[\tilde{Q}_{Y}(y)\right]^{\frac{1}{1+\eta}}\left[P_{Y|X}(y|x)\right]^{\frac{\eta}{1+\eta}}Q_{X}(x)}.
\end{alignat}
Thus, if we let ${\cal Z}={\cal X}\times{\cal Y}$ and consider the
measures $Q_{XY}$ and $\breve{Q}_{XY}\teq\tilde{Q}_{Y}^{\frac{1}{1+\eta}}P_{Y|X}^{\frac{\eta}{1+\eta}}Q_{X}$
\footnote{Note that this measure does not necessarily sum to $1$. %
} then the objective function is of the form of \eqref{eq: alternating minimization general form}.
Now, the feasible set for $Q_{XY}$ is 
\[
\left\{ Q_{XY}:\sum_{y\in{\cal Y}}Q_{XY}(x,y)=Q_{X}(x),D\left(Q_{XY}||P_{XY}\right)\leq\EE\right\} 
\]
which is a convex set. Now, using Corollary \ref{cor:Divergence output marginal minimization corollary (less than 1)}
of Lemma \ref{lem: Divergence output marginal minimization} (Appendix
\ref{sec: Useful Lemmas}), we have that the feasible region of $\tilde{Q}_{Y}$
can be extended from the simplex ${\cal Q}({\cal Y})$ to the set
\[
\tilde{{\cal Q}}({\cal Y})\teq\left\{ \tilde{Q}_{Y}:\sum_{y\in{\cal Y}}\tilde{Q}_{Y}(y)\leq1,\tilde{Q}_{Y}(y)\geq0\mbox{ for all }y\in{\cal Y}\right\} 
\]
which is also a convex set. Now, define the feasible set for the variables
$\breve{Q}_{XY}$ as 
\begin{multline*}
{\cal \breve{Q}}\teq\Bigl\{\breve{Q}_{XY}:\exists\tilde{Q}_{Y}\in\tilde{{\cal Q}}({\cal Y})\\
\mbox{ so that }\breve{Q}_{XY}(x,y)=\left[\tilde{Q}_{Y}(y)\right]^{\frac{1}{1+\eta}}\left[P_{Y|X}(y|x)\right]^{\frac{\eta}{1+\eta}}Q_{X}(x)\mbox{ for all }(x,y)\in{\cal X}\times{\cal Y}\Bigr\}.
\end{multline*}
We show that $\breve{{\cal Q}}$ is also a convex set. Let $\breve{Q}_{i,XY}(x,y)=\left[\tilde{Q}_{i,Y}(y)\right]^{\frac{1}{1+\eta}}\left[P_{Y|X}(y|x)\right]^{\frac{\eta}{1+\eta}}Q_{X}(x)$
for $\tilde{Q}_{i,Y}\in\tilde{{\cal Q}}({\cal Y})$, $i=0,1$, and
$0\leq\alpha\leq1$. Then, 
\begin{eqnarray}
\breve{Q}_{\alpha,XY} & \teq & (1-\alpha)\breve{Q}_{0,XY}+\alpha\breve{Q}_{1,XY}\\
 & = & P_{Y|X}^{\frac{\eta}{1+\eta}}Q_{X}\cdot\left((1-\alpha)\tilde{Q}_{0,Y}^{\frac{1}{1+\eta}}+\alpha\tilde{Q}_{1,Y}^{\frac{1}{1+\eta}}\right).
\end{eqnarray}
Thus, to show that $\breve{Q}_{\alpha,XY}\in{\cal \breve{Q}}$ all
is needed to prove is that $\tilde{Q}_{\alpha,Y}\teq\left((1-\alpha)\tilde{Q}_{0,Y}^{\frac{1}{1+\eta}}+\alpha\tilde{Q}_{1,Y}^{\frac{1}{1+\eta}}\right)^{1+\eta}\in{\cal \tilde{Q}}({\cal Y})$.
As positivity of $\tilde{Q}_{\alpha,Y}$ is clear, it remains to verify
that $\sum_{y\in{\cal Y}}\tilde{Q}_{\alpha,Y}(y)\leq1$. Indeed, we
have 
\begin{eqnarray}
\sum_{y\in{\cal Y}}\tilde{Q}_{\alpha,Y}(y) & = & \sum_{y\in{\cal Y}}\left((1-\alpha)\tilde{Q}_{0,Y}^{\frac{1}{1+\eta}}+\alpha\tilde{Q}_{1,Y}^{\frac{1}{1+\eta}}\right)^{1+\eta}\\
 & \trre[\leq,a] & \left[(1-\alpha)\left(\sum_{y\in{\cal Y}}\tilde{Q}_{0,Y}(y)\right)^{\frac{1}{1+\eta}}+\alpha\left(\sum_{y\in{\cal Y}}\tilde{Q}_{1,Y}(y)\right)^{\frac{1}{1+\eta}}\right]^{1+\eta}\\
 & \trre[\leq,b] & \left[(1-\alpha)+\alpha\right]^{1+\eta}\\
 & = & 1
\end{eqnarray}
where $(a)$ follows from a variant of Minkowski's inequality (Lemma
\ref{lem: Variant of Minkowski inequality} in Appendix \ref{sec: Useful Lemmas}),
and $(b)$ is from the fact that both $t^{\frac{1}{1+\eta}}$ and
$t^{1+\eta}$ are increasing functions of $t\in\mathbb{R}^{+}$ when
$\eta\geq0$, and $\tilde{Q}_{Y}\in{\cal \tilde{Q}}({\cal Y})$. Thus
the optimization problem \eqref{eq: v rb optimization Q_Y_tilde}
is of the form \eqref{eq: alternating minimization general form}
and an alternating minimization algorithm converges to the optimal,
unique, solution, which we denote by $(Q_{Y|X}^{*},\tilde{Q}_{Y}^{*})$. 

\emph{\uline{Part 2:}} First, suppose that $\tilde{Q}_{Y}$ is
given. In order to find the minimizer $Q_{Y|X}$ the Karush-Kuhn-Tucker
(KKT) conditions for convex problems \cite[Section 5.5.3]{boyd2004convex}
can be utilized. Ignoring positivity constraints for the moment, and
defining the Lagrangian 
\begin{eqnarray}
L(Q_{Y|X},\lambda,\mu_{x}) & = & \sum_{x\in{\cal X}}Q_{X}(x)D(Q_{Y|X}(\cdot|x)||\tilde{Q}_{Y})+\eta\sum_{x\in{\cal X}}Q_{X}(x)D\left(Q_{Y|X}(\cdot|x)||P_{Y|X}(\cdot|x)\right)\nonumber \\
 &  & +\lambda\cdot\sum_{x\in{\cal X}}Q_{X}(x)D\left(Q_{Y|X}(\cdot|x)||P_{Y|X}(\cdot|x)\right)+\sum_{x\in{\cal X}}\mu_{x}\sum_{y\in{\cal Y}}Q_{Y|X}(y|x)\\
 & = & \sum_{x\in{\cal X}}Q_{X}(x)\sum_{y\in{\cal Y}}Q_{Y|X}(y|x)\log\frac{\left[Q_{Y|X}(y|x)\right]^{1+\eta+\lambda}}{\tilde{Q}_{Y}\left[P_{Y|X}(y|x)\right]^{\eta+\lambda}}+\sum_{x\in{\cal X}}\mu_{x}\sum_{y\in{\cal Y}}Q_{Y|X}(y|x)\nonumber 
\end{eqnarray}
where $\lambda\geq0$ and $\mu_{x}\in\mathbb{R}$ for $x\in{\cal X}$.
Differentiating w.r.t. some $Q_{Y|X}(y'|x')$ for $x'\in{\cal X},y'\in{\cal Y}$
\[
\frac{\partial L}{\partial Q_{Y|X}(y'|x')}=Q_{X}(x')\left((1+\eta+\lambda)\cdot\left(\log Q_{Y|X}(y'|x')+1\right)+\log\frac{1}{\tilde{Q}_{Y}(y')\left[P_{Y|X}(y'|x')\right]^{\eta+\lambda}}\right)+\mu_{x'}
\]
and equating to zero we get 
\[
Q_{X}(x')\cdot\log\frac{\left[Q_{Y|X}(y'|x')\right]^{(1+\eta+\lambda)}}{\left[P_{Y|X}(y'|x')\right]^{\eta+\lambda}\tilde{Q}_{Y}(y')}+\mu'_{x'}=0
\]
where $\mu'_{x'}=\mu_{x'}+1+\eta+\lambda$. Thus, the argument of
the logarithm must not depend on $x$, and this implies that for any
$x\in{\cal X}$ such that $Q_{X}(x)\neq0$ we must have
\[
Q_{Y|X}^{*}(y|x)=\psi_{x}\left[P_{Y|X}(y|x)\right]^{\alpha}\left[\tilde{Q}_{Y}(y)\right]^{1-\alpha}
\]
for $\alpha=\frac{\eta+\lambda}{1+\eta+\lambda}$, where $\psi_{x}$
is a normalization constant. Clearly, from \eqref{eq: definition of alpha conditional probability}
we have $Q_{Y|X}^{*}=\mathbb{M}_{\s[g]}(P_{Y|X},\tilde{Q}_{Y},\alpha^{*})$.
The value of $Q_{Y|X}^{*}$ for $x\in{\cal X}$ with $Q_{X}(x)=0$
is immaterial as it does not affect the optimal value of the objective
function. Also, it is evident that the solution $Q_{Y|X}^{*}$ is
indeed positive. 

To find the optimal $Q_{Y|X}^{*}$, we need to choose $\alpha$ in
order to satisfy the constraint $D(Q_{Y|X}^{*}||P_{XY})\leq\EE$.
For this, the \emph{complementary slackness condition }\cite[Section 5.5.2]{boyd2004convex}\emph{
}implies that $\alpha$ should be chosen either to satisfy 
\[
D(Q_{Y|X}^{*}||P_{Y|X}|Q_{X})=\EE-D(Q_{X}||P_{X})
\]
and then $\frac{\eta}{1+\eta}<\alpha\leq1$, or $\alpha=\frac{\eta}{1+\eta}$
and then 
\[
D(Q_{Y|X}^{*}||P_{Y|X}|Q_{X})<\EE-D(Q_{X}||P_{X}).
\]
To find $\alpha^{*}$ that satisfies the complementary slackness\emph{
}condition, note that $D(Q_{Y|X}^{*}||P_{Y|X}|Q_{X})$ is a monotonically
decreasing function of $\alpha$. Indeed, it is easy to see that if
$\tilde{Q}_{Y}$ is initialized such that 
\[
\supp(\tilde{Q}_{Y})=\supp\left(\sum_{x\in{\cal X}}Q_{X}(x)P_{Y|X}(y|x)\right)
\]
 then this remains true for all iterations. Then, it follows from
Lemma \ref{lem: Monotinicty of divergence for exponential families}
(Appendix \ref{sec: Useful Lemmas}) that for any given $x\in{\cal X}$
such that $Q_{X}(x)\neq0$, we have that $D(Q_{Y|X}^{*}(\cdot|x)||P_{Y|X}(\cdot|x))$
is a decreasing function of $\alpha$, and thus their average $D(Q_{Y|X}^{*}||P_{Y|X}|Q_{X})$
is also a decreasing function of $\alpha$. Thus, if for $\alpha=\frac{\eta}{1+\eta}$
we have $D(Q_{Y|X}^{*}||P_{Y|X}|Q_{X})<\EE-D(Q_{X}||P_{X})$ then
$\alpha^{*}=\frac{\eta}{1+\eta}$. Otherwise, we have $D(\overline{Q}_{Y|X}^{\frac{\eta}{1+\eta}}||P_{Y|X}|Q_{X})>\EE-D(Q_{X}||P_{X})$
and $D(\overline{Q}_{Y|X}||P_{Y|X}|Q_{X})=0<\EE-D(Q_{X}||P_{X})$.
Thus, in the later case, a simple search finds $\alpha^{*}$. 

Second, assume that $Q_{Y|X}$ is given. The minimizer $\tilde{Q}_{Y}$
can be found using Lemma \ref{lem: Divergence output marginal minimization}
(Appendix \ref{sec: Useful Lemmas}) to be 
\begin{equation}
\tilde{Q}_{Y}(y)=\sum_{x\in{\cal X}}Q_{X}(x)Q_{Y|X}(y|x).\label{eq: optimal tilde Qy}
\end{equation}
It is easily seen that Algorithm \ref{alg:Alternating minimization algorithm v rb}
indeed implements the procedure described in this proof. 
\end{IEEEproof}

\begin{IEEEproof}[Proof of Lemma \ref{lem:Proof of alternating minimization algorithm Value rb}]
Introducing an auxiliary PMF $\tilde{Q}_{Y}$ and using Lemma \ref{lem: Divergence output marginal minimization}
(Appendix \ref{sec: Useful Lemmas}), we get that

\begin{multline*}
e_{\s[rb]}(P_{XY},\RR,\mathsf{E}_{r},t)\teq\min_{\tilde{Q}_{Y}}\min_{Q_{X}:D(Q_{X}||P_{X})\leq\mathsf{E}_{r}}\min_{Q_{Y|X}}\\
\left\{ D(Q_{X}||P_{X})+D(Q_{Y|X}||P_{Y|X}|Q_{X})+t\cdot\left[\RR-H(Q_{X})+D(Q_{Y|X}||\tilde{Q}_{Y}|Q_{X})\right]\right\} .
\end{multline*}
Now, Algorithm \ref{alg:Alternating minimization algorithm V rb}
is an alternating minimization algorithm, that keeps all parameters
but one fixed, and optimizes over the non-fixed parameter. Now, for
a given $t\geq0$, the objective function in \eqref{eq: minimization problem excess rate random binning}
is given by
\[
(1+t)\cdot\sum_{x,y}Q_{XY}(x,y)\log\frac{Q_{XY}(x,y)}{\left[P_{XY}(x,y)\right]^{\frac{1}{1+t}}\left[\tilde{Q}_{Y}(y)\right]^{\frac{t}{1+t}}}+t\RR.
\]
The same technique that was used in the proof of Lemma \ref{lem:Proof of alternating minimization algorithm v rb},
shows that this optimization problem is of the form \eqref{eq: alternating minimization general form}
(with additional constant $t\RR$). Thus, an alternating minimization
algorithm converges to the optimal solution. 

We now turn to the minimization of individual variables, assuming
that all other variables are fixed, for a given $t\geq0$. First,
consider the minimization over $Q_{XY}$, which itself can be separated
to an unconstrained minimization over $Q_{Y|X}$ and a constrained
minimization over $Q_{X}$. The minimizer $Q_{Y|X}^{*}$ can again
be found using similar Lagrange methods as in the proof of Lemma \ref{lem:Proof of alternating minimization algorithm v rb}.
The result is $Q_{Y|X}^{*}=\mathbb{M}_{\s[g]}(P_{Y|X},\tilde{Q}_{Y},\frac{1}{1+t})$
(for all $x\in{\cal X}$ such that $Q_{X}(x)\neq0$, and arbitrary
otherwise, since the value of $Q_{Y|X}^{*}(\cdot|x)$ for $x\in{\cal X}$
such that $Q_{X}(x)=0$ does not affect the value of the optimization
problem). For this optimal choice, using the definitions of $h_{1,t}(x)$
and $h_{2,t}(x)$ we obtain 
\[
\min_{\tilde{Q}_{Y}}\min_{Q_{X}:D(Q_{X}||P_{X})\leq\mathsf{E}_{r}}\left\{ D(Q_{X}||P_{X})+\sum_{x\in{\cal X}}Q_{X}(x)h_{1,t}(x)+t\cdot\left[\RR-H(Q_{X})+\sum_{x\in{\cal X}}Q_{X}(x)h_{2,t}(x)\right]\right\} .
\]
Next, we optimize over $Q_{X}$ using the KKT conditions. The Lagrangian
with $\lambda\text{\ensuremath{\geq}}0$ and $\mu$ is given by
\begin{eqnarray}
L(Q_{X},\lambda,\mu) & \triangleq & D(Q_{X}||P_{X})+\sum_{x\in{\cal X}}Q_{X}(x)h_{1,t}(x)+t\cdot\left[\RR-H(Q_{X})+\sum_{x\in{\cal X}}Q_{X}(x)h_{2,t}(x)\right]\nonumber \\
 &  & +\lambda\cdot D(Q_{X}||P_{X})+\mu\cdot\sum_{x\in{\cal X}}Q_{x}\\
 & = & t\cdot\RR+\sum_{x\in{\cal X}}Q_{X}(x)\left[\log\left(\frac{\left[Q_{X}(x)\right]^{1+t+\lambda}}{\left[P_{X}(x)\right]^{1+\lambda}}\cdot\exp(h_{1,t}(x)+t\cdot h_{2,t}(x))\right)+\mu\right].\nonumber 
\end{eqnarray}
Differentiating w.r.t. some $Q_{X}(x')$ for $x'\in{\cal X}$, we
get
\[
\frac{\partial L}{\partial Q_{X}(x')}=\log\left(\frac{\left[Q_{X}(x')\right]^{1+t+\lambda}}{\left[P_{X}(x')\right]^{1+\lambda}}\cdot\exp(h_{1,t}(x')+t\cdot h_{2,t}(x'))\right)+(1+t+\lambda)+\mu
\]
and equating to zero results in 
\[
Q_{X}^{*}(x)=\psi\cdot\left[P_{X}(x)\right]^{\frac{1+\lambda}{1+\lambda+t}}\cdot\exp\left(-\frac{1}{1+t+\lambda}\cdot h_{1,t}(x)-\frac{t}{1+t+\lambda}\cdot h_{2,t}(x)\right)
\]
where $\psi$ is a normalization constant. From the definition \eqref{eq: optimal prior excess rate},
it is evident that $Q_{X}^{*}=\mathbb{M}_{\s[h]}(P_{X},h_{1},h_{2},\lambda,t)$.
Using the complementary slackness condition \cite[Section 5.5.2]{boyd2004convex},
$\lambda$ should be found such that either $D(Q_{X}^{*}||P_{X})=\mathsf{E}_{r}$
or $\lambda=0$. From Lemma \ref{lem: Monotinicty of divergence for exponential families}
(Appendix \ref{sec: Useful Lemmas}) $D(Q_{X}^{*}||P_{X})$ is a monotonic
decreasing function of $\lambda$ and thus the above search is relatively
simple. To see that the conditions of Lemma \ref{lem: Monotinicty of divergence for exponential families}
are met, notice that initializing $\tilde{Q}_{Y}$ with support ${\cal Y}$
implies that in the first iteration $\supp(Q_{Y|X}^{*})=\supp(P_{Y|X})$
which assures that $h_{1,t}(x)$ and $h_{2,t}(x)$ are finite. As
$\supp(Q_{X}^{*})=\supp(P_{X})={\cal X}$ for all $\lambda>0$ and
$t\geq0$ then $\supp(\tilde{Q}_{Y})={\cal Y}$ for all iterations
(cf. \eqref{eq: optimal tilde_Qy for excess rate algorithm}). Thus,
for any $t\neq0$ we may express $Q_{X}^{*}$ as 
\[
Q_{X}^{*}(x)=\psi\cdot\left[P_{X}(x)\right]^{\frac{1+\lambda}{1+\lambda+t}}\cdot\left[\breve{P}_{X}(x)\right]^{\frac{t}{1+\lambda+t}}
\]
where 
\[
\breve{P}_{X}(x)\teq\breve{\psi}\cdot\exp\left(-\frac{h_{1,t}(x)}{t}-h_{2,t}(x)\right)
\]
and $\breve{\psi}$ is a normalization factor. Setting $\alpha=\frac{1+\lambda}{1+\lambda+t}$
we get that $D(Q_{X}^{*}||P_{X})$ is a decreasing function of $\alpha$.
Since $\alpha$ is a monotonically increasing function of $\lambda$
this implies that $D(Q_{X}^{*}||P_{X})$ is also a decreasing function
of $\lambda$. For $t=0$ we may write again 
\[
Q_{X}^{*}(x)=\psi\cdot P_{X}^{\frac{\lambda}{1+\lambda}}(x)\cdot\breve{P}_{X}^{\frac{1}{1+\lambda}}(x)
\]
where now 
\[
\breve{P}_{X}(x)\teq\breve{\psi}\cdot\exp\left(-h_{1,t}(x)+\log P_{X}(x)\right).
\]
Similar arguments show that $D(Q_{X}^{*}||P_{X})$ is a decreasing
function of $\lambda$.

The optimal $\tilde{Q}_{Y}^{*}$ for a given $t$ and $Q_{X},Q_{Y|X}$
is simply 
\begin{equation}
\tilde{Q}_{Y}^{*}(y)=\sum_{x\in{\cal X}}Q_{X}(x)Q_{Y|X}(y|x),\label{eq: optimal tilde_Qy for excess rate algorithm}
\end{equation}
using Lemma \ref{lem: Divergence output marginal minimization}.
\end{IEEEproof}

\begin{IEEEproof}[Proof of Lemma \ref{lem:Proof of alternating minimization algorithm value ex}]
Clearly \eqref{eq: value definition expurgated} can be written as
\[
v_{\s[ex]}(P_{XY},Q_{X},\EE)=\min_{\breve{Q}_{X\tilde{X}}\in{\cal G}}D(\breve{Q}_{X\tilde{X}}||Q_{X}\times Q_{X}),
\]
where ${\cal G}\teq{\cal G}_{1}\cap{\cal G}_{2}\cap{\cal G}_{3}$
and 
\[
{\cal G}_{1}\teq\{Q_{X\tilde{X}}:B(Q_{X\tilde{X}})=\EE+D(Q_{X}||P_{X})\},
\]
\[
{\cal G}_{2}\teq\{\breve{Q}_{X}=Q_{X}\},
\]
\[
{\cal G}_{3}\teq\{\breve{Q}_{\tilde{X}}=Q_{X}\}.
\]
It may be easily seen that $\{{\cal G}_{i}\}_{i=1}^{3}$ are\emph{
linear families}%
\footnote{A linear family of PMFs over the alphabet ${\cal X}$, is any set
of the form $\left\{ Q_{X}:\:\sum_{x\in{\cal X}}Q_{X}(x)f_{i}(x)=\alpha_{i},1\leq i\leq K\right\} $
for some given functions $\{f_{i}\}_{i=1}^{K}$ and constants $\{\alpha_{i}\}_{i=1}^{K}$. %
}\emph{. }In \cite[Theorem  5.1]{csiszar2004information}, the convergence
of an iterative algorithm for a minimization problem of the form 
\begin{equation}
\min_{\breve{Q}\in{\cal G}}D(\breve{Q}||Q),\label{eq: I-projection linear familiy}
\end{equation}
where ${\cal G}$ is the intersection of a finite number of linear
families, was proved. The minimizer of \eqref{eq: I-projection linear familiy}
is called the I-projection of $Q$ onto ${\cal G}$ and denoted $\breve{Q}^{*}$.
The algorithm is called \emph{iterative scaling} and works as follows:
First $\breve{Q}^{(0)}=Q$ is initialized. Then, $\breve{Q}^{(1)}$
is the I-projection of $\breve{Q}^{(0)}$ onto ${\cal G}_{1}$, $\breve{Q}^{(2)}$
is the I-projection of $\breve{Q}^{(1)}$ onto ${\cal G}_{2}$, and
so on, where for $n>L$, $\breve{Q}^{(n)}$ is the I-projection of
$\breve{Q}^{(n-1)}$ onto ${\cal G}_{n\mod L}$. Such a procedure
converges to $\breve{Q}^{*}$. 

Thus, to use the iterative scaling algorithm for the case at hand,
we initialize $\breve{Q}_{X\tilde{X}}^{(0)}(x,\tilde{x})=Q_{X}(x)Q_{X}(\tilde{x})$
for all $(x,\tilde{x})\in{\cal X}\times{\cal X}$, and then we need
to find for any given PMF $\tilde{Q}_{X\tilde{X}}$ the I-projections
\[
\min_{\breve{Q}_{X\tilde{X}}\in{\cal {\cal G}}_{i}}D(\breve{Q}_{X\tilde{X}}||\tilde{Q}_{X\tilde{X}})
\]
for $i=1,2,3$. In what follows we will perform the I-projection on
${\cal G}_{1}\cap{\cal G}_{2}$ jointly, and then on ${\cal G}_{3}$.
In this case, $\tilde{Q}_{X\tilde{X}}$ is of the form $Q_{X}\times\tilde{Q}_{\tilde{X}|X}$
for all iterations. 

First, for ${\cal G}_{1}\cap{\cal G}_{2}$, we need to solve 
\[
\min_{\breve{Q}_{\tilde{X}|X}:B(Q_{X}\times\breve{Q}_{\tilde{X}|X})+D(Q_{X}||P_{X})=\EE}D(Q_{X}\times\breve{Q}_{\tilde{X}|X}||Q_{X}\times\tilde{Q}_{\tilde{X}|X}).
\]
Ignoring positivity constraints for the moment, we define the Lagrangian
\begin{eqnarray}
L(\breve{Q}_{\tilde{X}|X},\lambda,\mu_{x}) & = & \sum_{x\in{\cal X},\tilde{x}\in{\cal X}}Q_{X}(x)\breve{Q}_{\tilde{X}|X}(\tilde{x}|x)\log\frac{\breve{Q}_{\tilde{X}|X}(\tilde{x}|x)}{\tilde{Q}_{\tilde{X}|X}(x,\tilde{x})}\nonumber \\
 &  & +\lambda\cdot\sum_{x\in{\cal X},\tilde{x}\in{\cal X}}Q_{X}(x)\breve{Q}_{\tilde{X}|X}(\tilde{x}|x)d_{P_{Y|X}}(x,\tilde{x})+\sum_{x\in{\cal X}}\mu_{x}\sum_{\tilde{x}\in{\cal X}}\breve{Q}_{\tilde{X}|X}(\tilde{x}|x)
\end{eqnarray}
and $\lambda,\mu_{x}\in\mathbb{R}$ for $x\in{\cal X}$. Differentiating
w.r.t. some $\breve{Q}_{\tilde{X}|X}(\tilde{x}'|x')$ for $x',\tilde{x}'\in{\cal X}$
\[
\frac{\partial L}{\partial\breve{Q}_{\tilde{X}|X}(\tilde{x}'|x')}=Q_{X}(x')\left(\log\frac{\breve{Q}_{\tilde{X}|X}(\tilde{x}'|x')}{\tilde{Q}_{\tilde{X}|X}(\tilde{x}'|x')}+1\right)+\lambda\cdot Q_{X}(x')d_{P_{Y|X}}(x',\tilde{x}')+\mu_{x'}
\]
and equating to zero we get 
\[
Q_{X}(x')\cdot\log\frac{\breve{Q}_{\tilde{X}|X}(\tilde{x}'|x')\cdot\exp\left[\lambda\cdot d_{P_{Y|X}}(x',\tilde{x}')\right]}{\tilde{Q}_{\tilde{X}|X}(\tilde{x}'|x')}+\mu'_{x'}=0
\]
where $\mu'_{x'}=\mu_{x'}+1$. Thus, the argument of the logarithm
must not depend on $x$, and this implies that for any $x\in{\cal X}$
such that $Q_{X}(x)\neq0$ we must have
\[
\breve{Q}_{\tilde{X}|X}^{*}(\tilde{x}|x)=\psi_{x}\tilde{Q}_{\tilde{X}|X}(\tilde{x}|x)\exp\left[-\lambda\cdot d_{P_{Y|X}}(x,\tilde{x})\right]
\]
where $\psi_{x}$ is a normalization constant, such that $\sum_{\tilde{x}\in{\cal X}}\breve{Q}_{\tilde{X}|X}^{*}(\tilde{x}|x)=1$,
for any $x\in\supp(Q_{X})$, namely $\breve{Q}_{\tilde{X}|X}^{*}=\mathbb{M}_{\s[B]}(\tilde{Q}_{\tilde{X}|X},P_{Y|X},\lambda)$.
The value of $\breve{Q}_{\tilde{X}|X}^{*}(\tilde{x}|x)$ for $x\in{\cal X}$
with $Q_{X}(x)=0$ is immaterial as it does not affect the optimal
value of the objective function. Also, it is evident that the solution
$\breve{Q}_{\tilde{X}|X}^{*}$ is indeed positive. Finally, we need
to find $\lambda\in\mathbb{R}$ such that the constraint $B(Q_{X}\times\breve{Q}_{\tilde{X}|X})+D(Q_{X}||P_{X})=\EE$
is satisfied, namely 
\begin{eqnarray}
\EE+D(Q_{X}||P_{X}) & = & \sum_{x\in{\cal X},\tilde{x}\in{\cal X}}Q_{X}(x)\breve{Q}_{\tilde{X}|X}^{*}(\tilde{x}|x)d_{P_{Y|X}}(x,\tilde{x})\\
 & = & \sum_{x\in{\cal X},\tilde{x}\in{\cal X}}\psi_{x}\tilde{Q}_{\tilde{X}|X}(\tilde{x}|x)\exp(-\lambda\cdot d_{P_{Y|X}}(x,\tilde{x}))d_{P_{Y|X}}(x,\tilde{x}).
\end{eqnarray}
Second, the linear family ${\cal G}_{2}$ induces a simple constraint
on the $\tilde{X}$-marginal of $\breve{Q}_{X\tilde{X}}$. In this
case, the \emph{lumping property} \cite[Lemma  4.1 and Section  5.1]{csiszar2004information}
implies that the I-projection onto ${\cal G}_{2}$ is given by $\breve{Q}_{X\tilde{X}}^{*}=\mathbb{M}_{\s[l]}(\tilde{Q}_{X\tilde{X}})$
, which evidently satisfies $\breve{Q}_{X\tilde{X}}^{*}\in{\cal G}_{2}$.

It is easily seen that Algorithm \ref{alg:Iterative scaling algorithm value ex}
indeed implements the procedure described in this proof.
\end{IEEEproof}

\section{\label{sec:The-Fixed-Composition}}

\begin{figure}
\begin{centering}
\includegraphics[scale=0.5]{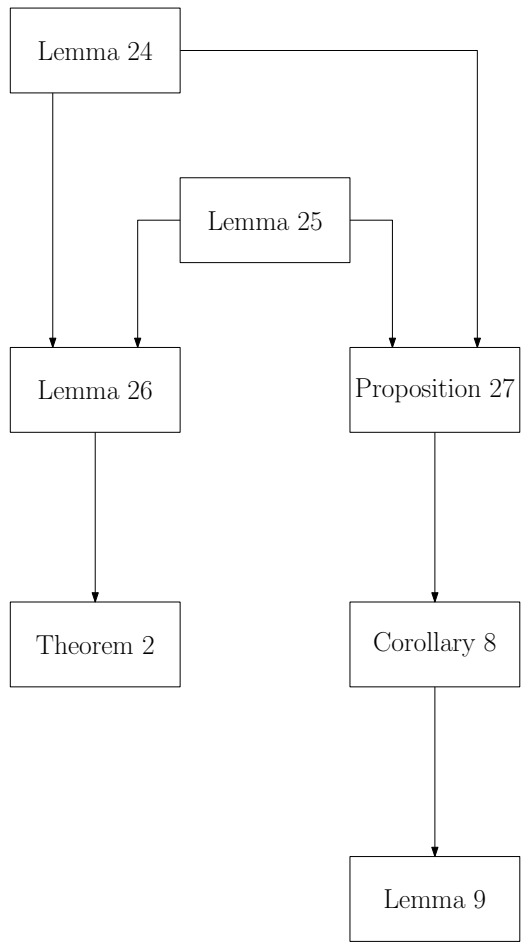}
\par\end{centering}

\protect\caption{Dependency graph of various proofs. \label{fig:Dependency-graph}}

\end{figure}

In this appendix, we will focus only on codes for the channel $W$.
As mentioned in subsection \ref{sub:Channel-Coding defs}, $\underline{E}_{e}^{*}(\RR,Q{}_{X},W)$
and $\overline{E}_{e}^{*}(\RR,Q{}_{X},W)$ are not fully known. Nonetheless,
some of their general properties can be obtained, and these are useful
for SW codes. Notice that in channel coding, the type of the fixed
composition code is under the control of the code designer. Indeed,
according to the definition of the infimum fixed-composition reliability
function, to achieve $\underline{E}_{e}^{*}(\RR,Q{}_{X},W)$, one
needs to find a sequence of types $Q_{X}^{(n)}\to Q_{X}$ with required
error probability, but it is not required that $Q_{X}^{(n)}=Q_{X}$
for all $n$. In Theorem \ref{thm: Error exponent bounds, single type},
we transform channel codes into SW codes, and conditioning on the
event that the source block belongs to the type $Q_{X}$, we would
like to achieve conditional error exponent of $\underline{E}_{e}^{*}(\RR,Q{}_{X},W)$
for the SW code. Thus, for SW coding, the type is determined by the
source, and there is no flexibility to choose a `nearby' type, since
the type of the source block is not under the control of the code
designer. The next three lemmas address this issue. Lemma \ref{lem: type modification}
shows that if two types are close, then by concatenating a relatively
short vector (of length linear in the block length, with small coefficient)
one can modify a vector from the first type to \emph{exactly} the
second type. Lemma \ref{lem: blocklength fillling Lemma } shows that
if a sequence of channel codes exists with good error probability
for block lengths which are not too far apart, then there is also
a sequence of channel codes for \emph{all }block lengths sufficiently
large, with essentially the same performance. Lemma \ref{lem: exact type codebook}
states that the infimum reliability function can be achieved with
codes of type $Q_{X}$\emph{ exactly}, when the block length is sufficiently
large. Finally, Proposition \ref{prop: property of channel reliability}
discusses the monotonicity and continuity of $\underline{E}_{e}^{*}(\RR,Q_{X},W)$
and $\overline{E}_{e}^{*}(\RR,Q_{X},W)$, which is of importance since
we will be interested in the inverses of these functions. We first
formulate all the Lemmas and the Proposition mentioned above, and
only afterwards provide their proofs. To assist the reader, Figure
\ref{fig:Dependency-graph} displays the connections between the proofs.
In this graph, the proof of any given result, depends on the results
displayed above it. For example, the proof of Lemma 26 requires the
result of Lemma 24. 

\begin{lem}
\label{lem: type modification}Assume that $Q\in{\cal Q}({\cal X}),\overline{Q}\in\interior{\cal Q}({\cal X})$
and $||Q-\overline{Q}||=\epsilon$. Also, let $Q^{(n)}\in{\cal P}_{n}({\cal X})$
be such that $Q^{(n)}\to Q$. Then, there exists $\mathbf{z}^{(n)}\in{\cal X}^{l_{n}}$
such that if $\mathbf{x}\in{\cal T}_{n}(Q^{(n)})$ then $\hat{Q}_{(\mathbf{x},\mathbf{z}^{(n)})}\to\overline{Q}$
as $n\to\infty$, and $\lim_{\epsilon\downarrow0}\lim_{n\to\infty}\frac{l_{n}}{n}=0$. 

Moreover, assume, in addition, that $\overline{Q}\in{\cal P}({\cal X})$.
Then, there exists $\mathbf{z}^{(n)}\in{\cal X}^{l_{n}}$ and $\overline{\epsilon}>0$,
such that for any $0<\epsilon<\overline{\epsilon}$ if $\mathbf{x}\in{\cal T}_{n}(Q^{(n)})$
then $\hat{Q}_{(\mathbf{x},\mathbf{z}^{(n)})}=\overline{Q}$ for $n$
sufficiently large, and $\lim_{\epsilon\downarrow0}\lim_{n\to\infty}\frac{l_{n}}{n}=0$. 
\end{lem}
The lengths $n+l_{n}$ in Lemma \ref{lem: type modification} need
not be increasing. However, since $n+l_{n}\to\infty$ as $n\to\infty$,
a strictly increasing sub-sequence may be extracted with the desired
property. So, henceforth we will assume that $n+l_{n}$ is an increasing
sequence. 
\begin{lem}
\label{lem: blocklength fillling Lemma }Let ${\cal C}$ be a sequence
of fixed composition codes ${\cal C}_{n_{k}}\subseteq{\cal T}_{n_{k}}(Q_{X}^{(k)})$
such that $Q_{X}^{(k)}\to Q_{X}$, and 
\[
\liminf_{k\to\infty}\frac{1}{n_{k}}\log|{\cal C}_{n_{k}}|\geq\RR,
\]
\[
\liminf_{k\to\infty}-\frac{1}{n_{k}}\log p_{e}({\cal C}_{n_{k}})=\mathsf{E}_{e}.
\]
Then, if $\limsup_{n\to\infty}\frac{(n_{k+1}-n_{k})}{n_{k}}=\delta$
there exists $\epsilon>0$ and a sequence ${\cal C}'$ of fixed composition
codes ${\cal C}'_{m}\subseteq{\cal T}_{m}(Q_{X}^{(m)})$ such that
$Q_{X}^{(m)}\to Q_{X}$, and 
\[
\liminf_{m\to\infty}\frac{1}{m}\log|{\cal C}'_{m}|\geq\RR-\epsilon
\]
\[
{\cal E}_{c}^{-}({\cal C}')\geq\mathsf{E}_{e}-\epsilon,
\]
where $\lim_{\delta\downarrow0}\epsilon=0$.
\end{lem}

\begin{lem}
\label{lem: exact type codebook}Let $Q_{X}\in{\cal P}({\cal X})\cap\interior{\cal Q}({\cal X})$,
and let $n_{0}\in\mathbb{N}$ be the minimal block length such that
$Q{}_{X}\in{\cal P}_{n_{0}}({\cal X})$. Let ${\cal C}$ be a sequence
of fixed composition codes ${\cal {\cal C}}_{n}\subseteq{\cal T}_{n}(Q_{X}^{(n)})$
such that $Q_{X}^{(n)}\to Q_{X}$, and $\liminf_{n\to\infty}\frac{1}{n}\log|{\cal C}_{n}|\geq\RR$.
Then, for any $\epsilon>0$, there exists a sequence of fixed composition
channel codes ${\cal C}'_{m}\subseteq{\cal T}_{m\cdot n_{0}}(Q_{X})$
such that $\liminf_{m\to\infty}\frac{1}{mn_{0}}\log|{\cal C}'_{m}|\geq\RR-\epsilon$,
and $\liminf_{m\to\infty}-\frac{1}{mn_{0}}\log p_{e}({\cal C}'_{m})\geq{\cal E}_{c}^{-}({\cal C})-\epsilon$.
\end{lem}

\begin{prop}
\label{prop: property of channel reliability}For a given channel
$W$, the functions $\underline{E}_{e}^{*}(\RR,Q_{X},W)$ and $\overline{E}_{e}^{*}(\RR,Q_{X},W)$
are:
\begin{enumerate}
\item Strictly decreasing functions of $\RR$ in the interval $(C_{0}(Q_{X},W),I(Q_{X}\times W))$
. 
\item Continuous functions of $(\RR,Q_{X})$ for $\RR\in(C_{0}(Q_{X},W),I(Q_{X}\times W))$. 
\end{enumerate}
\end{prop}
\begin{IEEEproof}[Proof of Lemma \ref{lem: type modification}]
We will prove this fact by induction over the alphabet size $|{\cal X}|$,
where without loss of generality (w.l.o.g.) we denote ${\cal X}=\{0,1,\ldots,|{\cal X}|-1\}$.
If $Q=\overline{Q}$ then the Lemma is trivial, thus we assume $Q\neq\overline{Q}$
and so there exists a letter $x^{*}$, such that $\overline{Q}(x^{*})>Q(x^{*}).$
W.l.o.g., we assume $x^{*}=0$. Then, for any sufficiently large $n$,
we also get $\overline{Q}(0)>Q^{(n)}(0)$.

For ${\cal X}=\{0,1\}$ letting 
\begin{alignat}{1}
l_{n} & =\left\lceil n\cdot\frac{\overline{Q}(0)-Q{}^{(n)}(0)}{1-\overline{Q}(0)}\right\rceil \label{eq: length of fixed vector alphabet of 2}\\
 & =n\cdot\frac{\overline{Q}(0)-Q{}^{(n)}(0)}{1-\overline{Q}(0)}+\alpha_{n}
\end{alignat}
for some $0\leq\alpha_{n}<1$, and choosing $\mathbf{z}^{(n)}=\mathbf{0}\in{\cal X}^{l_{n}}$,
we get
\begin{alignat}{1}
\hat{Q}_{(\mathbf{x},\mathbf{z}^{(n)})}(0) & =\frac{n\cdot Q{}^{(n)}(0)+l_{n}}{n+l_{n}}\label{eq: resulting type after fixed vector alphabet of 2}\\
 & =\frac{Q{}^{(n)}(0)+\nicefrac{\left(\overline{Q}(0)-Q{}^{(n)}(0)\right)}{\left(1-\overline{Q}(0)\right)}+\nicefrac{\alpha_{n}}{n}}{1+\nicefrac{\left(\overline{Q}(0)-Q{}^{(n)}(0)\right)}{\left(1-\overline{Q}(0)\right)}+\nicefrac{\alpha_{n}}{n}}\\
 & =\frac{\overline{Q}(0)(1-Q{}^{(n)}(0))+\left(\nicefrac{\alpha_{n}}{n}\right)(1-\overline{Q}(0))}{1-Q{}^{(n)}(0)+\left(\nicefrac{\alpha_{n}}{n}\right)(1-\overline{Q}(0))}\label{eq: resulting type after fixed vector alphabet of 2 second}\\
 & \to\overline{Q}(0)
\end{alignat}
as $n\to\infty$, and clearly also $\hat{Q}_{(\mathbf{x},\mathbf{z}^{(n)})}(1)\to\overline{Q}(1)$.
In addition, for all sufficiently large $n$ and any $\epsilon'>\epsilon$
\begin{alignat}{1}
l_{n} & =n\cdot\frac{\overline{Q}(0)-Q{}^{(n)}(0)}{1-\overline{Q}(0)}+\alpha_{n}\\
 & \leq n\cdot\frac{|\overline{Q}(0)-Q{}^{(n)}(0)|}{1-\overline{Q}(0)}+\alpha_{n}\\
 & \leq n\cdot\frac{\epsilon'/2}{1-\overline{Q}(0)}+\alpha_{n}\label{eq: resulting length of fixed vector alphabet of 2}
\end{alignat}
since $\epsilon=2\left[\overline{Q}(0)-Q(0)\right]$. So, $\lim_{\epsilon\downarrow0}\lim_{n\to\infty}\frac{l_{n}}{n}=0$
and the statement of the lemma is proved for $|{\cal X}|=2$. Now,
assume that the statement of the lemma holds for $|{\cal X}|\geq2$
and consider an alphabet $\overline{{\cal X}}=\{0,1,\ldots,|{\cal X}|\}$
of size $|{\cal X}|+1$. For this case too, we may assume w.l.o.g.
that $\overline{Q}(0)\geq Q(0)$, and for every sufficiently large
$n$, $\overline{Q}(0)\geq Q^{(n)}(0)$. Now, let ${\cal X}_{1}=\{1,\ldots,|{\cal X}|+1\}$,
and consider the distributions $\tilde{Q},\grave{Q}\in\interior{\cal Q}({\cal X}_{1})$
given for all $x\in{\cal X}_{1}$ by
\begin{equation}
\tilde{Q}(x)\teq\frac{Q(x)}{1-Q(0)},
\end{equation}
\begin{equation}
\tilde{Q}^{(n)}(x)\teq\frac{Q^{(n)}(x)}{1-Q^{(n)}(0)},
\end{equation}
and 
\begin{equation}
\grave{Q}(x)\teq\frac{\overline{Q}(x)}{1-\overline{Q}(0)},\label{eq: conditional no '0' of target type}
\end{equation}
namely, the conditional distribution of $X$ given that $X\neq0$.
We then also have $||\tilde{Q}-\grave{Q}||=\epsilon'$ where $\epsilon'\downarrow0$
as $\epsilon\downarrow0$. For any $\mathbf{x}$, let $\chi_{0}(\mathbf{x})$
be the vector obtained by \emph{deleting} all the $'0'$ components
of $\mathbf{x}$, e.g. $\chi_{0}([1,0,2])=[1,2]$. From the induction
assumption, since $|{\cal X}_{1}|=|{\cal X}|$, there exists a sequence
of vectors $\mathbf{z}_{1}^{(n)}\in{\cal X}_{1}^{m_{n}}$ such that
for any given $\mathbf{x}\in{\cal T}_{n}(Q^{(n)})$, $\hat{Q}_{\chi_{0}\left((\mathbf{x},\mathbf{z}_{1}^{(n)})\right)}(x)\to\grave{Q}(x)$
as $n\to\infty$ for all $x\in{\cal X}_{1}$, and $\lim_{\epsilon'\downarrow0}\lim_{n\to\infty}\frac{m_{n}}{n(1-Q^{(n)}(0))}=0$.
Thus, we have found a sequence of vectors such that when concatenated
to $\mathbf{x}$, the relative frequency of each $x\in{\cal X}_{1}$
in the concatenated vector tends to $\grave{Q}(x)$. It remains to
assure that the relative frequency of the letter $0\in{\cal X}_{1}$
in the final vector will also tend to $\overline{Q}(0)$. Now, 
\begin{equation}
\overline{Q}(0)>Q^{(n)}(0)>\frac{n}{n+m_{n}}Q^{(n)}(0)=\hat{Q}_{(\mathbf{x},\mathbf{z}_{1}^{(n)})}(0)
\end{equation}
for sufficiently large $n$, and then for sufficiently small $\epsilon>0$,
we have $\overline{Q}(0)-\hat{Q}_{(\mathbf{x},\mathbf{z}_{1}^{(n)})}(0)<2\epsilon$.
Thus, just as for the case of $|{\cal X}|=2$, one can concatenate
another sequence of vectors $\mathbf{z}_{0}^{(n)}=\mathbf{0}\in{\cal X}^{m'_{n}}$
such that for any $\mathbf{x}\in{\cal T}_{n}(Q^{(n)})$, the relative
frequency of $x=0$ in $(\mathbf{x},\mathbf{z}_{1}^{(n)},\mathbf{z}_{0}^{(n)})$
tends to $\overline{Q}{}_{X}(0)$ as $n\to\infty$, and $\lim_{\epsilon\downarrow0}\lim_{n\to\infty}\frac{m'_{n}}{n}=0$.
Consequently, $\hat{Q}_{(\mathbf{x},\mathbf{z}_{1}^{(n)},\mathbf{z}_{0}^{(n)})}\to\overline{Q}$
as $n\to\infty$. The total length of the concatenated fixed sequence
$l_{n}=m_{n}+m'_{n}$ satisfies $\lim_{\epsilon\downarrow0}\lim_{n\to\infty}\frac{l_{n}}{n}=0$.

Next, we consider the case in which $\overline{Q}\in{\cal P}_{n_{0}}({\cal X})$,
where $n_{0}>1$ is the minimal block length satisfying this property.
We will use the following two simple facts:
\begin{itemize}
\item \emph{Fact 1:} If $Q,\overline{Q}\in{\cal P}_{n}({\cal X})$ and $||Q-\overline{Q}||<\frac{2}{n}$
then $Q=\overline{Q}$. 
\item \emph{Fact 2:} Let $\mathbf{x}\in\{0,1\}^{n}$ and assume that for
some $\overline{Q}$, we have $||\hat{Q}_{\mathbf{x}}-\overline{Q}||\leq\frac{2}{n}$.
Also, denote $\hat{Q}_{\mathbf{x}}(0)=\frac{k}{n}$ and assume w.l.o.g.
that $\hat{Q}_{\mathbf{x}}(0)=\frac{k}{n}<\overline{Q}(0)$, where
consequently $k<n$. Then $||\hat{Q}_{(\mathbf{x},0)}-\overline{Q}||\leq\frac{2}{n+1}$.
To see this, notice that 
\begin{equation}
k+1-(n+1)\overline{Q}(0)<k+1-(n+1)\frac{k}{n}<1\label{eq: ineq with 1}
\end{equation}
and since $||\hat{Q}_{\mathbf{x}}-\overline{Q}||\leq\frac{2}{n}$
then 
\[
\overline{Q}(0)-\frac{k}{n}\leq\frac{1}{n}
\]
and also 
\begin{equation}
k+1-(n+1)\overline{Q}(0)\geq-\frac{k+1}{n}\geq-1.\label{eq: ineq with -1}
\end{equation}
Thus, 
\begin{alignat}{1}
||\hat{Q}_{(\mathbf{x},0)}-\overline{Q}|| & =2\cdot\left|\frac{k+1}{n+1}-\overline{Q}(0)\right|\\
 & =\frac{2}{n+1}\cdot\left|k+1-(n+1)\overline{Q}(0)\right|\\
 & \trre[\leq,a]\frac{2}{n+1}
\end{alignat}
where $(a)$ follows from \eqref{eq: ineq with 1} and \eqref{eq: ineq with -1}. 
\end{itemize}
Now, for $|{\cal X}|=2$, recall the construction before \eqref{eq: length of fixed vector alphabet of 2}
and onward, and let the resulting size of the vector $(\mathbf{x},\mathbf{z}^{(n)})$
be $n+m_{n}$. Denote 
\[
\gamma_{n}\teq\frac{1-\overline{Q}(0)}{1-Q^{(n)}(0)}.
\]
From the assumption $\overline{Q}(0)>Q(0)$, we get $\gamma=\lim_{n\to\infty}\gamma_{n}<1$.
Let $\gamma'$ satisfy $\gamma<\gamma'<1$. Then, on the one hand
for $n$ sufficiently large, we have from \eqref{eq: resulting type after fixed vector alphabet of 2 second}
\begin{alignat}{1}
\hat{Q}_{(\mathbf{x},\mathbf{z}^{(n)})}(0) & =\left(\overline{Q}(0)+\frac{\alpha_{n}\gamma_{n}}{n}\right)\cdot\frac{1}{1+\left(\nicefrac{\alpha_{n}\gamma_{n}}{n}\right)}\\
 & \leq\overline{Q}(0)+\frac{\gamma_{n}\alpha_{n}}{n}\\
 & <\overline{Q}(0)+\frac{\gamma'}{n}
\end{alignat}
and on the other hand, 
\begin{alignat}{1}
\hat{Q}_{(\mathbf{x},\mathbf{z}^{(n)})}(0) & =\left(\overline{Q}(0)+\frac{\alpha_{n}\gamma_{n}}{n}\right)\cdot\frac{1}{1+\left(\nicefrac{\alpha_{n}\gamma_{n}}{n}\right)}\\
 & \geq\left(\overline{Q}(0)+\frac{\alpha_{n}\gamma_{n}}{n}\right)\cdot\left(1-\gamma_{n}\frac{\alpha_{n}}{n}\right)\\
 & =\overline{Q}(0)+(1-\overline{Q}(0))\cdot\gamma_{n}\frac{\alpha_{n}}{n}-\left[\frac{\gamma_{n}\alpha_{n}}{n}\right]^{2}\\
 & \geq\overline{Q}(0)-\left[\frac{\gamma_{n}\alpha_{n}}{n}\right]^{2}\\
 & >\overline{Q}(0)-\frac{\gamma'}{n}.
\end{alignat}
Thus, for sufficiently large $n$, 
\begin{equation}
||\hat{Q}_{(\mathbf{x},\mathbf{z}^{(n)})}-\overline{Q}||<\frac{2\gamma'}{n}.\label{eq: types are very close with length n}
\end{equation}
Moreover, from \eqref{eq: resulting length of fixed vector alphabet of 2},
for sufficiently large $n$, we have $\frac{m_{n}}{n}\leq\delta'$
where $\delta'>\delta$ and 
\[
\delta=\frac{\epsilon/2}{1-\overline{Q}(0)}.
\]
Then, since 
\begin{alignat}{1}
\gamma\cdot(1+\delta) & =\frac{1-\overline{Q}(0)}{1-Q(0)}\cdot\left(1+\frac{\epsilon/2}{1-\overline{Q}(0)}\right)\\
 & =\frac{1}{1-Q(0)}\cdot\left(1-\overline{Q}(0)+\epsilon/2\right)\\
 & =1
\end{alignat}
for sufficiently large $n$, \eqref{eq: types are very close with length n}
implies 
\begin{alignat}{1}
||\hat{Q}_{(\mathbf{x},\mathbf{z}^{(n)})}-\overline{Q}|| & <\frac{2\gamma'}{n}\\
 & <\frac{2\gamma'}{n(1+\delta)\gamma}\\
 & \leq\frac{2\gamma'}{(n+m_{n})\gamma}.
\end{alignat}
Since this is true for any $\gamma'>\gamma$ then 
\begin{eqnarray}
||\hat{Q}_{(\mathbf{x},\mathbf{z}^{(n)})}-\overline{Q}|| & \leq & \frac{2}{n+m_{n}}.\label{eq: after one concatnation type is close enough}
\end{eqnarray}

Let $k^{*}(n)\teq\left\lceil \frac{n+m_{n}}{n_{0}}\right\rceil $
and $m'_{n}=n_{0}\cdot k^{*}(n)-n$ and construct $\overline{\mathbf{z}}^{(n)}\in{\cal X}^{m'_{n}}$
iteratively in the following way: 
\begin{enumerate}
\item Initialize $\overline{\mathbf{z}}^{(n)}$ with the empty string.
\item For $i=1$ to $i=m'_{n}$: If $\hat{Q}_{(\mathbf{x},\mathbf{z}^{(n)},\overline{\mathbf{z}}^{(n)})}(0)<\overline{Q}(0)$
then set $\overline{\mathbf{z}}^{(n)}\leftarrow(\overline{\mathbf{z}}^{(n)},0)$
and otherwise set $\overline{\mathbf{z}}^{(n)}\leftarrow(\overline{\mathbf{z}}^{(n)},1)$. 
\end{enumerate}
From the second general fact above and \eqref{eq: after one concatnation type is close enough}
we have 
\begin{equation}
||\hat{Q}_{(\mathbf{x},\mathbf{z}^{(n)},\overline{\mathbf{z}}^{(n)})}-\overline{Q}||\leq\frac{2}{n+m_{n}+m'_{n}}.\label{eq: after two concatnation type is very close}
\end{equation}
If the inequality in \eqref{eq: after two concatnation type is very close}
is strict, then since ($n+m_{n}+m'_{n})\mod n_{0}=0$, clearly $\hat{Q}_{(\mathbf{x},\mathbf{z}^{(n)},\overline{\mathbf{z}}^{(n)})}=\overline{Q}$
and by setting $l_{n}=m_{n}+m'_{n}$ and the fact that $m'_{n}<n_{0}$,
we get the desired result. Otherwise, if equality is obtained in \eqref{eq: after two concatnation type is very close}
then let $\overline{\overline{\mathbf{z}}}^{(n)}\in{\cal X}^{n_{0}}$
such that $\overline{\overline{\mathbf{z}}}^{(n)}\in{\cal T}_{n_{0}}(\overline{\overline{Q}})$
where 
\[
\overline{\overline{Q}}(0)=\overline{Q}(0)+\frac{1}{n_{0}}
\]
if $\hat{Q}_{(\mathbf{x},\mathbf{z}^{(n)},\overline{\mathbf{z}}^{(n)})}(0)<\overline{Q}(0)$
and 
\[
\overline{\overline{Q}}(0)=\overline{Q}(0)-\frac{1}{n_{0}}
\]
if $\hat{Q}_{(\mathbf{x},\mathbf{z}^{(n)},\overline{\mathbf{z}}^{(n)})}(0)>\overline{Q}(0)$.
It is easily verified that $\hat{Q}_{(\mathbf{x},\mathbf{z}^{(n)},\overline{\mathbf{z}}^{(n)},\overline{\overline{\mathbf{z}}}^{(n)})}=\overline{Q}$.
By setting $l_{n}=m_{n}+m'_{n}+n_{0}$ and the fact that $m'_{n}<n_{0}$
we get the desired result. 

Finally, for $|{\cal X}|>2$, notice that from \eqref{eq: conditional no '0' of target type},
if $\overline{Q}\in{\cal P}({\cal X})$ then also $\grave{Q}(x)\in{\cal P}({\cal X})$.
Then, the same proof by induction can be used, just as for the first
statement of the lemma.
\end{IEEEproof}

\begin{IEEEproof}[Proof of Lemma \ref{lem: blocklength fillling Lemma }]
For any given block length $m$, let $k(m)$ be such that $n_{k(m)}<m<n_{k(m)+1}$,
and thus for any $\delta'>\delta$, and sufficiently large $m$ we
have $m-n_{k(m)}\leq\delta'n_{k(m)}$. Next, find the vector $\mathbf{w}{}^{(m)}\in{\cal X}^{m-n_{k(m)}}$
with empirical distribution closet possible to $Q{}_{X}$ (in the
variation distance norm). Denote this empirical distribution by $\grave{Q}{}^{(m)}$.
The code ${\cal C}'_{m}$ will be constructed by concatenating the
fixed vector $\mathbf{w}^{(m)}$ to the codewords $\mathbf{x}\in{\cal C}_{n_{k(m)}}$.
The rate of this code satisfies
\begin{eqnarray}
\liminf_{m\to\infty}\frac{\log|{\cal C}'_{m}|}{m} & = & \liminf_{m\to\infty}\frac{n_{k(m)}}{m}\cdot\frac{\log|{\cal C}{}_{n_{k}}|}{n_{k(m)}}\\
 & \geq & \frac{\RR}{1+\delta'}\\
 & \geq & \RR-\epsilon'
\end{eqnarray}
where $\lim_{\delta'\downarrow0}\epsilon'=0$. Since the error probability
of ${\cal C}'_{m}$ is equal to the error probability of ${\cal C}_{n_{k}}$,
its error exponent is
\begin{eqnarray}
\liminf_{m\to\infty}-\frac{1}{m}\log p_{e}({\cal C}'_{m}) & = & \liminf_{m\to\infty}-\frac{n_{k(m)}}{m}\frac{1}{n_{k(m)}}\log p_{e}({\cal C}'_{m})\\
 & \geq & \frac{\mathsf{E}_{e}}{1+\delta'}\\
 & \geq & \mathsf{E}_{e}-\epsilon''
\end{eqnarray}
where $\lim_{\delta'\downarrow0}\epsilon''=0$. Choosing $\epsilon=\max\{\epsilon',\epsilon''\}$
we get $\lim_{\delta'\downarrow0}\epsilon=0$. To conclude, it only
remains to prove that if $Q'{}_{X}^{(m)}$ is the type of the codewords
of ${\cal C}'_{m}$ then $Q'{}_{X}^{(m)}\to Q{}_{X}$ as $m\to\infty$.
To show this, notice that since $\grave{Q}{}^{(m)}\in{\cal P}_{m-n_{k(m)}}({\cal X})$
is the closest type to $Q_{X}$ then%
\footnote{A non-optimal choice for $\grave{Q}{}^{(m)}$ is to round the first
$m-n_{k(m)}-1$ components of $Q_{X}$ to multiples of $\frac{1}{m-n_{k(m)}}$.
This results the variation distance in the bound.%
} 
\begin{equation}
||\grave{Q}{}^{(m)}-Q{}_{X}||\leq\frac{|{\cal X}|}{m-n_{k(m)}}\label{eq: the closet vector to a type minimal distance}
\end{equation}
and so
\[
\frac{m-n_{k(m)}}{m}\cdot||\grave{Q}{}^{(m)}-Q{}_{X}||\leq\frac{|{\cal X}|}{m}.
\]
Also 
\[
Q'{}_{X}^{(m)}=\frac{n_{k(m)}}{m}\cdot Q{}_{X}^{(k(m))}+\frac{m-n_{k(m)}}{m}\cdot\grave{Q}{}^{(m)}
\]
and by convexity of the variation distance (${\cal L}_{1}$ norm),
for $m$ sufficiently large 
\begin{eqnarray}
||Q'{}_{X}^{(m)}-Q{}_{X}|| & \leq & \frac{n_{k(m)}}{m}\cdot||Q{}_{X}^{(k(m))}-Q{}_{X}||+\frac{m-n_{k(m)}}{m}\cdot||\grave{Q}{}^{(m)}-Q{}_{X}||\\
 & \leq & \frac{n_{k(m)}}{m}\cdot||Q{}_{X}^{(k(m))}-Q{}_{X}||+\frac{|{\cal X}|}{m}\\
 & \leq & \frac{1}{1+\delta'}\cdot||Q{}_{X}^{(k(m))}-Q{}_{X}||+\frac{|{\cal X}|}{m}
\end{eqnarray}
and since $Q{}_{X}^{(k(m))}\to Q_{X}$ as $m\to\infty$ also $Q'{}_{X}^{(m)}\to Q{}_{X}$. 
\end{IEEEproof}

\begin{IEEEproof}[Proof of Lemma \ref{lem: exact type codebook}]
Let $\epsilon>0$ be given. By setting $\overline{Q}{}_{X}=Q_{X}$
in the second part of Lemma \ref{lem: type modification}, there exists
a sequence of vectors $\mathbf{z}^{(n)}\in{\cal X}^{l_{n}}$, such
that if $\mathbf{x}\in{\cal T}_{n}(Q_{X}^{(n)})$, the concatenation
$(\mathbf{x},\mathbf{z}^{(n)})\in{\cal T}_{n+l_{n}}(Q{}_{X})$, and
$\limsup_{n\to\infty}\frac{l_{n}}{n}=0$. Let us now construct a sequence
of channel codes ${\cal C}'$ were ${\cal C}'_{k_{n}}\subseteq{\cal T}_{k_{n}}(Q'{}_{X}^{(n)})$
is of length $k_{n}=l_{n}+n$, and such that $Q'{}_{X}^{(n)}=Q{}_{X}$
for sufficiently large $n$, by concatenating to the codewords $\mathbf{x}\in{\cal C}_{n}$
the fixed vector $\mathbf{z}^{(n)}$. Now, the rate of the sequence
of the new codes ${\cal C}'$ is 
\begin{alignat}{1}
\liminf_{n\to\infty}\frac{\log|{\cal C}'_{k_{n}}|}{k_{n}} & =\liminf_{n\to\infty}\frac{n}{k_{n}}\cdot\frac{\log|{\cal C}{}_{k_{n}}|}{n}\\
 & =\liminf_{n\to\infty}\frac{n}{n(1+\frac{l_{n}}{n})}\cdot\frac{\log|{\cal C}{}_{k_{n}}|}{n}\\
 & =\RR
\end{alignat}
and since the error probability of ${\cal C}'_{k_{n}}$ is equal to
the error probability of ${\cal C}_{n}$, its error exponent is 
\begin{alignat}{1}
{\cal E}_{c}^{-}({\cal C}') & =\liminf_{n\to\infty}-\frac{1}{k_{n}}\log p_{e}({\cal C}'_{k_{n}}).\\
 & =\liminf_{n\to\infty}-\frac{n}{k_{n}}\cdot\frac{1}{n}\log p_{e}({\cal C}{}_{n})\\
 & ={\cal E}_{c}^{-}({\cal C}).
\end{alignat}
Since for any $\mathbf{x}\in{\cal T}_{n}(Q_{X}^{(n)})$, we have $(\mathbf{x},\mathbf{z}^{(n)})\in{\cal T}_{k_{n}}(Q{}_{X})$,
$k_{n}\mod n_{0}=0$. Now, let us focus on the block length $m\cdot n_{0}$
for any $m$ sufficiently large. There must exist $n_{1}(m)$ and
$n_{2}(m)$ such that $k_{n_{1}(m)}\leq m\cdot n_{0}\leq k_{n_{2}(m)}$%
\footnote{The reason for not choosing $n_{2}(m)=n_{1}(m)+1$ is because the
lengths $k_{n}=l_{n}+n$, guaranteed by Lemma \ref{lem: type modification}
are not necessarily increasing.%
}. If $k_{n_{1}(m)}<m\cdot n_{0}<k_{n_{2}(m)}$ then as in Lemma \ref{lem: blocklength fillling Lemma },
one can concatenate a fixed vector $\overline{\mathbf{z}}^{(m)}\in{\cal X}^{m\cdot n_{0}-k_{n_{1}(m)}}$
to the codewords of ${\cal C}'_{k_{n}}$ such that $\overline{\mathbf{z}}^{(m)}\in{\cal T}_{m\cdot n_{0}-k_{n_{1}(m)}}(Q_{X})$,
and obtain a codebook ${\cal C}''_{m}\in{\cal T}_{m}(Q_{X}).$ Since
\[
\limsup_{n\to\infty}\frac{k_{n+1}-k_{n}}{k_{n}}=0
\]
then also 
\[
\limsup_{n\to\infty}\frac{k_{n_{2}(m)}-k_{n_{1}(m)}}{k_{n_{2}(m)}}=0
\]
the effect of $\overline{\mathbf{z}}^{(m)}$ on the rate and error
exponent is negligible, using similar arguments to to the ones used
in the proof of Lemma \ref{lem: blocklength fillling Lemma }.
\end{IEEEproof}

\begin{IEEEproof}[Proof of Proposition \ref{prop: property of channel reliability}]
Since for any $Q_{X},Q_{X}'\in{\cal Q}({\cal X})$ and $\RR,\RR'$
\begin{multline}
|\overline{E}_{e}^{*}(\RR,Q_{X},W)-\overline{E}_{e}^{*}(\RR',Q'_{X},W)|\leq|\overline{E}_{e}^{*}(\RR,Q_{X},W)-\overline{E}_{e}^{*}(\RR,Q'_{X},W)|\\
+|\overline{E}_{e}^{*}(\RR,Q'_{X},W)-\overline{E}_{e}^{*}(\RR',Q'_{X},W)|\label{eq: continuity in R,Q_X may be seperated}
\end{multline}
then continuity in $\RR$ and in $Q_{X}$ may be proved separately.
The same is true for $\underline{E}_{e}^{*}(\RR,Q_{X},W)$. Accordingly,
the proof is divided into two steps: In the first step we will prove
continuity and monotonicity in $\RR$, and the first property of the
proposition will be proved. Then, in the second step, we will prove
continuity in $Q'_{X}$, which combined with the first property and
\eqref{eq: continuity in R,Q_X may be seperated}, proves the second
property of the proposition. Notice that if $\RR\in(C_{0}(Q_{X},W),I(Q_{X}\times W))$
then for $Q'_{X}$ sufficiently close to $Q_{X}$ also we have $\RR\in(C_{0}(Q'_{X},W),I(Q'_{X}\times W))$
using the continuity of $C_{0}(Q{}_{X},W)$%
\footnote{The continuity of $C_{0}(Q{}_{X},W)$ in $Q_{X}$ can be rigorously
proved using the same techniques used in the rest of the proof.%
} and $I(Q{}_{X}\times W)$ in $Q_{X}$. 

\emph{\uline{Step I:}} The fact that $\overline{E}_{e}^{*}(\RR,Q_{X},W)$
is continuous and strictly decreasing in $\RR$ was established by
Haroutunian \cite{Haroutunian_paper,Haroutunian}. Since the proof
is based on the sphere packing bound, which is also valid for the
infimum error exponent \cite[Problem 10.7(b)]{csiszar2011information}
and the list decoding inequality \cite[Problem 10.30]{csiszar2011information},
\cite[Lemma 3.8.2]{viterbi2009principles} which is valid for any
given block length, then the same proof is also valid for $\underline{E}_{e}^{*}(\RR,Q_{X},W)$.

\emph{\uline{Step II:}} Let us begin with continuity of $\overline{E}_{e}^{*}(\RR,Q_{X},W)$
in ${\cal Q}({\cal X})$. Let $\epsilon>0$ be given such that for
any $Q'_{X}$ which satisfies $||Q'_{X}-Q_{X}||\leq\epsilon/2$ also
$\supp(Q_{X})\subseteq\supp(Q'_{X})$. By definition, for any given
$\delta_{1}\in[0,\RR]$ there exist a sequence of channel codes ${\cal C}$
where ${\cal C}_{n}\subseteq{\cal T}_{n}(Q_{X}^{(n)})$ such that
$Q_{X}^{(n)}\to Q_{X}$, and $\liminf_{n\to\infty}\frac{\log|{\cal C}_{n}|}{n}\geq\RR-\delta_{1}$,
as well as ${\cal E}_{c}^{+}({\cal C})\geq\overline{E}_{e}^{*}(\RR,Q_{X},W)-\delta_{1}$.
Also, for sufficiently large $n$ 
\[
||Q_{X}^{(n)}-Q'_{X}||\leq||Q_{X}^{(n)}-Q{}_{X}||+||Q_{X}-Q'{}_{X}||\leq\epsilon.
\]
Using Lemma \ref{lem: type modification} (for the alphabet $\supp(Q_{X})$),
one can find $\delta>0$ and a sequence of vectors $\mathbf{z}^{(n)}\in{\cal X}^{k_{n}-n}$
such that for all $\mathbf{x}\in{\cal T}_{n}(Q{}_{X}^{(n)})$, the
concatenation $(\mathbf{x},\mathbf{z}^{(n)})\in{\cal T}_{k_{n}}(Q'{}_{X}^{(n)})$
where $Q'{}_{X}^{(n)}\to Q'{}_{X}$ as $n\to\infty$, and $\limsup_{n\to\infty}\frac{l_{n}}{n}\leq\delta$.
Let us now construct a sequence of channel codes ${\cal C}'$ where
${\cal C}'_{k_{n}}\subseteq{\cal T}_{k_{n}}(Q'{}_{X}^{(n)})$, of
length $k_{n}=l_{n}+n$ such that $Q'{}_{X}^{(n)}\to Q'_{X}$, by
concatenating $\mathbf{z}^{(n)}$ to the codewords $\mathbf{x}\in{\cal C}_{n}$.
Now, the rate of the sequence of the new codes ${\cal C}'$ is 
\begin{alignat}{1}
\liminf_{n\to\infty}\frac{\log|{\cal C}'_{k_{n}}|}{k_{n}} & =\liminf_{n\to\infty}\frac{n}{k_{n}}\cdot\frac{\log|{\cal C}{}_{k_{n}}|}{n}\\
 & \geq\liminf_{n\to\infty}\frac{n}{(1+2\delta)n}\cdot\frac{\log|{\cal C}{}_{k_{n}}|}{n}\\
 & \geq\frac{\RR-\delta_{1}}{1+2\delta}\label{eq: proof channel reliability part III Rate is good enough}
\end{alignat}
and since the error probability of ${\cal C}'_{k_{n}}$ is at least
as small as the error probability of ${\cal C}_{n}$ then the error
exponent is 
\begin{alignat}{1}
{\cal E}_{c}^{+}({\cal C}') & =\limsup_{n\to\infty}-\frac{1}{k_{n}}\log p_{e}({\cal C}'_{k_{n}})\\
 & =\limsup_{n\to\infty}-\frac{n}{k_{n}}\cdot\frac{1}{n}\log p_{e}({\cal C}{}_{n})\\
 & \geq\frac{1}{(1+2\delta)}\limsup_{n\to\infty}-\frac{1}{n}\log p_{e}({\cal C}{}_{n})\\
 & =\frac{{\cal E}_{c}^{+}({\cal C})}{(1+2\delta)}\\
 & \geq\frac{1}{(1+2\delta)}\left[\overline{E}_{e}^{*}(\RR,Q_{X},W)-\delta_{1}\right].\label{eq: proof channel reliability part III exponent is good enough}
\end{alignat}
As $\delta_{1}>0$ may be chosen arbitrary small then by definition
\[
\overline{E}_{e}^{*}(\RR,Q'_{X},W)\geq\frac{1}{(1+2\delta)}\overline{E}_{e}^{*}\left(\frac{\RR}{1+2\delta},Q_{X},W\right).
\]
Since one can exchange the roles of $Q_{X}$ and $Q'_{X}$ in the
above construction then also
\[
\overline{E}_{e}^{*}(\RR,Q{}_{X},W)\geq\frac{1}{(1+2\delta)}\overline{E}_{e}^{*}\left(\frac{\RR}{1+2\delta},Q'_{X},W\right).
\]
and the continuity of $\overline{E}_{e}^{*}(\RR,Q{}_{X},W)$ immediately
follows from the continuity of $\overline{E}_{e}^{*}(\RR,Q{}_{X},W)$
in 
\[
\RR\in(C_{0}(Q_{X},W),I(Q_{X}\times W)),
\]
and the fact that, from Lemma \ref{lem: type modification}, $\lim_{\epsilon\downarrow0}\delta=0$. 

To prove the same property for $\underline{E}_{e}^{*}(\RR,Q{}_{X},W)$,
notice that the same construction can be used to modify a sequence
of codes ${\cal C}$ which achieve $\underline{E}_{e}^{*}(\RR,Q_{X},W)$,
into a new sequence of codes ${\cal C}'$, by concatenating a fixed
vector $\mathbf{z}^{(n)}\in{\cal X}^{k_{n}-n}$ with type that tends
to $Q'_{X}$ to the codewords of ${\cal C}_{n}$, and obtain error
probability close to that of ${\cal C}_{n}$. However, the resulting
codes only exist for lengths $k_{n}$, while in order to achieve $\underline{E}_{e}^{*}(\RR,Q'{}_{X},W)$,
a code should be constructed for any block length (due to the limit
infimum in its definition). Nonetheless, since
\[
\limsup_{n\to\infty}\frac{k_{n+1}-k_{n}}{k_{n}}\leq\limsup_{n\to\infty}\frac{k_{n+1}-n}{n}\leq\delta
\]
one can invoke Lemma \ref{lem: blocklength fillling Lemma } and construct
from ${\cal C}'_{k_{n}}$ a new sequence of codes ${\cal C}''_{m}$
for any block length $m$ which achieve $\underline{E}_{e}^{*}(\RR,Q'{}_{X},W)-\eta$
where $\lim_{\delta\to0}\eta=0$. The property is obtained by taking
$\epsilon\downarrow0$ which results $\delta\downarrow0$. 
\end{IEEEproof}

\section{\label{sec:Tightness-of-the-Random-Binning}}

In this appendix, we define the ensemble of type-dependent, variable-rate
random binning SW codes, and analyze its exact error exponent. While
in essence the proof techniques used for fixed-rate coding in \cite{csiszar1980towards}
can be generalized, we provide a somewhat simpler proof, which also
shows that the resulting expression is tight, and not just a lower
bound on the actual random binning exponent (a result analogous to
\cite{Gallager1973_tight} for the random channel coding error exponent).
The proof is based on the following lemma. The lemma is of importance
as it verifies the asymptotic tightness of the union bound, even for
the union of \emph{exponentially} many events.
\begin{lem}[Tightness of the union bound]
\label{lem: Union bound tightness}Let ${\cal A}_{1},{\cal A}_{2}\ldots,{\cal A}_{K}$
be pairwise independent events from a probability space. Then
\[
\P\left\{ \bigcup_{k=1}^{K}{\cal A}_{k}\right\} \geq\frac{1}{2}\cdot\min\left\{ 1,\sum_{k=1}^{K}\P\left({\cal A}_{k}\right)\right\} 
\]
\end{lem}
\begin{IEEEproof}
See \cite[Lemma A.2, pp. 109]{Shulman03}.
\end{IEEEproof}
A random ensemble of SW codes is defined by a random sequence of encoders-decoders
$(S_{n},\Sigma_{n})$ with probability $\P(S_{n}=s_{n},\Sigma_{n}=\sigma_{n})$,
and the average error probability over the random ensemble of codes
is defined as $\overline{p}_{e,n}=\E\left[p_{e}({\cal S}_{n})\right].$
The \emph{random-binning error exponent} is defined as
\begin{equation}
\bar{E}_{e}\triangleq\liminf_{n\to\infty}-\frac{1}{n}\log\overline{p}_{e,n}.\label{eq: error exponent definition}
\end{equation}

We analyze the ensemble performance of type-dependent, variable-rate
SW codes, which are defined as follows: 
\begin{itemize}
\item \emph{Codebook generation:} For a given rate function $\rho(Q_{X})$,
generate $e^{n\rho(Q_{X})}$ bins for every $Q_{X}\in{\cal P}_{n}({\cal X})$
and map each bin into a different binary string of length $n\cdot\rho(Q_{X})$
nats. Next, assign to each $\mathbf{x}\in{\cal X}^{n}$ a bin by independent
random selection with a uniform distribution over all the bins of
type class ${\cal T}_{n}(\hat{Q}_{\mathbf{x}})$. Then, assign an
index to each type $Q_{X}\in{\cal P}_{n}({\cal X})$. The above data
is revealed to both the encoder and the decoder off-line.
\item \emph{Encoding:} Upon observing $\mathbf{x}$, determine its type
class ${\cal T}_{n}(\hat{Q}_{\mathbf{x}})$. Send to the decoder its
type index, concatenated with its bin index (for the current type
class ${\cal T}_{n}(\hat{Q}_{\mathbf{x}})$).
\item \emph{Decoding:} First, recover the type class ${\cal T}_{n}(\hat{Q}_{\mathbf{x}})$
of $\mathbf{x}$. Then, we consider two options:

\begin{itemize}
\item Maximum likelihood (ML): Choose $\tilde{\mathbf{x}}\in s_{n}^{-1}(s_{n}(\mathbf{x}))$
that maximizes $P_{\mathbf{X}|\mathbf{Y}}(\tilde{\mathbf{x}}|\mathbf{y})$.
Since all $\tilde{\mathbf{x}}\in s_{n}^{-1}(s_{n}(\mathbf{x}))$ are
in the same type class, they have the same probability $P_{\mathbf{X}}(\tilde{\mathbf{x}})$,
so this decoding rule is equivalent to maximizing $P_{\mathbf{Y}|\mathbf{X}}(\mathbf{y}|\tilde{\mathbf{x}})$.
\item Minimum conditional entropy (MCE): Choose $\tilde{\mathbf{x}}\in s_{n}^{-1}(s_{n}(\mathbf{x}))$
that minimizes $H(\hat{Q}_{\tilde{\mathbf{x}}|\mathbf{y}}|\hat{Q}_{\mathbf{y}})$.
Since all $\tilde{\mathbf{x}}\in s_{n}^{-1}(s_{n}(\mathbf{x}))$ have
the same empirical entropy $H(\hat{Q}_{\mathbf{\tilde{x}}})$, so
this decoding rule is equivalent to well-known, maximum mutual information
(MMI) decoder (see, e.g., \cite[Section IV.B]{reliable_comm_review}).
The MCE decoder is equivalent to a decoder that estimates the unknown
PMF $P_{XY}$ for any candidate source block (generalized likelihood
ratio test). 
\end{itemize}
\end{itemize}
It is well known that the ML decoder, which depends on the source
statistics $P_{\mathbf{X}\mathbf{Y}}$, minimizes the error probability.
By contrast, the MCE decoder does not use $P_{\mathbf{X}\mathbf{Y}}$
at all. In the next theorem, we evaluate the random binning error
exponent of the ML decoder, and show that the MCE decoder also achieves
the same exponent, and thus it is a \emph{universal} decoder. This
exponent was initially derived in \cite{csiszar1980towards} (for
both decoders), but the proof here is simpler, and also shows that
the lower bound on the ML error exponent is tight for all rates. 
\begin{thm}
\label{thm:err exp}Let $\rho(\cdot)$ be a given rate function, and
let the ensemble of SW codes be as defined in Section \ref{sub:System Model}.
Then for both the ML decoder and the MCE decoder, the limit in \eqref{eq: error exponent definition}
exists and equals 
\begin{equation}
\bar{E}_{e}=\min_{Q_{XY}}\left\{ D(Q_{XY}||P_{XY})+\left[\rho(Q_{X})-H(Q_{X|Y}|Q_{Y})\right]_{+}\right\} .\label{eq: error expoenent expression in the theorem}
\end{equation}
\end{thm}
\begin{IEEEproof}
Suppose that $(\mathbf{x},\mathbf{y})$ was emitted from the source
and its joint type is $Q_{XY}=\hat{Q}_{\mathbf{xy}}$. Let the marginals
and conditional types be $Q_{X}=\hat{Q}_{\mathbf{x}}$, $Q_{Y}=\hat{Q}_{\mathbf{y}}$,
and $Q_{X|Y}=\hat{Q}_{\mathbf{x}|\mathbf{y}}$. 

For the ML decoder, let 
\[
\Lambda_{o}(\mathbf{x},\mathbf{y})\triangleq\left\{ \tilde{\mathbf{x}}\in{\cal X}^{n}:P_{\mathbf{X}|\mathbf{Y}}(\tilde{\mathbf{x}}|\mathbf{y})\geq P_{\mathbf{X}|\mathbf{Y}}(\mathbf{x}|\mathbf{y})\right\} .
\]
The conditional error probability, averaged over the random choice
of binning is 
\begin{alignat}{1}
\bar{\Pi}{}_{e,o}(\mathbf{x},\mathbf{y}) & \triangleq\P\left\{ \bigcup_{\tilde{\mathbf{x}}\in\{\Lambda_{o}(\mathbf{x},\mathbf{y})\cap{\cal T}_{n}(Q_{X})\}}S_{n}(\tilde{\mathbf{x}})=S_{n}(\mathbf{x})\right\} \\
 & \geq\frac{1}{2}\min\left\{ 1,e^{-n\rho(Q_{X})}\cdot\left|\Lambda_{o}(\mathbf{x},\mathbf{y})\cap{\cal T}_{n}(Q_{X})\right|\right\} \\
 & \geq\frac{1}{2}\min\left\{ 1,e^{-n\rho(Q_{X})}\cdot\left|{\cal T}_{n}(Q_{X|Y})\right|\right\} \\
 & \doteq\min\left\{ 1,\exp\left[-n\left(\rho(Q_{X})-H(Q_{X|Y}|Q_{Y})\right)\right]\right\} \\
 & =\exp\left[-n\left[\rho(Q_{X})-H(Q_{X|Y}|Q_{Y})\right]_{+}\right]
\end{alignat}
where the first inequality is due to Lemma \ref{lem: Union bound tightness},
and the fact that the bin indices are drawn independently in a given
type class, and the second inequality is because for any pair $\left(\tilde{\mathbf{x}},\mathbf{y}\right)\in{\cal T}_{n}(Q_{X|Y})$,
we have that $\tilde{\mathbf{x}}\in{\cal T}_{n}(Q_{X})$ and that
$P_{\mathbf{X}|\mathbf{Y}}(\tilde{\mathbf{x}}|\mathbf{y})=P_{\mathbf{X}|\mathbf{Y}}(\mathbf{x}|\mathbf{y})$. 

For the MCE decoder, let 
\[
\Lambda_{u}(\mathbf{x},\mathbf{y})\triangleq\left\{ \tilde{\mathbf{x}}\in{\cal X}^{n}:H\left(\hat{Q}_{\tilde{\mathbf{x}}|\mathbf{y}}|Q_{Y}\right)\leq H\left(Q_{X|Y}|Q_{Y}\right)\right\} .
\]
Similarly,
\begin{alignat}{1}
\bar{\Pi}_{e,u}(\mathbf{x},\mathbf{y}) & \triangleq\P\left\{ \bigcup_{\tilde{\mathbf{x}}\in\{\Lambda_{u}(\mathbf{x},\mathbf{y})\cap{\cal T}_{n}(Q_{X})\}}S_{n}(\tilde{\mathbf{x}})=S_{n}(\mathbf{x})\right\} \\
 & \leq\min\left\{ 1,e^{-n\rho(Q_{X})}\cdot\left|\Lambda_{u}(\mathbf{x},\mathbf{y})\cap{\cal T}_{n}(Q_{X})\right|\right\} \\
 & \leq\min\left\{ 1,e^{-n\rho(Q_{X})}\cdot\left|\Lambda_{u}(\mathbf{x},\mathbf{y})\right|\right\} \\
 & \dot{\leq}\min\left\{ 1,\exp\left[-n\left(\rho(Q_{X})-H\left(Q_{X|Y}|Q_{Y}\right)\right)\right]\right\} \\
 & =\exp\left[-n\left[\rho(Q_{X})-H\left(Q_{X|Y}|Q_{Y}\right)\right]_{+}\right],
\end{alignat}
where the first inequality is by the union bound, and the following
equality is because the number of sequences in any conditional type
that belongs to $\Lambda_{u}(\mathbf{x},\mathbf{y})$ is exponentially
upper bounded by $e^{nH(Q_{X|Y}|Q_{Y})}$ and the number of joint
types is polynomial $|{\cal P}_{n}({\cal X}\times{\cal Y})|\leq(n+1)^{|{\cal X}||{\cal Y}|}$
. 

It can be seen that on the exponential scale, the lower bound on $\bar{\Pi}{}_{e,o}(\mathbf{x},\mathbf{y})$
and the upper bound on $\bar{\Pi}_{e,u}(\mathbf{x},\mathbf{y})$ are
identical. Thus, when taking expectation w.r.t. the i.i.d. source
$P_{XY}$, the resulting asymptotic bounds on the error probability
are identical (lower bound for the ML decoder, and upper bound for
the MCE decoder). Moreover, since the ML decoder minimizes the error
probability, taking expectation w.r.t. $P_{XY}$ we get 
\begin{alignat}{1}
\E\left\{ \exp\left[-n\left[\rho(\hat{Q}_{\mathbf{X}})-H(\hat{Q}_{\mathbf{X}|\mathbf{Y}}|\hat{Q}_{\mathbf{Y}})\right]_{+}\right]\right\}  & \dot{\leq}\E\left\{ \bar{\Pi}_{e,o}(\mathbf{X},\mathbf{Y})\right\} \\
 & \leq\E\left\{ \bar{\Pi}_{e,u}(\mathbf{X},\mathbf{Y})\right\} \\
 & \dot{\leq}\E\left\{ \exp\left[-n\left[\rho(\hat{Q}_{\mathbf{X}})-H(\hat{Q}_{\mathbf{X}|\mathbf{Y}}|\hat{Q}_{\mathbf{Y}})\right]_{+}\right]\right\} 
\end{alignat}
so the asymptotic average error probability of both the ML decoder
and the MCE decoder is 
\begin{alignat}{1}
\overline{p}_{e,n} & \doteq\E\left\{ \exp\left[-n\left[\rho(\hat{Q}_{\mathbf{X}})-H(\hat{Q}_{\mathbf{X}|\mathbf{Y}}|\hat{Q}_{\mathbf{Y}})\right]_{+}\right]\right\} \\
 & =\sum_{Q_{XY}\in{\cal P}_{n}({\cal X}\times{\cal Y})}\P\left(\hat{Q}_{\mathbf{XY}}=Q_{XY}\right)\cdot\exp\left[-n\left[\rho(Q_{X})-H(Q_{X|Y}|Q_{Y})\right]_{+}\right]\\
 & \doteq\sum_{Q_{XY}\in{\cal P}_{n}({\cal X}\times{\cal Y})}\exp\left[-n\cdot D(Q_{XY}||P_{XY})-n\left[\rho(Q_{X})-H(Q_{X|Y}|Q_{Y})\right]_{+}\right]\\
 & \doteq\exp\left[-n\cdot\min_{Q_{XY}\in{\cal P}_{n}({\cal X}\times{\cal Y})}\left\{ D(Q_{XY}||P_{XY})+\left[\rho(Q_{X})-H(Q_{X|Y}|Q_{Y})\right]_{+}\right\} \right]
\end{alignat}
where the last inequality is again because $|{\cal P}_{n}({\cal X}\times{\cal Y})|\leq(n+1)^{|{\cal X}||{\cal Y}|}$.
Since the optimal value of the minimization problem inside the exponent
is clearly finite, and the minimization argument is a continuous function,
then 
\[
\overline{p}_{e,n}\doteq\exp\left[-n\cdot\min_{Q_{XY}}\left\{ D(Q_{XY}||P_{XY})+\left[\rho(Q_{X})-H(Q_{X|Y}|Q_{Y})\right]_{+}\right\} \right].
\]

\end{IEEEproof}

\section{\label{sec: wakly correlated sources}}

Consider the case of very weakly correlated sources%
\footnote{In channel coding, this is referred to as \emph{``very noisy channel''}
\cite[Section 3.4]{viterbi2009principles}. %
}, namely
\[
P_{Y|X}(y|x)=P_{Y}(y)\cdot(1+\epsilon_{xy})
\]
where for all $x\in{\cal X}$ we have $\sum_{y\in{\cal Y}}\epsilon_{xy}=0$
and $|\epsilon_{xy}|\ll1$ for all $(x,y)\in({\cal X},{\cal Y})$.
Consider again the minimization problem in \eqref{eq: random binning rate function expression}
\[
\min_{Q_{Y|X}:D(Q_{X}\times Q_{Y|X}||P_{XY})\leq\EE}\left\{ I(Q_{X}\times Q_{Y|X})+D(Q_{Y|X}||P_{Y|X}|Q_{X})\right\} 
\]
which from Lemma \ref{lem: Divergence output marginal minimization}
is equivalent to 
\[
\min_{\tilde{Q}_{Y}}\min_{Q_{Y|X}:D(Q_{X}\times Q_{Y|X}||P_{XY})\leq\EE}\left\{ D(Q_{Y|X}||\tilde{Q}_{Y}|Q_{X})+D\left(Q_{Y|X}||P_{Y|X}|Q_{X}\right)\right\} .
\]
Now, notice that if $X$ and $Y$ are independent, then the optimal
solution is $\tilde{Q}_{Y}^{*}=Q_{Y|X}=P_{Y}$ for all $x\in{\cal X}$
and both divergences vanish. A continuity argument then implies that
for the low dependence case, the two divergences at the optimal solution
are close to $0$. Therefore, we can use the following Euclidean approximation
\cite[Theorem 4.1]{csiszar2004information}: For two PMFs $P_{X},Q_{X}$
such that $\supp(P_{x})={\cal X}$ and $Q_{X}\approx P_{X}$ we have
that 
\[
D(Q_{X}||P_{X})\approx\frac{1}{2}\chi^{2}(Q_{X},P_{X})\triangleq\frac{1}{2}\sum_{x\in{\cal X}}\frac{(Q_{X}(x)-P_{X}(x))^{2}}{P_{X}(x)}.
\]
 Moreover, for another PMF $\tilde{P}_{X}$, if $\tilde{P}_{X}\approx P_{X}$
then 
\begin{equation}
D(Q_{X}||P_{X})\approx\frac{1}{2}\sum_{x\in{\cal X}}\frac{(Q_{X}(x)-P_{X}(x))^{2}}{\tilde{P}_{X}(x)}\label{eq: Divergence Euclidian approximation}
\end{equation}
 which also shows that $D(P_{X}||Q_{X})\approx D(Q_{X}||P_{X})$.
Now, the objective function of the minimization problem can be approximated
as
\begin{align}
 & D(Q_{Y|X}||\tilde{Q}_{Y}|Q_{X})+D(Q_{Y|X}||P_{Y|X}|Q_{X})\nonumber \\
 & \approx\frac{1}{2}\E_{Q_{X}}\left\{ \sum_{y\in{\cal Y}}\frac{(Q_{Y|X}(y|X)-\tilde{Q}_{Y}(y))^{2}}{\tilde{Q}_{Y}(y)}+\sum_{y\in{\cal Y}}\frac{\left(Q_{Y|X}(y|X)-P_{Y|X}(y|x)\right)^{2}}{P_{Y|X}(y|X)}\right\} \\
 & \approx\frac{1}{2}\E_{Q_{X}}\left\{ \sum_{y\in{\cal Y}}\frac{(Q_{Y|X}(y|X)-\tilde{Q}_{Y}(y))^{2}+\left(Q_{Y|X}(y|X)-P_{Y|X}(y|X)\right)^{2}}{P_{Y}(y)}\right\} 
\end{align}
and similarly the constraint $Q_{Y|X}\in{\cal A}_{\s[rb]}$ is approximated
by 
\begin{equation}
\frac{1}{2}\cdot\E_{Q_{X}}\left\{ \sum_{y\in{\cal Y}}\frac{\left(Q_{Y|X}(y|X)-P_{Y|X}(y|X)\right)^{2}}{P_{Y}(y)}\right\} \leq\EE-D(Q_{X}||P_{X}).\label{eq:optimal rate weak correlation constraint}
\end{equation}
The Lagrangian for a given $\tilde{Q}_{Y}$ (ignoring positivity constraints
for the moment) is 
\begin{eqnarray}
L\left(Q_{Y|X},\lambda,\mu_{x}\right) & =\frac{1}{2}\cdot & \sum_{x\in{\cal X}}Q_{X}(x)\sum_{y\in{\cal Y}}\frac{(Q_{Y|X}-\tilde{Q}_{Y}(y))^{2}+(1+\lambda)\left(Q_{Y|X}(y|x)-P_{Y|X}(y|x)\right)^{2}}{P_{Y}(y)}\nonumber \\
 &  & +\sum_{x\in{\cal X}}\mu_{x}\sum_{y\in{\cal Y}}Q_{Y|X}(y|x)
\end{eqnarray}
with $\lambda>0$ and $\mu_{x}\in\mathbb{R}$ for $x\in{\cal X}$.
Differentiating w.r.t. some $Q_{Y|X}(y'|x')$ for $x'\in X,y'\in{\cal Y}$
we have 
\[
\frac{\partial L}{\partial Q_{Y|X}(y'|x')}=\frac{1}{2}\cdot Q_{X}(x')\cdot\frac{2\left(Q_{Y|X}(y'|x')-\tilde{Q}_{Y}(y')\right)+2(1+\lambda)\left(Q_{Y|X}(y'|x')-P_{Y|X}(y'|x')\right)}{P_{Y}(y')}+\mu_{x'}
\]
and equating the derivative to zero in this case is equivalent to
\[
\frac{Q_{Y|X}(y'|x')-\tilde{Q}_{Y}(y')+(1+\lambda)\left[Q_{Y|X}(y'|x')-P_{Y|X}(y'|x')\right]}{P_{Y}(y')}+\mu'_{x'}=0.
\]
Thus, for some $\lambda>0$
\[
Q_{Y|X}^{*}(y|x)=\frac{1+\lambda}{2+\lambda}P_{Y|X}(y|x)+\frac{1}{2+\lambda}\tilde{Q}_{Y}(y)-\frac{\mu'_{x}}{2+\lambda}P_{Y}(y).
\]
It can be easily seen that $\mu'_{x}=0$ for all $x\in{\cal X}$ so
\[
Q_{Y|X}^{*}=\alpha P_{Y|X}+(1-\alpha)\tilde{Q}_{Y}
\]
for some $\alpha=\frac{1+\lambda}{2+\lambda}$, where $\alpha$ is
either chosen to satisfy the constraint or $\alpha=\nicefrac{1}{2}$.
It is evident that indeed the solution satisfies the positivity constraints.
Now, for any given $\alpha$ the resulting value of the optimization
problem is 
\[
\left[\frac{\alpha^{2}+(\alpha-1)^{2}}{2}\right]\sum_{x\in{\cal X}}Q_{X}(x)\sum_{y\in{\cal Y}}\frac{\left[P_{Y|X}(y|x)-\tilde{Q}_{Y}(y)\right]^{2}}{P_{Y}(y)}
\]
and by differentiating w.r.t. some $\tilde{Q}_{Y}(y')$ for $y'\in{\cal Y}$
we have 
\[
\left[\frac{\alpha^{2}+(\alpha-1)^{2}}{2}\right]\cdot\sum_{x\in{\cal X}}\frac{Q_{X}(x)}{P_{Y}(y')}\left[-2\left(P_{Y|X}(y'|x)-\tilde{Q}_{Y}(y')\right)\right]
\]
and equating to zero we obtain that the optimal solution is 
\[
\tilde{Q}_{Y}^{*}(y)=\sum_{x\in{\cal X}}Q_{X}(x)P_{Y|X}(y|x).
\]
Notice that the optimal solution $\tilde{Q}_{Y}^{*}$ does not depend
on $\alpha$. Thus, for a given $\EE\geq D(Q_{X}||P_{X})$ the optimal
value of $\alpha$ is given by $\alpha^{*}\approx\max(\tilde{\alpha},\nicefrac{1}{2})$
where $\tilde{\alpha}$ achieves equality in \eqref{eq:optimal rate weak correlation constraint},
\[
\tilde{\alpha}=1-\sqrt{\frac{\EE-D(Q_{X}||P_{X})}{\frac{1}{2}\sum_{x\in{\cal X}}Q_{X}(x)\sum_{y\in{\cal Y}}\frac{\left(P_{Y|X}(y|x)-\tilde{Q}_{Y}^{*}(y)\right)^{2}}{P_{Y}(y)}}}\approx1-\sqrt{\frac{\EE-D(Q_{X}||P_{X})}{D(P_{Y|X}||\tilde{Q}_{Y}^{*}|Q_{X})}}
\]
using again \eqref{eq: Divergence Euclidian approximation}. Then,
in the case of very weakly correlated sources, the optimal rate function
can be approximated by
\begin{equation}
\rho_{\s[rb]}(Q_{X},\EE)\approx\EE+H(Q_{X})-D(Q_{X}||P_{X})-\left[\alpha^{*2}+(\alpha^{*}-1)^{2}\right]D(P_{Y|X}||\tilde{Q}_{Y}^{*}|Q_{X})\label{eq: optimal rate approximation low dependency}
\end{equation}
where $\alpha^{*}$ is given analytically as a function of $\EE$.
In addition, similar approximations for the unconstrained minimization
problem \eqref{eq: random binning optimal rate function affine part}
show that 
\[
D(Q'_{Y|X}||P_{Y|X}|Q_{X})\approx\frac{1}{4}D(P_{Y|X}||\tilde{Q}_{Y}^{*}|Q_{X}).
\]
Thus, for $D(Q_{X}||P_{X})\leq\EE\leq D(Q_{X}||P_{X})+\frac{1}{4}D(P_{Y|X}||\tilde{Q}_{Y}^{*}|Q_{X})$
we have $\tilde{\alpha}\leq\nicefrac{1}{2}$ and by substituting $\tilde{\alpha}$
in \eqref{eq: optimal rate approximation low dependency} we obtain
\begin{eqnarray}
\rho_{\s[rb]}(Q_{X},\EE) & \approx & H(Q_{X})-\EE+D(Q_{X}||P_{X})\nonumber \\
 &  & -D(P_{Y|X}||\tilde{Q}_{Y}^{*}|Q_{X})+2\sqrt{D(P_{Y|X}||\tilde{Q}_{Y}^{*}|Q_{X})\left(\EE-D(Q_{X}||P_{X})\right)}.\label{eq: random binning rate function weakly correlated}
\end{eqnarray}
For $\rho_{\s[sp]}(Q_{X},\EE)$ the analysis is similar, and in this
case $\alpha^{*}\in(0,1)$, so we obtain the exact same expression
as in \eqref{eq: random binning rate function weakly correlated},
but this time it is valid for the entire range $D(Q_{X}||P_{X})\leq\EE\leq D(Q_{X}\times(Q_{X}\times P_{Y|X})_{Y}||P_{XY})$. 

Notice that $D(P_{Y|X}||\tilde{Q}_{Y}^{*}|Q_{X})$ is the mutual information
of the joint distribution $Q_{X}\times P_{Y|X}$ and thus is a measure
of the independence between $X$ and $Y$. As $D(P_{Y|X}||\tilde{Q}_{Y}^{*}|Q_{X})\to0$
then $X$ and $Y$ become ``more'' independent, and then the rate
function $\rho_{\s[rb]}(Q_{X},\EE)$ is affine for almost the entire
range $\EE\geq D(Q_{X}||P_{X})$. Indeed, in this case, the main error
event is associated with ``bad binning'', i.e. at least two source
blocks of the same type are mapped to the same bin by the random binning
procedure.

\section{\label{sec: Useful Lemmas}}

In this appendix, we provide several useful lemmas. 
\begin{lem}
\label{lem:divergence increasing for a ray}Let $P,Q$ be two PMFs
over some alphabet ${\cal X}$ such that $\supp(P)=\supp(Q)={\cal X}$,
$P\neq Q$, and 
\[
Q_{\alpha}\teq(1-\alpha)P+\alpha Q.
\]
Also, let $\alpha_{max}=\max\{\alpha:Q_{\alpha}\in{\cal Q}({\cal X})\}$.
Then, $D(Q_{\alpha}||P)$ is a strictly increasing function of $\alpha$
for $\alpha\in(0,\alpha_{max})$.\end{lem}
\begin{IEEEproof}
Let $0<\alpha_{1}<\alpha_{2}\leq\alpha_{max}$. Then,
\begin{alignat}{1}
Q_{\alpha_{1}} & =(1-\alpha_{1})P+\alpha_{1}Q\\
 & =\frac{\alpha_{1}}{\alpha_{2}}\left(\frac{\alpha_{2}}{\alpha_{1}}-\alpha_{2}+1-\frac{\alpha_{1}}{\alpha_{1}}\right)P+\frac{\alpha_{1}}{\alpha_{2}}\alpha_{2}Q\\
 & =\frac{\alpha_{1}}{\alpha_{2}}\left(1-\alpha_{2}+\frac{\alpha_{2}-\alpha_{1}}{\alpha_{1}}\right)P+\frac{\alpha_{1}}{\alpha_{2}}\alpha_{2}Q\\
 & =\frac{\alpha_{1}}{\alpha_{2}}\left[(1-\alpha_{2})P+\alpha_{2}Q\right]+\frac{(\alpha_{2}-\alpha_{1})}{\alpha_{2}}P\\
 & =\frac{\alpha_{1}}{\alpha_{2}}Q_{\alpha_{2}}+\left(1-\frac{\alpha_{1}}{\alpha_{2}}\right)P
\end{alignat}
thus $Q_{\alpha_{1}}$is a convex combination of $Q_{\alpha_{2}}$
and $P$ with coefficient $\gamma\teq\frac{\alpha_{1}}{\alpha_{2}}$,
and $0<\gamma<1$. Now, since divergence is strictly convex function
then
\begin{alignat}{1}
D(Q_{\alpha_{1}}||P) & =D(\gamma Q_{\alpha_{2}}+(1-\gamma)P||P)\\
 & <\gamma D(Q_{\alpha_{2}}||P)+(1-\gamma)D(P||P)\\
 & =\gamma D(Q_{\alpha_{2}}||P)\\
 & <D(Q_{\alpha_{2}}||P)
\end{alignat}
and thus $D(Q_{\alpha}||P)$ is strictly increasing in $\alpha$.\end{IEEEproof}
\begin{lem}
\label{lem: concave of min problem}Let $f_{i}(z):\mathbb{R}^{N}\to\mathbb{R}$
be convex functions for $i=1,2$. Consider the optimization problem
\[
W(E)=\min_{f_{1}(z)\leq E}f_{2}(z).
\]
assuming that the constraint is feasible for some interval $E\in{\cal J}$.
Then $W(E)$ is a convex function of $E$ in \textup{${\cal J}$ }\textup{\emph{and
}}\textup{$E-W(E)$ }\textup{\emph{is a concave function}}\emph{ $E$
in }\textup{\emph{${\cal J}$}}\textup{. }\end{lem}
\begin{IEEEproof}
This is a standard result. For example, in \cite[Theorem 3]{blahut1974hypothesis},
this theorem is proved for the case that $f_{1}$ and $f_{2}$ are
information divergences. The proof may be used verbatim for any convex
functions.\end{IEEEproof}
\begin{lem}
\label{lem: Divergence output marginal minimization}Let $P_{X}\times P_{Y|X}$
be a given joint distribution over ${\cal X}\times{\cal Y}$. Then
the distribution $Q_{Y}$ that minimizes $D(P_{X}\times P_{Y|X}||P_{X}\times Q_{Y})$
is the marginal distribution $Q_{Y}^{*}$ corresponding to $P_{Y|X}$
namely, $Q_{Y}^{*}(y)=\sum_{x}P_{X}(x)P_{Y|X}(y|x)$.\end{lem}
\begin{IEEEproof}
See \cite[Lemma 10.8.1]{Cover:2006:EIT:1146355}.\end{IEEEproof}
\begin{cor}
\label{cor:Divergence output marginal minimization corollary (less than 1)}Let
$P_{X}\times P_{Y|X}$ be a given joint distribution over ${\cal X}\times{\cal Y}$.
Then the vector $Q_{Y}\in\mathbb{R}^{|{\cal Y}|}$ that minimizes
$D(P_{X}\times P_{Y|X}||P_{X}\times Q_{Y})$%
\footnote{Notice that the divergence is well defined even if $\{Q_{Y}\}$ do
not sum exactly to $1$.%
} under the constraint $\sum_{y\in{\cal Y}}Q_{Y}(y)\leq1$ and $Q_{Y}(y)\geq0$
for all $y\in{\cal Y}$, is $Q_{Y}^{*}(y)=\sum_{x}P_{X}(x)P_{Y|X}(y|x).$\end{cor}
\begin{IEEEproof}
Suppose that the minimizer vector $Q_{Y}^{*}$ satisfies $\sum_{y\in{\cal Y}}Q_{Y}^{*}(y)<1$.
Then for some $y'\in{\cal Y}$, we can increase $Q_{Y}^{*}(y')$ by
$1-\sum_{y\in{\cal Y}}Q_{Y}^{*}(y)>0$ and obtain $\bar{Q}_{Y}$ which
satisfies $\sum_{y\in{\cal Y}}\overline{Q}_{Y}(y)=1$. But,
\begin{alignat}{1}
D(P_{X}\times P_{Y|X}||P_{X}\times Q_{Y}^{*}) & =\sum_{x,y}P_{XY}(x,y)\log\frac{P_{XY}(x,y)}{P_{X}(x)Q_{Y}^{*}(y)}\\
 & =\sum_{x,y\neq y'}P_{XY}(x,y)\log\frac{P_{XY}(x,y)}{P_{X}(x)Q_{Y}^{*}(y)}\nonumber \\
 & +\sum_{x}P_{XY}(x,y')\log\frac{P_{XY}(x,y')}{P_{X}(x)Q_{Y}^{*}(y')}\\
 & >\sum_{x,y\neq y'}P_{XY}(x,y)\log\frac{P_{XY}(x,y)}{P_{X}(x)\overline{Q}_{Y}(y)}\nonumber \\
 & +\sum_{x}P_{XY}(x,y')\log\frac{P_{XY}(x,y')}{P_{X}(x)\overline{Q}_{Y}(y')}
\end{alignat}
and this contradicts the fact that $Q_{Y}^{*}$ is a minimizer. Thus,
we must have $\sum_{y\in{\cal Y}}Q_{Y}^{*}(y)=1$. In this case, Lemma
\ref{lem: Divergence output marginal minimization} shows that the
optimal solution is $Q_{Y}^{*}(y)=\sum_{x}P_{X}(x)P_{Y|X}(y|x).$\end{IEEEproof}
\begin{lem}
\label{lem: Divergence output marginal minimization backwards}Let
$P_{X}\times P_{Y|X}$ be a given joint distribution over ${\cal X}\times{\cal Y}$.
Then the distribution $Q_{Y}$ that minimizes $D(P_{X}\times Q_{Y}||P_{X}\times P_{Y|X})$
is the marginal distribution $Q_{Y}^{*}$ corresponding to $P_{Y|X}$
namely, $Q_{Y}^{*}(y)=\sum_{x}P_{X}(x)P_{Y|X}(y|x)$.\end{lem}
\begin{IEEEproof}
We have
\begin{alignat}{1}
D(P_{X}\times Q_{Y}||P_{X}\times P_{Y|X}) & =D(Q_{Y}||P_{Y|X}|P_{X})\\
 & =-\sum_{y}Q_{Y}(y)\sum_{x}P_{X}(x)\log\frac{P_{Y|X}(y|x)}{Q_{Y}(y)}\\
 & \geq-\sum_{y}Q_{Y}(y)\log\frac{\sum_{x}P_{X}(x)P_{Y|X}(y|x)}{Q_{Y}(y)}\\
 & =-\sum_{y}Q_{Y}(y)\log\frac{Q_{Y}^{*}(y)}{Q_{Y}(y)}\\
 & =D(Q_{Y}||Q_{Y}^{*})\\
 & \geq0
\end{alignat}
and equality is obtained for $Q_{Y}=Q_{Y}^{*}$.
\end{IEEEproof}
This following lemma is stated and proved in \cite[Section 3A.1, inequality (k)]{viterbi2009principles}. 
\begin{lem}[A variant of Minkowski's inequality]
\label{lem: Variant of Minkowski inequality} Let $0\leq\lambda\le1$,
let $Q_{X}$ be a PMF over a finite alphabet ${\cal X}$, and let
$\{a_{x}(i)\}$ be a set of non-negative numbers for $1\leq i\leq I$
and $x\in{\cal X}$ . Then, 
\[
\sum_{i=1}^{I}\left(\sum_{x\in{\cal X}}^{{\cal }}Q_{X}(x)a_{x}(i)^{\lambda}\right)^{\nicefrac{1}{\lambda}}\leq\left(\sum_{x\in{\cal X}}^{{\cal }}Q_{X}(x)\left(\sum_{i=1}^{I}a_{x}(i)\right)^{\lambda}\right)^{\nicefrac{1}{\lambda}}
\]

\end{lem}

\begin{lem}
\label{lem: Monotinicty of divergence for exponential families}Let
$P_{1},P_{2}$ be two PMFs over some alphabet ${\cal X}$, such that
$\supp(P_{2})\subseteq\supp(P_{1})$. Define
\[
Q_{\alpha}(x)\teq\psi_{\alpha}P_{1}^{\alpha}(x)P_{2}^{1-\alpha}(x),
\]
where $\alpha\in\left[0,1\right]$ and $\psi_{\alpha}$ is a normalization
factor such that $Q_{\alpha}\in{\cal Q}({\cal X})$. Then, $D\left(Q_{\alpha}||P_{1}\right)$
is a continuous function of $\alpha$ whose limit, as $\alpha\to0$,
is $D(Q'||P)$ where
\[
P_{2}'(x)=\begin{cases}
\psi'\cdot P_{2}(x) & P_{1}(x)>0\\
0 & P_{1}(x)=0
\end{cases}
\]
for some normalization factor $\psi'$. Moreover, $D\left(Q_{\alpha}||P_{1}\right)$
is monotonic, strictly decreasing function of $\alpha$ unless $P_{2}'=P_{1}$. \end{lem}
\begin{IEEEproof}
This is \cite[Problem 2.14]{csiszar2011information} but for completeness,
we provide a proof here based on \cite{blahut1974hypothesis}. First,
notice that $P_{1}(x)=0\Rightarrow Q_{\alpha}(x)=0$ and thus all
$x\in{\cal X}$ such that $P_{1}(x)=0$ are immaterial to the divergence,
assuming the regular convention, that any summand of the form $0\cdot\frac{0}{0}$
is $0$. Thus it may be assumed w.l.o.g. that $\supp(P_{1})={\cal X}$
and $P_{2}'=P_{2}$.

\emph{\uline{Continuity:}} Since $\supp(P_{1})={\cal X}$ then
$D\left(Q_{\alpha}||P_{1}\right)$ is a continuous function of $Q_{\alpha}$
in ${\cal Q}({\cal X})$. As $Q_{\alpha}$ is a continuous function
of $\alpha$ we get that $D\left(Q_{\alpha}||P_{1}\right)$ is a continuous
function of $\alpha$. 

\emph{\uline{Limit for $\alpha\to0$:}} Since $\supp(P_{1})={\cal X}$
we get that $\supp(Q_{\alpha})=\supp(P_{2})$. It is easily seen that
as $\alpha\to0$ we have $Q_{\alpha}(x)\to P_{2}(x)$.

\emph{\uline{Monotonicity:}} Consider the following optimization
problem
\[
W(E)=\min_{D(Q||P_{2})\leq E}D(Q||P_{1}).
\]
Standard Lagrange techniques, as used in this paper, show that the
optimal solution is 
\[
Q(x)=\psi P_{1}^{\frac{1}{1+\lambda}}(x)P_{2}^{\frac{\lambda}{1+\lambda}}(x)
\]
where $\lambda\geq0$ is either chosen such that the constraint is
satisfied with equality, or $\lambda=0$. When $\lambda>0$ defining
$\alpha\teq\frac{1}{1+\lambda}$ we get $W(E)=D(Q_{\alpha}||P_{1})$.
Thus, if we show that $W(E)$ is a monotonic increasing function of
$\lambda$, then the proof is finished because $\alpha$ is an increasing
function of $\lambda$. To this end, notice that:
\begin{enumerate}
\item $W(E)$ is a strictly decreasing function of $E$.
\item Using Lemma \ref{lem: concave of min problem}, $W(E)$ is a strictly
convex function of $E$ which implies that $\frac{dW(E)}{dE}$ is
a strictly increasing function of $E$. 
\item We have that 
\[
\frac{dW(E)}{dE}=-\lambda.
\]

To see this relation, suppose that $\lambda$ is chosen to satisfy
the constraint $E$. Then, we get
\begin{alignat}{1}
W(E) & =D\left(Q_{\alpha}||P_{1}\right)\\
 & =\sum_{x\in{\cal X}}Q_{\alpha}(x)\cdot\log\frac{Q_{\alpha}(x)}{P_{1}(x)}\\
 & =\sum_{x\in{\cal X}}Q_{\alpha}(x)\cdot\log\frac{P_{2}^{\lambda}(x)}{Q_{\alpha}^{\lambda}(x)}+\sum_{x\in{\cal X}}Q_{\alpha}(x)\cdot\log\frac{Q_{\alpha}^{1+\lambda}(x)}{P_{1}(x)\cdot P_{2}^{\lambda}(x)}\\
 & =-\lambda E-(\lambda+1)\log(\psi)\\
 & =-\lambda E-(\lambda+1)\log\left(\sum_{x\in{\cal X}}P_{1}^{\frac{1}{1+\lambda}}(x)P_{2}^{\frac{\lambda}{1+\lambda}}(x)\right).
\end{alignat}
When differentiating we obtain
\[
\frac{dW(E)}{dE}=-\lambda-E\frac{d\lambda}{dE}-\frac{d\lambda}{dE}\cdot\frac{d}{d\lambda}\left[(\lambda+1)\log\left(\sum_{x\in{\cal X}}P_{1}^{\frac{1}{1+\lambda}}(x)P_{2}^{\frac{\lambda}{1+\lambda}}(x)\right)\right],
\]
and because $\frac{d}{d\lambda}\left[(\lambda+1)\log\left(\sum_{x\in{\cal X}}P_{1}^{\frac{1}{1+\lambda}}(x)P_{2}^{\frac{\lambda}{1+\lambda}}(x)\right)\right]=-E$
we obtain the desired result. 

\end{enumerate}

These properties imply that as $E$ increases $W(E)$ decreases and
$\frac{dW(E)}{dE}=-\lambda$ increases. This results that $W(E)$
is a monotonic increasing function of $\lambda$, and concludes the
proof.

Strict monotonicity can be verified by noticing that all monotonicity
relations are strict.

\end{IEEEproof}

\bibliographystyle{plain}
\bibliography{Slepian_Wolf}

\end{document}